\documentclass[twoside]{IEEEtran}
\pdfoutput=1 

\usepackage{xspace}
\usepackage[nosort]{cite}        
\usepackage{url} 
\usepackage[cmex10]{amsmath}
\usepackage{bbm}
\usepackage{graphicx}
\usepackage{paralist}
\usepackage{fancyref}
\usepackage{svninfo}
\usepackage{graphicx}
\usepackage{bm}
\usepackage[printonlyused]{acronym}
\usepackage[stretch=16,shrink=16,step=4]{microtype}

\newlength{\figwidth}
\makeatletter
\def\@IEEEinterspaceratioM{0.265}
\def\@IEEEinterspaceMINratioM{0.1651}
\def\@IEEEinterspaceMAXratioM{0.38}

\def\@IEEEinterspaceratioB{0.31}
\def\@IEEEinterspaceMINratioB{0.19}
\def\@IEEEinterspaceMAXratioB{0.38}
\@IEEEtunefonts
\makeatother

\usepackage{amsmath}
\usepackage{amssymb}
\usepackage{amsfonts}
\usepackage{amsthm}
\usepackage{mathrsfs}
\usepackage{xspace}
\usepackage{bm}
\usepackage{fancyref}
\usepackage{textcomp}
\usepackage[Symbol]{upgreek}

\newcommand{\safemath}[2]{\newcommand{#1}{\ensuremath{#2}\xspace}}

\newcommand{\ssa}{\mathsf{a}}
\newcommand{\ssb}{\mathsf{b}}
\newcommand{\ssc}{\mathsf{c}}
\newcommand{\ssd}{\mathsf{d}}
\newcommand{\sse}{\mathsf{e}}
\newcommand{\ssf}{\mathsf{f}}
\newcommand{\ssg}{\mathsf{g}}
\newcommand{\ssh}{\mathsf{h}}
\newcommand{\ssi}{\mathsf{i}}
\newcommand{\ssj}{\mathsf{j}}
\newcommand{\ssk}{\mathsf{k}}
\newcommand{\ssl}{\mathsf{l}}
\newcommand{\ssm}{\mathsf{m}}
\newcommand{\ssn}{\mathsf{n}}
\newcommand{\sso}{\mathsf{o}}
\newcommand{\ssp}{\mathsf{p}}
\newcommand{\ssq}{\mathsf{q}}
\newcommand{\ssr}{\mathsf{r}}
\newcommand{\sss}{\mathsf{s}}
\newcommand{\sst}{\mathsf{t}}
\newcommand{\ssu}{\mathsf{u}}
\newcommand{\ssv}{\mathsf{v}}
\newcommand{\ssw}{\mathsf{w}}
\newcommand{\ssx}{\mathsf{x}}
\newcommand{\ssy}{\mathsf{y}}
\newcommand{\ssz}{\mathsf{z}}

\safemath{\bmsa}{\bm{\ssa}}
\safemath{\bmsb}{\bm{\ssb}}
\safemath{\bmsc}{\bm{\ssc}}
\safemath{\bmsd}{\bm{\ssd}}
\safemath{\bmse}{\bm{\sse}}
\safemath{\bmsf}{\bm{\ssf}}
\safemath{\bmsg}{\bm{\ssg}}
\safemath{\bmsh}{\bm{\ssh}}
\safemath{\bmsi}{\bm{\ssi}}
\safemath{\bmsj}{\bm{\ssj}}
\safemath{\bmsk}{\bm{\ssk}}
\safemath{\bmsl}{\bm{\ssl}}
\safemath{\bmsm}{\bm{\ssm}}
\safemath{\bmsn}{\bm{\ssn}}
\safemath{\bmso}{\bm{\sso}}
\safemath{\bmsp}{\bm{\ssp}}
\safemath{\bmsq}{\bm{\ssq}}
\safemath{\bmsr}{\bm{\ssr}}
\safemath{\bmss}{\bm{\sss}}
\safemath{\bmst}{\bm{\sst}}
\safemath{\bmsu}{\bm{\ssu}}
\safemath{\bmsv}{\bm{\ssv}}
\safemath{\bmsw}{\bm{\ssw}}
\safemath{\bmsx}{\bm{\ssx}}
\safemath{\bmsy}{\bm{\ssy}}
\safemath{\bmsz}{\bm{\ssz}}

\bmdefine{\bmualphad}{\upalpha}
\bmdefine{\bmubetad}{\upbeta}
\bmdefine{\bmuchid}{\upchi}
\bmdefine{\bmudeltad}{\updelta}
\bmdefine{\bmuepsilond}{\upepsilon}
\bmdefine{\bmuvarepsilond}{\upvarepsilon}
\bmdefine{\bmuetad}{\upeta}
\bmdefine{\bmugammad}{\upgamma}
\bmdefine{\bmuiotad}{\upiota}
\bmdefine{\bmukappad}{\upkappa}
\bmdefine{\bmulambdad}{\uplambda}
\bmdefine{\bmumud}{\upmu}
\bmdefine{\bmunud}{\upnu}
\bmdefine{\bmuomegad}{\upomega}
\bmdefine{\bmuphid}{\upphi}
\bmdefine{\bmuvarphid}{\upvarphi}
\bmdefine{\bmupid}{\uppi}
\bmdefine{\bmuvarpid}{\upvarpi}
\bmdefine{\bmupsid}{\uppsi}
\bmdefine{\bmurhod}{\uprho}
\bmdefine{\bmuvarrhod}{\upvarrho}
\bmdefine{\bmusigmad}{\upsigma}
\bmdefine{\bmuvarsigmad}{\upvarsigma}
\bmdefine{\bmutaud}{\uptau}
\bmdefine{\bmuthetad}{\uptheta}
\bmdefine{\bmuvarthetad}{\upvartheta}
\bmdefine{\bmuupsilond}{\upupsilon}
\bmdefine{\bmuxid}{\upxi}
\bmdefine{\bmuzetad}{\upzeta}

\safemath{\bmua}{\mathbf{a}}
\safemath{\bmub}{\mathbf{b}}
\safemath{\bmuc}{\mathbf{c}}
\safemath{\bmud}{\mathbf{d}}
\safemath{\bmue}{\mathbf{e}}
\safemath{\bmuf}{\mathbf{f}}
\safemath{\bmug}{\mathbf{g}}
\safemath{\bmuh}{\mathbf{h}}
\safemath{\bmui}{\mathbf{i}}
\safemath{\bmuj}{\mathbf{j}}
\safemath{\bmuk}{\mathbf{k}}
\safemath{\bmul}{\mathbf{l}}
\safemath{\bmum}{\mathbf{m}}
\safemath{\bmun}{\mathbf{n}}
\safemath{\bmuo}{\mathbf{o}}
\safemath{\bmup}{\mathbf{p}}
\safemath{\bmuq}{\mathbf{q}}
\safemath{\bmur}{\mathbf{r}}
\safemath{\bmus}{\mathbf{s}}
\safemath{\bmut}{\mathbf{t}}
\safemath{\bmuu}{\mathbf{u}}
\safemath{\bmuv}{\mathbf{v}}
\safemath{\bmuw}{\mathbf{w}}
\safemath{\bmux}{\mathbf{x}}
\safemath{\bmuy}{\mathbf{y}}
\safemath{\bmuz}{\mathbf{z}}

\safemath{\bmualpha}{\bmualphad}
\safemath{\bmubeta}{\bmubetad}
\safemath{\bmuchi}{\bumchid}
\safemath{\bmudelta}{\bmudeltad}
\safemath{\bmuepsilon}{\bmuepsilond}
\safemath{\bmuvarepsilon}{\bmuvarepsilond}
\safemath{\bmueta}{\bmuetad}
\safemath{\bmugamma}{\bmugammad}
\safemath{\bmuiota}{\bmuiotad}
\safemath{\bmukappa}{\bmukappad}
\safemath{\bmulambda}{\bmulambdad}
\safemath{\bmumu}{\bmumud}
\safemath{\bmunu}{\bmunud}
\safemath{\bmuomega}{\bmuomegad}
\safemath{\bmuphi}{\bmuphid}
\safemath{\bmuvarphi}{\bmuvarphid}
\safemath{\bmupi}{\bmupid}
\safemath{\bmuvarpi}{\bmuvarpid}
\safemath{\bmupsi}{\bmupsid}
\safemath{\bmurho}{\bmurhod}
\safemath{\bmuvarrho}{\bmuvarrhod}
\safemath{\bmusigma}{\bmusigmad}
\safemath{\bmuvarsigma}{\bmuvarsigmad}
\safemath{\bmutau}{\bmutaud}
\safemath{\bmutheta}{\bmuthetad}
\safemath{\bmuvartheta}{\bmuvarthetad}
\safemath{\bmuupsilon}{\bmuupsilond}
\safemath{\bmuxi}{\bmuxid}
\safemath{\bmuzeta}{\bmuzetad}

\bmdefine{\bmiad}{a}
\bmdefine{\bmibd}{b}
\bmdefine{\bmicd}{c}
\bmdefine{\bmidd}{d}
\bmdefine{\bmied}{e}
\bmdefine{\bmifd}{f}
\bmdefine{\bmigd}{g}
\bmdefine{\bmihd}{h}
\bmdefine{\bmiid}{i}
\bmdefine{\bmijd}{j}
\bmdefine{\bmikd}{k}
\bmdefine{\bmild}{l}
\bmdefine{\bmimd}{m}
\bmdefine{\bmind}{n}
\bmdefine{\bmiod}{o}
\bmdefine{\bmipd}{p}
\bmdefine{\bmiqd}{q}
\bmdefine{\bmird}{r}
\bmdefine{\bmisd}{s}
\bmdefine{\bmitd}{t}
\bmdefine{\bmiud}{u}
\bmdefine{\bmivd}{v}
\bmdefine{\bmiwd}{w}
\bmdefine{\bmixd}{x}
\bmdefine{\bmiyd}{y}
\bmdefine{\bmizd}{z}

\bmdefine{\bmialphad}{\alpha}
\bmdefine{\bmibetad}{\beta}
\bmdefine{\bmichid}{\chi}
\bmdefine{\bmideltad}{\delta}
\bmdefine{\bmiepsilond}{\epsilon}
\bmdefine{\bmivarepsilond}{\varepsilon}
\bmdefine{\bmietad}{\eta}
\bmdefine{\bmigammad}{\gamma}
\bmdefine{\bmiiotad}{\iota}
\bmdefine{\bmikappad}{\kappa}
\bmdefine{\bmivarkappad}{\varkappa}
\bmdefine{\bmilambdad}{\lambda}
\bmdefine{\bmimud}{\mu}
\bmdefine{\bminud}{\nu}
\bmdefine{\bmiomegad}{\omega}
\bmdefine{\bmiphid}{\phi}
\bmdefine{\bmivarphid}{\varphi}
\bmdefine{\bmipid}{\pi}
\bmdefine{\bmivarpid}{\varpi}
\bmdefine{\bmipsid}{\psi}
\bmdefine{\bmirhod}{\rho}
\bmdefine{\bmivarrhod}{\varrho}
\bmdefine{\bmisigmad}{\sigma}
\bmdefine{\bmivarsigmad}{\varsigma}
\bmdefine{\bmitaud}{\tau}
\bmdefine{\bmithetad}{\theta}
\bmdefine{\bmivarthetad}{\vartheta}
\bmdefine{\bmiupsilond}{\upsilon}
\bmdefine{\bmixid}{\xi}
\bmdefine{\bmizetad}{\zeta}

\safemath{\bmia}{\bmiad}
\safemath{\bmib}{\bmibd}
\safemath{\bmic}{\bmicd}
\safemath{\bmid}{\bmidd}
\safemath{\bmie}{\bmied}
\safemath{\bmif}{\bmifd}
\safemath{\bmig}{\bmigd}
\safemath{\bmih}{\bmihd}
\safemath{\bmii}{\bmiid}
\safemath{\bmij}{\bmijd}
\safemath{\bmik}{\bmikd}
\safemath{\bmil}{\bmild}
\safemath{\bmim}{\bmimd}
\safemath{\bmin}{\bmind}
\safemath{\bmio}{\bmiod}
\safemath{\bmip}{\bmipd}
\safemath{\bmiq}{\bmiqd}
\safemath{\bmir}{\bmird}
\safemath{\bmis}{\bmisd}
\safemath{\bmit}{\bmitd}
\safemath{\bmiu}{\bmiud}
\safemath{\bmiv}{\bmivd}
\safemath{\bmiw}{\bmiwd}
\safemath{\bmix}{\bmixd}
\safemath{\bmiy}{\bmiyd}
\safemath{\bmiz}{\bmizd}

\safemath{\bmialpha}{\bmialphad}
\safemath{\bmibeta}{\bmibetad}
\safemath{\bmichi}{\bmichid}
\safemath{\bmidelta}{\bmideltad}
\safemath{\bmiepsilon}{\bmiepsilond}
\safemath{\bmivarepsilon}{\bmivarepsilond}
\safemath{\bmieta}{\bmietad}
\safemath{\bmigamma}{\bmigammad}
\safemath{\bmiiota}{\bmiiotad}
\safemath{\bmikappa}{\bmikappad}
\safemath{\bmivarkappa}{\bmivarkappad}
\safemath{\bmilambda}{\bmilambdad}
\safemath{\bmimu}{\bmimud}
\safemath{\bminu}{\bminud}
\safemath{\bmiomega}{\bmiomegad}
\safemath{\bmiphi}{\bmiphid}
\safemath{\bmivarphi}{\bmivarphid}
\safemath{\bmipi}{\bmipid}
\safemath{\bmivarpi}{\bmivarpid}
\safemath{\bmipsi}{\bmipsid}
\safemath{\bmirho}{\bmirhod}
\safemath{\bmivarrho}{\bmivarrhod}
\safemath{\bmisigma}{\bmisigmad}
\safemath{\bmivarsigma}{\bmivarsigmad}
\safemath{\bmitau}{\bmitaud}
\safemath{\bmitheta}{\bmithetad}
\safemath{\bmivartheta}{\bmivarthetad}
\safemath{\bmiupsilon}{\bmiupsilond}
\safemath{\bmixi}{\bmixid}
\safemath{\bmizeta}{\bmizetad}

\bmdefine{\bmuDeltad}{\Updelta}
\bmdefine{\bmuGammad}{\Upgamma}
\bmdefine{\bmuLambdad}{\Uplambda}
\bmdefine{\bmuOmegad}{\Upomega}
\bmdefine{\bmuPhid}{\Upphi}
\bmdefine{\bmuPid}{\Uppi}
\bmdefine{\bmuPsid}{\Uppsi}
\bmdefine{\bmuSigmad}{\Upsigma}
\bmdefine{\bmuThetad}{\Uptheta}
\bmdefine{\bmuUpsilond}{\Upupsilon}
\bmdefine{\bmuXid}{\Upxi}

\safemath{\bmuA}{\mathbf{A}}
\safemath{\bmuB}{\mathbf{B}}
\safemath{\bmuC}{\mathbf{C}}
\safemath{\bmuD}{\mathbf{D}}
\safemath{\bmuE}{\mathbf{E}}
\safemath{\bmuF}{\mathbf{F}}
\safemath{\bmuG}{\mathbf{G}}
\safemath{\bmuH}{\mathbf{H}}
\safemath{\bmuI}{\mathbf{I}}
\safemath{\bmuJ}{\mathbf{J}}
\safemath{\bmuK}{\mathbf{K}}
\safemath{\bmuL}{\mathbf{L}}
\safemath{\bmuM}{\mathbf{M}}
\safemath{\bmuN}{\mathbf{N}}
\safemath{\bmuO}{\mathbf{O}}
\safemath{\bmuP}{\mathbf{P}}
\safemath{\bmuQ}{\mathbf{Q}}
\safemath{\bmuR}{\mathbf{R}}
\safemath{\bmuS}{\mathbf{S}}
\safemath{\bmuT}{\mathbf{T}}
\safemath{\bmuU}{\mathbf{U}}
\safemath{\bmuV}{\mathbf{V}}
\safemath{\bmuW}{\mathbf{W}}
\safemath{\bmuX}{\mathbf{X}}
\safemath{\bmuY}{\mathbf{Y}}
\safemath{\bmuZ}{\mathbf{Z}}

\safemath{\bmuZero}{\mathbf{0}}
\safemath{\bmuOne}{\mathbf{1}}

\safemath{\bmuDelta}{\bmuDeltad}
\safemath{\bmuGamma}{\bmuGammad}
\safemath{\bmuLambda}{\bmuLambdad}
\safemath{\bmuOmega}{\bmuOmegad}
\safemath{\bmuPhi}{\bmuPhid}
\safemath{\bmuPi}{\bmuPid}
\safemath{\bmuPsi}{\bmuPsid}
\safemath{\bmuSigma}{\bmuSigmad}
\safemath{\bmuTheta}{\bmuThetad}
\safemath{\bmuUpsilon}{\bmuUpsilond}
\safemath{\bmuXi}{\bmuXid}

\bmdefine{\bmiAd}{A}
\bmdefine{\bmiBd}{B}
\bmdefine{\bmiCd}{C}
\bmdefine{\bmiDd}{D}
\bmdefine{\bmiEd}{E}
\bmdefine{\bmiFd}{F}
\bmdefine{\bmiGd}{G}
\bmdefine{\bmiHd}{H}
\bmdefine{\bmiId}{I}
\bmdefine{\bmiJd}{J}
\bmdefine{\bmiKd}{K}
\bmdefine{\bmiLd}{L}
\bmdefine{\bmiMd}{M}
\bmdefine{\bmiOd}{N}
\bmdefine{\bmiPd}{O}
\bmdefine{\bmiQd}{P}
\bmdefine{\bmiRd}{R}
\bmdefine{\bmiSd}{S}
\bmdefine{\bmiTd}{T}
\bmdefine{\bmiUd}{U}
\bmdefine{\bmiVd}{V}
\bmdefine{\bmiWd}{W}
\bmdefine{\bmiXd}{X}
\bmdefine{\bmiYd}{Y}
\bmdefine{\bmiZd}{Z}

\bmdefine{\bmiDeltad}{\Delta}
\bmdefine{\bmiGammad}{\Gamma}
\bmdefine{\bmiLambdad}{\Lambda}
\bmdefine{\bmiOmegad}{\Omega}
\bmdefine{\bmiPhid}{\Phi}
\bmdefine{\bmiPid}{\Pi}
\bmdefine{\bmiPsid}{\Psi}
\bmdefine{\bmiSigmad}{\Sigma}
\bmdefine{\bmiThetad}{\Theta}
\bmdefine{\bmiUpsilond}{\Upsilon}
\bmdefine{\bmiXid}{\Xi}

\safemath{\bmiA}{\bmiAd}
\safemath{\bmiB}{\bmiBd}
\safemath{\bmiC}{\bmiCd}
\safemath{\bmiD}{\bmiDd}
\safemath{\bmiE}{\bmiEd}
\safemath{\bmiF}{\bmiFd}
\safemath{\bmiG}{\bmiGd}
\safemath{\bmiH}{\bmiHd}
\safemath{\bmiI}{\bmiId}
\safemath{\bmiJ}{\bmiJd}
\safemath{\bmiK}{\bmiKd}
\safemath{\bmiL}{\bmiLd}
\safemath{\bmiM}{\bmiMd}
\safemath{\bmiN}{\bmiNd}
\safemath{\bmiO}{\bmiOd}
\safemath{\bmiP}{\bmiPd}
\safemath{\bmiQ}{\bmiQd}
\safemath{\bmiR}{\bmiRd}
\safemath{\bmiS}{\bmiSd}
\safemath{\bmiT}{\bmiTd}
\safemath{\bmiU}{\bmiUd}
\safemath{\bmiV}{\bmiVd}
\safemath{\bmiW}{\bmiWd}
\safemath{\bmiX}{\bmiXd}
\safemath{\bmiY}{\bmiYd}
\safemath{\bmiZ}{\bmiZd}

\safemath{\bmiDelta}{\bmiDeltad}
\safemath{\bmiGamma}{\bmiGammad}
\safemath{\bmiLambda}{\bmiLambdad}
\safemath{\bmiOmega}{\bmiOmegad}
\safemath{\bmiPhi}{\bmiPhid}
\safemath{\bmiPi}{\bmiPid}
\safemath{\bmiPsi}{\bmiPsid}
\safemath{\bmiSigma}{\bmiSigmad}
\safemath{\bmiTheta}{\bmiThetad}
\safemath{\bmiUpsilon}{\bmiUpsilond}
\safemath{\bmiXi}{\bmiXid}

\safemath{\evA}{\mathcal{A}}
\safemath{\evB}{\mathcal{B}}
\safemath{\evC}{\mathcal{C}}
\safemath{\evD}{\mathcal{D}}
\safemath{\evE}{\mathcal{E}}
\safemath{\evF}{\mathcal{F}}
\safemath{\evG}{\mathcal{G}}
\safemath{\evH}{\mathcal{H}}
\safemath{\evI}{\mathcal{I}}
\safemath{\evJ}{\mathcal{J}}
\safemath{\evK}{\mathcal{K}}
\safemath{\evL}{\mathcal{L}}
\safemath{\evM}{\mathcal{M}}
\safemath{\evN}{\mathcal{N}}
\safemath{\evO}{\mathcal{O}}
\safemath{\evP}{\mathcal{P}}
\safemath{\evQ}{\mathcal{Q}}
\safemath{\evR}{\mathcal{R}}
\safemath{\evS}{\mathcal{S}}
\safemath{\evT}{\mathcal{T}}
\safemath{\evU}{\mathcal{U}}
\safemath{\evV}{\mathcal{V}}
\safemath{\evW}{\mathcal{W}}
\safemath{\evX}{\mathcal{X}}
\safemath{\evY}{\mathcal{Y}}
\safemath{\evZ}{\mathcal{Z}}

\safemath{\setA}{\mathcal{A}}
\safemath{\setB}{\mathcal{B}}
\safemath{\setC}{\mathcal{C}}
\safemath{\setD}{\mathcal{D}}
\safemath{\setE}{\mathcal{E}}
\safemath{\setF}{\mathcal{F}}
\safemath{\setG}{\mathcal{G}}
\safemath{\setH}{\mathcal{H}}
\safemath{\setI}{\mathcal{I}}
\safemath{\setJ}{\mathcal{J}}
\safemath{\setK}{\mathcal{K}}
\safemath{\setL}{\mathcal{L}}
\safemath{\setM}{\mathcal{M}}
\safemath{\setN}{\mathcal{N}}
\safemath{\setO}{\mathcal{O}}
\safemath{\setP}{\mathcal{P}}
\safemath{\setQ}{\mathcal{Q}}
\safemath{\setR}{\mathcal{R}}
\safemath{\setS}{\mathcal{S}}
\safemath{\setT}{\mathcal{T}}
\safemath{\setU}{\mathcal{U}}
\safemath{\setV}{\mathcal{V}}
\safemath{\setW}{\mathcal{W}}
\safemath{\setX}{\mathcal{X}}
\safemath{\setY}{\mathcal{Y}}
\safemath{\setZ}{\mathcal{Z}}
\safemath{\emptySet}{\varnothing}

\safemath{\colA}{\mathscr{A}}
\safemath{\colB}{\mathscr{B}}
\safemath{\colC}{\mathscr{C}}
\safemath{\colD}{\mathscr{D}}
\safemath{\colE}{\mathscr{E}}
\safemath{\colF}{\mathscr{F}}
\safemath{\colG}{\mathscr{G}}
\safemath{\colH}{\mathscr{H}}
\safemath{\colI}{\mathscr{I}}
\safemath{\colJ}{\mathscr{J}}
\safemath{\colK}{\mathscr{K}}
\safemath{\colL}{\mathscr{L}}
\safemath{\colM}{\mathscr{M}}
\safemath{\colN}{\mathscr{N}}
\safemath{\colO}{\mathscr{O}}
\safemath{\colP}{\mathscr{P}}
\safemath{\colQ}{\mathscr{Q}}
\safemath{\colR}{\mathscr{R}}
\safemath{\colS}{\mathscr{S}}
\safemath{\colT}{\mathscr{T}}
\safemath{\colU}{\mathscr{U}}
\safemath{\colV}{\mathscr{V}}
\safemath{\colW}{\mathscr{W}}
\safemath{\colX}{\mathscr{X}}
\safemath{\colY}{\mathscr{Y}}
\safemath{\colZ}{\mathscr{Z}}

\safemath{\opA}{\mathbb{A}}
\safemath{\opB}{\mathbb{B}}
\safemath{\opC}{\mathbb{C}}
\safemath{\opD}{\mathbb{D}}
\safemath{\opE}{\mathbb{E}}
\safemath{\opF}{\mathbb{F}}
\safemath{\opG}{\mathbb{G}}
\safemath{\opH}{\mathbb{H}}
\safemath{\opI}{\mathbb{I}}
\safemath{\opJ}{\mathbb{J}}
\safemath{\opK}{\mathbb{K}}
\safemath{\opL}{\mathbb{L}}
\safemath{\opM}{\mathbb{M}}
\safemath{\opN}{\mathbb{N}}
\safemath{\opO}{\mathbb{O}}
\safemath{\opP}{\mathbb{P}}
\safemath{\opQ}{\mathbb{Q}}
\safemath{\opR}{\mathbb{R}}
\safemath{\opS}{\mathbb{S}}
\safemath{\opT}{\mathbb{T}}
\safemath{\opU}{\mathbb{U}}
\safemath{\opV}{\mathbb{V}}
\safemath{\opW}{\mathbb{W}}
\safemath{\opX}{\mathbb{X}}
\safemath{\opY}{\mathbb{Y}}
\safemath{\opZ}{\mathbb{Z}}
\safemath{\opZero}{\mathbb{O}}
\safemath{\identityop}{\opI}

\safemath{\sca}{a}
\safemath{\scb}{b}
\safemath{\scc}{c}
\safemath{\scd}{d}
\safemath{\sce}{e}
\safemath{\scf}{f}
\safemath{\scg}{g}
\safemath{\sch}{h}
\safemath{\sci}{i}
\safemath{\scj}{j}
\safemath{\sck}{k}
\safemath{\scl}{l}
\safemath{\scm}{m}
\safemath{\scn}{n}
\safemath{\sco}{o}
\safemath{\scp}{p}
\safemath{\scq}{q}
\safemath{\scr}{r}
\safemath{\scs}{s}
\safemath{\sct}{t}
\safemath{\scu}{u}
\safemath{\scv}{v}
\safemath{\scw}{w}
\safemath{\scx}{x}
\safemath{\scy}{y}
\safemath{\scz}{z}

\safemath{\scA}{A}
\safemath{\scB}{B}
\safemath{\scC}{C}
\safemath{\scD}{D}
\safemath{\scE}{E}
\safemath{\scF}{F}
\safemath{\scG}{G}
\safemath{\scH}{H}
\safemath{\scI}{I}
\safemath{\scJ}{J}
\safemath{\scK}{K}
\safemath{\scL}{L}
\safemath{\scM}{M}
\safemath{\scN}{N}
\safemath{\scO}{O}
\safemath{\scP}{P}
\safemath{\scQ}{Q}
\safemath{\scR}{R}
\safemath{\scS}{S}
\safemath{\scT}{T}
\safemath{\scU}{U}
\safemath{\scV}{V}
\safemath{\scW}{W}
\safemath{\scX}{X}
\safemath{\scY}{Y}
\safemath{\scZ}{Z}

\safemath{\scalpha}{\alpha}
\safemath{\scbeta}{\beta}
\safemath{\scchi}{\chi}
\safemath{\scdelta}{\delta}
\safemath{\scepsilon}{\epsilon}
\safemath{\scvarepsilon}{\varepsilon}
\safemath{\sceta}{\eta}
\safemath{\scgamma}{\gamma}
\safemath{\sciota}{\iota}
\safemath{\sckappa}{\kappa}
\safemath{\scvarkappa}{\varkappa}
\safemath{\sclambda}{\lambda}
\safemath{\scmu}{\mu}
\safemath{\scnu}{\nu}
\safemath{\scomega}{\omega}
\safemath{\scphi}{\phi}
\safemath{\scvarphi}{\varphi}
\safemath{\scpi}{\pi}
\safemath{\scvarpi}{\varpi}
\safemath{\scpsi}{\psi}
\safemath{\scrho}{\rho}
\safemath{\scvarrho}{\varrho}
\safemath{\scsigma}{\sigma}
\safemath{\scvarsigma}{\varsigma}
\safemath{\sctau}{\tau}
\safemath{\sctheta}{\theta}
\safemath{\scvartheta}{\vartheta}
\safemath{\scupsilon}{\upsilon}
\safemath{\scxi}{\xi}
\safemath{\sczeta}{\zeta}

\safemath{\veca}{\mathrm{a}}
\safemath{\vecb}{\mathrm{b}}
\safemath{\vecc}{\mathrm{c}}
\safemath{\vecd}{\mathrm{d}}
\safemath{\vece}{\mathrm{e}}
\safemath{\vecf}{\mathrm{f}}
\safemath{\vecg}{\mathrm{g}}
\safemath{\vech}{\mathrm{h}}
\safemath{\veci}{\mathrm{i}}
\safemath{\vecj}{\mathrm{j}}
\safemath{\veck}{\mathrm{k}}
\safemath{\vecl}{\mathrm{l}}
\safemath{\vecm}{\mathrm{m}}
\safemath{\vecn}{\mathrm{n}}
\safemath{\veco}{\mathrm{o}}
\safemath{\vecp}{\mathrm{p}}
\safemath{\vecq}{\mathrm{q}}
\safemath{\vecr}{\mathrm{r}}
\safemath{\vecs}{\mathrm{s}}
\safemath{\vect}{\mathrm{t}}
\safemath{\vecu}{\mathrm{u}}
\safemath{\vecv}{\mathrm{v}}
\safemath{\vecw}{\mathrm{w}}
\safemath{\vecx}{\mathrm{x}}
\safemath{\vecy}{\mathrm{y}}
\safemath{\vecz}{\mathrm{z}}
\safemath{\veczero}{\mathrm{0}}
\safemath{\vecone}{\mathrm{1}}

\safemath{\vecalpha}{\upalpha}
\safemath{\vecbeta}{\upbeta}
\safemath{\vecchi}{\upchi}
\safemath{\vecdelta}{\updelta}
\safemath{\vecepsilon}{\upepsilon}
\safemath{\vecvarepsilon}{\upvarepsilon}
\safemath{\veceta}{\upeta}
\safemath{\vecgamma}{\upgamma}
\safemath{\veciota}{\upiota}
\safemath{\veckappa}{\upkappa}
\safemath{\veclambda}{\uplambda}
\safemath{\vecmu}{\text{\textmu}}
\safemath{\vecnu}{\upnu}
\safemath{\vecomega}{\upomega}
\safemath{\vecphi}{\upphi}
\safemath{\vecvarphi}{\upvarphi}
\safemath{\vecpi}{\uppi}
\safemath{\vecvarpi}{\upvarpi}
\safemath{\vecpsi}{\uppsi}
\safemath{\vecrho}{\uprho}
\safemath{\vecvarrho}{\upvarrho}
\safemath{\vecsigma}{\upsigma}
\safemath{\vecvarsigma}{\upvarsigma}
\safemath{\vectau}{\uptau}
\safemath{\vectheta}{\uptheta}
\safemath{\vecvartheta}{\upvartheta}
\safemath{\vecupsilon}{\upupsilon}
\safemath{\vecxi}{\upxi}
\safemath{\veczeta}{\upzeta}

\safemath{\vecac}{a}
\safemath{\vecbc}{b}
\safemath{\veccc}{c}
\safemath{\vecdc}{d}
\safemath{\vecec}{e}
\safemath{\vecfc}{f}
\safemath{\vecgc}{g}
\safemath{\vechc}{h}
\safemath{\vecic}{i}
\safemath{\vecjc}{j}
\safemath{\veckc}{k}
\safemath{\veclc}{l}
\safemath{\vecmc}{m}
\safemath{\vecnc}{n}
\safemath{\vecoc}{o}
\safemath{\vecpc}{p}
\safemath{\vecqc}{q}
\safemath{\vecrc}{r}
\safemath{\vecsc}{s}
\safemath{\vectc}{t}
\safemath{\vecuc}{u}
\safemath{\vecvc}{v}
\safemath{\vecwc}{w}
\safemath{\vecxc}{x}
\safemath{\vecyc}{y}
\safemath{\veczc}{z}

\safemath{\matA}{\mathrm{A}}
\safemath{\matB}{\mathrm{B}}
\safemath{\matC}{\mathrm{C}}
\safemath{\matD}{\mathrm{D}}
\safemath{\matE}{\mathrm{E}}
\safemath{\matF}{\mathrm{F}}
\safemath{\matG}{\mathrm{G}}
\safemath{\matH}{\mathrm{H}}
\safemath{\matI}{\mathrm{I}}
\safemath{\matJ}{\mathrm{J}}
\safemath{\matK}{\mathrm{K}}
\safemath{\matL}{\mathrm{L}}
\safemath{\matM}{\mathrm{M}}
\safemath{\matN}{\mathrm{N}}
\safemath{\matO}{\mathrm{O}}
\safemath{\matP}{\mathrm{P}}
\safemath{\matQ}{\mathrm{Q}}
\safemath{\matR}{\mathrm{R}}
\safemath{\matS}{\mathrm{S}}
\safemath{\matT}{\mathrm{T}}
\safemath{\matU}{\mathrm{U}}
\safemath{\matV}{\mathrm{V}}
\safemath{\matW}{\mathrm{W}}
\safemath{\matX}{\mathrm{X}}
\safemath{\matY}{\mathrm{Y}}
\safemath{\matZ}{\mathrm{Z}}
\safemath{\matzero}{\mathrm{0}}

\safemath{\matDelta}{\Updelta}
\safemath{\matGamma}{\Upgammma}
\safemath{\matLambda}{\Uplambda}
\safemath{\matOmega}{\Upomega}
\safemath{\matPhi}{\Upphi}
\safemath{\matPi}{\Uppi}
\safemath{\matPsi}{\Uppsi}
\safemath{\matSigma}{\Upsigma}
\safemath{\matTheta}{\Uptheta}
\safemath{\matUpsilon}{\Upupsilon}
\safemath{\matXi}{\Upxi}

\safemath{\matidentity}{\matI}
\safemath{\vecunit}{\vece} 
\safemath{\matone}{\matO}

\safemath{\matAc}{a}
\safemath{\matBc}{b}
\safemath{\matCc}{c}
\safemath{\matDc}{d}
\safemath{\matEc}{e}
\safemath{\matFc}{f}
\safemath{\matGc}{g}
\safemath{\matHc}{h}
\safemath{\matIc}{i}
\safemath{\matJc}{j}
\safemath{\matKc}{k}
\safemath{\matLc}{l}
\safemath{\matMc}{m}
\safemath{\matNc}{n}
\safemath{\matOc}{o}
\safemath{\matPc}{p}
\safemath{\matQc}{q}
\safemath{\matRc}{r}
\safemath{\matSc}{s}
\safemath{\matTc}{t}
\safemath{\matUc}{u}
\safemath{\matVc}{v}
\safemath{\matWc}{w}
\safemath{\matXc}{x}
\safemath{\matYc}{y}
\safemath{\matZc}{z}

\safemath{\rnda}{\mathsf{a}}
\safemath{\rndb}{\mathsf{b}}
\safemath{\rndc}{\mathsf{c}}
\safemath{\rndd}{\mathsf{d}}
\safemath{\rnde}{\mathsf{e}}
\safemath{\rndf}{\mathsf{f}}
\safemath{\rndg}{\mathsf{g}}
\safemath{\rndh}{\mathsf{h}}
\safemath{\rndi}{\mathsf{i}}
\safemath{\rndj}{\mathsf{j}}
\safemath{\rndk}{\mathsf{k}}
\safemath{\rndl}{\mathsf{l}}
\safemath{\rndm}{\mathsf{m}}
\safemath{\rndn}{\mathsf{n}}
\safemath{\rndo}{\mathsf{o}}
\safemath{\rndp}{\mathsf{p}}
\safemath{\rndq}{\mathsf{q}}
\safemath{\rndr}{\mathsf{r}}
\safemath{\rnds}{\mathsf{s}}
\safemath{\rndt}{\mathsf{t}}
\safemath{\rndu}{\mathsf{u}}
\safemath{\rndv}{\mathsf{v}}
\safemath{\rndw}{\mathsf{w}}
\safemath{\rndx}{\mathsf{x}}
\safemath{\rndy}{\mathsf{y}}
\safemath{\rndz}{\mathsf{z}}

\safemath{\rndA}{\bmiA}
\safemath{\rndB}{\bmiB}
\safemath{\rndC}{\bmiC}
\safemath{\rndD}{\bmiD}
\safemath{\rndE}{\bmiE}
\safemath{\rndF}{\bmiF}
\safemath{\rndG}{\bmiG}
\safemath{\rndH}{\bmiH}
\safemath{\rndI}{\bmiI}
\safemath{\rndJ}{\bmiJ}
\safemath{\rndK}{\bmiK}
\safemath{\rndL}{\bmiL}
\safemath{\rndM}{\bmiM}
\safemath{\rndN}{\bmiN}
\safemath{\rndO}{\bmiO}
\safemath{\rndP}{\bmiP}
\safemath{\rndQ}{\bmiQ}
\safemath{\rndR}{\bmiR}
\safemath{\rndS}{\bmiS}
\safemath{\rndT}{\bmiT}
\safemath{\rndU}{\bmiU}
\safemath{\rndV}{\bmiV}
\safemath{\rndW}{\bmiW}
\safemath{\rndX}{\bmiX}
\safemath{\rndY}{\bmiY}
\safemath{\rndZ}{\bmiZ}

\safemath{\rndalpha}{\bmialpha}
\safemath{\rndbeta}{\bmibeta}
\safemath{\rndchi}{\bmichi}
\safemath{\rnddelta}{\bmidelta}
\safemath{\rndepsilon}{\bmiepsilon}
\safemath{\rndvarepsilon}{\bmivarepsilon}
\safemath{\rndeta}{\bmieta}
\safemath{\rndgamma}{\bmigamma}
\safemath{\rndiota}{\bmiiota}
\safemath{\rndkappa}{\bmikappa}
\safemath{\rndlambda}{\bmilambda}
\safemath{\rndmu}{\bmimu}
\safemath{\rndnu}{\bminu}
\safemath{\rndomega}{\bmiomega}
\safemath{\rndphi}{\bmiphi}
\safemath{\rndvarphi}{\bmivarphi}
\safemath{\rndpi}{\bmipi}
\safemath{\rndvarpi}{\bmivarpi}
\safemath{\rndpsi}{\bmipsi}
\safemath{\rndrho}{\bmirho}
\safemath{\rndvarrho}{\bmivarrho}
\safemath{\rndsigma}{\bmisigma}
\safemath{\rndvarsigma}{\bmivarsigma}
\safemath{\rndtau}{\bmitau}
\safemath{\rndtheta}{\bmitheta}
\safemath{\rndvartheta}{\bmivartheta}
\safemath{\rndupsilon}{\bmiupsilon}
\safemath{\rndxi}{\bmixi}
\safemath{\rndzeta}{\bmizeta}

\safemath{\rveca}{\mathbf{a}}
\safemath{\rvecb}{\mathbf{b}}
\safemath{\rvecc}{\mathbf{c}}
\safemath{\rvecd}{\mathbf{d}}
\safemath{\rvece}{\mathbf{e}}
\safemath{\rvecf}{\mathbf{f}}
\safemath{\rvecg}{\mathbf{g}}
\safemath{\rvech}{\mathbf{h}}
\safemath{\rveci}{\mathbf{i}}
\safemath{\rvecj}{\mathbf{j}}
\safemath{\rveck}{\mathbf{k}}
\safemath{\rvecl}{\mathbf{l}}
\safemath{\rvecm}{\mathbf{m}}
\safemath{\rvecn}{\mathbf{n}}
\safemath{\rveco}{\mathbf{o}}
\safemath{\rvecp}{\mathbf{p}}
\safemath{\rvecq}{\mathbf{q}}
\safemath{\rvecr}{\mathbf{r}}
\safemath{\rvecs}{\mathbf{s}}
\safemath{\rvect}{\mathbf{t}}
\safemath{\rvecu}{\mathbf{u}}
\safemath{\rvecv}{\mathbf{v}}
\safemath{\rvecw}{\mathbf{w}}
\safemath{\rvecx}{\mathbf{x}}
\safemath{\rvecy}{\mathbf{y}}
\safemath{\rvecz}{\mathbf{z}}

\safemath{\rvecalpha}{\bmualpha}
\safemath{\rvecbeta}{\bmubeta}
\safemath{\rvecchi}{\bmuchi}
\safemath{\rvecdelta}{\bmudelta}
\safemath{\rvecepsilon}{\bmuepsilon}
\safemath{\rvecvarepsilon}{\bmuvarepsilon}
\safemath{\rveceta}{\bmueta}
\safemath{\rvecgamma}{\bmugamma}
\safemath{\rveciota}{\bmuiota}
\safemath{\rveckappa}{\bmukappa}
\safemath{\rveclambda}{\bmulambda}
\safemath{\rvecmu}{\bmumu}
\safemath{\rvecnu}{\bmunu}
\safemath{\rvecomega}{\bmuomega}
\safemath{\rvecphi}{\bmuphi}
\safemath{\rvecvarphi}{\bmuvarphi}
\safemath{\rvecpi}{\bmupi}
\safemath{\rvecvarpi}{\bmuvarpi}
\safemath{\rvecpsi}{\bmupsi}
\safemath{\rvecrho}{\bmurho}
\safemath{\rvecvarrho}{\bmuvarrho}
\safemath{\rvecsigma}{\bmusigma}
\safemath{\rvecvarsigma}{\bmuvarsigma}
\safemath{\rvectau}{\bmutau}
\safemath{\rvectheta}{\bmutheta}
\safemath{\rvecvartheta}{\bmuvartheta}
\safemath{\rvecupsilon}{\bmuupsilon}
\safemath{\rvecxi}{\bmuxi}
\safemath{\rveczeta}{\bmuzeta}

\safemath{\rvecac}{\rnda}
\safemath{\rvecbc}{\rndb}
\safemath{\rveccc}{\rndc}
\safemath{\rvecdc}{\rndd}
\safemath{\rvecec}{\rnde}
\safemath{\rvecfc}{\rndf}
\safemath{\rvecgc}{\rndg}
\safemath{\rvechc}{\rndh}
\safemath{\rvecic}{\rndi}
\safemath{\rvecjc}{\rndj}
\safemath{\rveckc}{\rndk}
\safemath{\rveclc}{\rndl}
\safemath{\rvecmc}{\rndm}
\safemath{\rvecnc}{\rndn}
\safemath{\rvecoc}{\rndo}
\safemath{\rvecpc}{\rndp}
\safemath{\rvecqc}{\rndq}
\safemath{\rvecrc}{\rndr}
\safemath{\rvecsc}{\rnds}
\safemath{\rvectc}{\rndt}
\safemath{\rvecuc}{\rndu}
\safemath{\rvecvc}{\rndv}
\safemath{\rvecwc}{\rndw}
\safemath{\rvecxc}{\rndx}
\safemath{\rvecyc}{\rndy}
\safemath{\rveczc}{\rndz}

\safemath{\rmatA}{\mathbf{A}}
\safemath{\rmatB}{\mathbf{B}}
\safemath{\rmatC}{\mathbf{C}}
\safemath{\rmatD}{\mathbf{D}}
\safemath{\rmatE}{\mathbf{E}}
\safemath{\rmatF}{\mathbf{F}}
\safemath{\rmatG}{\mathbf{G}}
\safemath{\rmatH}{\mathbf{H}}
\safemath{\rmatI}{\mathbf{I}}
\safemath{\rmatJ}{\mathbf{J}}
\safemath{\rmatK}{\mathbf{K}}
\safemath{\rmatL}{\mathbf{L}}
\safemath{\rmatM}{\mathbf{M}}
\safemath{\rmatN}{\mathbf{N}}
\safemath{\rmatO}{\mathbf{O}}
\safemath{\rmatP}{\mathbf{P}}
\safemath{\rmatQ}{\mathbf{Q}}
\safemath{\rmatR}{\mathbf{R}}
\safemath{\rmatS}{\mathbf{S}}
\safemath{\rmatT}{\mathbf{T}}
\safemath{\rmatU}{\mathbf{U}}
\safemath{\rmatV}{\mathbf{V}}
\safemath{\rmatW}{\mathbf{W}}
\safemath{\rmatX}{\mathbf{X}}
\safemath{\rmatY}{\mathbf{Y}}
\safemath{\rmatZ}{\mathbf{Z}}
\safemath{\rmatDelta}{\bmuDelta}
\safemath{\rmatGamma}{\bmuGamma}
\safemath{\rmatLambda}{\bmuLambda}
\safemath{\rmatOmega}{\bmuOmega}
\safemath{\rmatPhi}{\bmuPhi}
\safemath{\rmatPi}{\bmuPi}
\safemath{\rmatPsi}{\bmuPsi}
\safemath{\rmatSigma}{\bmuSigma}
\safemath{\rmatTheta}{\bmuTheta}
\safemath{\rmatUpsilon}{\bmuUpsilon}
\safemath{\rmatXi}{\bmuXi}

\safemath{\rmatAc}{\rnda}
\safemath{\rmatBc}{\rndb}
\safemath{\rmatCc}{\rndc}
\safemath{\rmatDc}{\rndd}
\safemath{\rmatEc}{\rnde}
\safemath{\rmatFc}{\rndf}
\safemath{\rmatGc}{\rndg}
\safemath{\rmatHc}{\rndh}
\safemath{\rmatIc}{\rndi}
\safemath{\rmatJc}{\rndj}
\safemath{\rmatKc}{\rndk}
\safemath{\rmatLc}{\rndl}
\safemath{\rmatMc}{\rndm}
\safemath{\rmatNc}{\rndn}
\safemath{\rmatOc}{\rndo}
\safemath{\rmatPc}{\rndp}
\safemath{\rmatQc}{\rndq}
\safemath{\rmatRc}{\rndr}
\safemath{\rmatSc}{\rnds}
\safemath{\rmatTc}{\rndt}
\safemath{\rmatUc}{\rndu}
\safemath{\rmatVc}{\rndv}
\safemath{\rmatWc}{\rndw}
\safemath{\rmatXc}{\rndx}
\safemath{\rmatYc}{\rndy}
\safemath{\rmatZc}{\rndz}

\newenvironment{textbmatrix}{	\setlength{\arraycolsep}{2.5pt}%
								\big[\begin{matrix}}{\end{matrix}\big]%
								\raisebox{0.08ex}{\vphantom{M}}}

 \def\btm{\begin{textbmatrix}}
 \def\etm{\end{textbmatrix}}

\newcommand{\lefto}{\mathopen{}\left}

\DeclareMathOperator{\diag}{diag}			
\DeclareMathOperator{\adj}{adj}				
\DeclareMathOperator{\kron}{\otimes}			
\DeclareMathOperator{\Exop}{\opE}			
\DeclareMathOperator{\spn}{span}			 	

\DeclareMathOperator{\landauO}{\mathcal{O}}

\safemath{\fun}{\scf}						

\safemath{\vrbl}{x}						
\safemath{\altvrbl}{y}						
\safemath{\aaltvrbl}{z}						

\safemath{\vvrbl}{\vecx}						
\safemath{\altvvrbl}{\vecy}						
\safemath{\aaltvvrbl}{\vecz}						

\safemath{\altfun}{\scg}
\safemath{\aaltfun}{\sch}
\safemath{\bel}{\sce}					
\safemath{\altbel}{\sce}					
\safemath{\frel}{g}					
\safemath{\altfrel}{g}					
\safemath{\dfrel}{\tilde{g}}					
\safemath{\altdfrel}{\tilde{g}}					

\safemath{\mat}{\matA}						
\safemath{\matc}{\matAc}						
\safemath{\altmat}{\matB}						
\safemath{\altmatc}{\matBc}						

\safemath{\vectr}{\vecu}						
\safemath{\vectrc}{\vecuc}						
\safemath{\altvectr}{\vecv}						
\safemath{\aaltvectr}{\vect}						
\safemath{\altvectrc}{\vecvc}						
\safemath{\genvar}{u}						
\safemath{\altgenvar}{v}						

\safemath{\rvectr}{\rvecu}						
\safemath{\rvectrc}{\rvecuc}						
\safemath{\raltvectr}{\rvecv}						
\safemath{\raaltvectr}{\rvect}						
\safemath{\raltvectrc}{\rvecvc}						
\safemath{\rgenvar}{\rndu}						
\safemath{\raltgenvar}{\rndv}						

\newcommand{\nullspace}{\setN}	 			
\newcommand{\simplifiedmathchoice}[2]{\mathchoice{#1}{#2}{#2}{#2}}

\newcommand{\Ex}[2]{\simplifiedmathchoice{\ensuremath{\Exop_{#1}\lefto[#2\right]}}{\ensuremath{\Exop_{#1}\bigl[#2\bigr]}}} 	
\newcommand{\abs}[1]{\mathchoice{{\left\lvert#1\right\rvert}}{{\bigl\lvert#1\bigr\rvert}}{{\left\lvert#1\right\rvert}}{{\left\lvert#1\right\rvert}}}		
\newcommand{\card}[1]{\lvert#1\rvert}			
\newcommand{\Union}{\bigcup}
\newcommand{\intersect}{\cap}				

\newcommand{\vecnorm}[1]{\lVert#1\rVert}		
\newcommand{\conj}[1]{\ensuremath{#1^{*}}} 	
\newcommand{\tp}[1]{\ensuremath{#1^{\mathsf{T}}}} 		
\newcommand{\herm}[1]{\ensuremath{#1^{\mathsf{H}}}} 	
\newcommand{\inv}[1]{\ensuremath{#1^{-1}}} 	

\safemath{\dirac}{\delta}					
\safemath{\diracp}{\dirac(\time)}			
\safemath{\krond}{\dirac}					
\safemath{\indfun}{I}						
\safemath{\stepfun}{u}						

\safemath{\upto}{\uparrow}
\safemath{\downto}{\downarrow}
\safemath{\iu}{\mathrm{i}}							
\safemath{\maj}{\succ}
\newcommand{\dftmat}[1]{\matF_{#1}}			
\safemath{\mdft}{\dftmat{}}					
\safemath{\runity}{\beta}					
\safemath{\eval}{\lambda}					

\safemath{\veval}{\veclambda}				
\safemath{\littleo}{\sco}					

\let\im\undefined
\safemath{\re}{\Re}				
\safemath{\im}{\Im}				

\safemath{\euclidspace}{\complexset}			
\safemath{\confunspace}{\setC}				
\newcommand{\banachseqspace}[1]{l^{#1}}		
\safemath{\hilseqspace}{\banachseqspace{2}}	
\newcommand{\banachfunspace}[1]{\setL^{#1}}	
\safemath{\hilfunspace}{\banachfunspace{2}}	
\safemath{\hilfunspacep}{\hilfunspace(\complexset)}	
\safemath{\schwarzspace}{\setS}				

\newcommand{\hadj}[1]{#1^{\star}}			

\safemath{\SNR}{\rho} 				
\safemath{\SINR}{\text{\sc sinr}} 				
\safemath{\No}{N_0}							
\safemath{\Es}{E_s}							
\safemath{\Eb}{E_b}							
\safemath{\EbNo}{\frac{\Eb}{\No}}
\safemath{\EsNo}{\frac{\Es}{\No}}
\safemath{\NoVar}{\variance}                 

\let\time\undefined
\safemath{\time}{\sct}						
\safemath{\dtime}{\sck}						
\safemath{\delay}{\sctau}					
\safemath{\ddelay}{\scl}						
\safemath{\doppler}{\scnu}					
\safemath{\ddoppler}{\scm}					
\safemath{\freq}{\scf}						
\safemath{\dfreq}{\scn}						
\safemath{\Dtime}{\Delta\time}
\safemath{\Dfreq}{\Delta\freq}
\safemath{\Ddtime}{\dtime}
\safemath{\Ddfreq}{\dfreq}
\safemath{\bandwidth}{\scB}
\safemath{\maxdoppler}{\doppler_{0}}			
\safemath{\maxdelay}{\delay_{0}}				
\safemath{\spread}{\Delta_{\CHop}}			
\DeclareMathOperator{\CHop}{\ensuremath{\opH}} 
\safemath{\kernel}{\rndk_{\CHop}}			
\safemath{\kernelp}{\kernel(\time,\time')}	
\safemath{\tvir}{\rndh_{\CHop}}				
\safemath{\tvirp}{\tvir(\time,\delay)}		
\safemath{\tvirc}{\conj{\rndh}_{\CHop}}		
\safemath{\tvtf}{\rndl_{\CHop}}				
\safemath{\tvtfp}{\tvtf(\time,\freq)}			
\safemath{\tvtfc}{\conj{\rndl}_{\CHop}}		
\safemath{\spf}{\rnds_{\CHop}}				
\safemath{\spfp}{\spf(\doppler,\delay)}		
\safemath{\spfc}{\conj{\rnds}_{\CHop}}		
\safemath{\bff}{\rndb_{\CHop}}				
\safemath{\bffp}{\bff(\doppler,\freq)}		

\safemath{\irc}{\scr_{\rndh}}				
\safemath{\tfc}{\scr_{\rndl}}				
\safemath{\spc}{\scr_{\rnds}}				
\safemath{\bfc}{\scr_{\rndb}}				

\safemath{\scaf}{\scc_{\rnds}}				
\safemath{\scafp}{\scaf(\doppler,\delay)}		
\safemath{\ccf}{\scc_{\rndl}}				
\safemath{\ccfp}{\ccf(\Dtime,\Dfreq)}			
\safemath{\cic}{\scc_{\rndh}}				
\safemath{\cicp}{\cic(\Dtime,\delay)}			

\safemath{\mi}{I}							
\safemath{\capacity}{C}					

\DeclareMathOperator{\Prob}{\opP}		
\newcommand{\pdf}[1]{\scf_{#1}}			

\newcommand{\diffent}{h}				
\safemath{\normal}{\mathcal{N}}			
\safemath{\jpg}{\mathcal{CN}}			
\safemath{\uniform}{\mathcal{U}}				
\safemath{\mchain}{\leftrightarrow}		
\newcommand{\given}{\,\vert\,}				

\safemath{\dB}{\,\mathrm{dB}}
\safemath{\dBm}{\,\mathrm{dBm}}
\safemath{\Hz}{\,\mathrm{Hz}}
\safemath{\kHz}{\,\mathrm{kHz}}
\safemath{\MHz}{\,\mathrm{MHz}}
\safemath{\GHz}{\,\mathrm{GHz}}
\safemath{\s}{\,\mathrm{s}}
\safemath{\ms}{\,\mathrm{ms}}
\safemath{\mus}{\,\mathrm{\text{\textmu}s}}
\safemath{\ns}{\,\mathrm{ns}}
\safemath{\ps}{\,\mathrm{ps}}
\safemath{\meter}{\,\mathrm{m}}
\safemath{\mm}{\,\mathrm{mm}}
\safemath{\cm}{\,\mathrm{cm}}
\safemath{\m}{\,\mathrm{m}}
\safemath{\W}{\,\mathrm{W}}
\safemath{\mW}{\, \mathrm{mW}}
\safemath{\J}{\,\mathrm{J}}
\safemath{\K}{\,\mathrm{K}}
\safemath{\bit}{\,\mathrm{bit}}
\safemath{\nat}{\,\mathrm{nat}}

\safemath{\define}{\triangleq}					

\safemath{\equivalent}{\sim}
\safemath{\distas}{\sim}					
\safemath{\sdiff}{\Delta}				
\safemath{\setdiff}{\setminus}				

\safemath{\reals}{\mathbb R}
\safemath{\positivereals}{\reals^{+}}
\safemath{\integers}{\mathbb Z}
\safemath{\posint}{\integers^{+}}
\safemath{\naturals}{\mathbb N}
\safemath{\posnaturals}{\naturals^{+}}
\safemath{\complexset}{\mathbb C}
\safemath{\rationals}{\mathbb Q}

\newcommand{\natseg}[2]{[#1 \text{\phantom{\tiny{.}}:\phantom{\tiny{.}}} #2]} 
\newcommand{\matsegs}[2]{#1_{ #2 }} 
\newcommand{\vecseg}[2]{#1_{#2}} 
\newcommand{\col}[2]{#1_{\{#2\}}} 

\safemath{\iSet}{\setI}
\safemath{\rel}{\bowtie}					
\safemath{\eqrel}{\sim}					
\safemath{\rlord}{\leq}					
\safemath{\slord}{<}						
\safemath{\rpord}{\preceq}				
\safemath{\rrpord}{\succeq}				
\safemath{\spord}{\prec}					
\safemath{\sig}{\sigma}					
\safemath{\metric}{d}					

\safemath{\setfun}{\Phi}					
\safemath{\measure}{\mu}					
\safemath{\altmeasure}{\lambda}					
\newcommand{\outerm}[1]{#1^{\star}}		
\newcommand{\innerm}[1]{#1_{\star}}		
\safemath{\omeasure}{\outerm{\measure}}		
\safemath{\imeasure}{\innerm{\measure}}		
\safemath{\aecol}{\colS^{\star}_{\measure}} 
\safemath{\emeasure}{\bar{\measure}_{0}}	
\safemath{\rmeasure}{\tilde{\measure}}	
\safemath{\bmeasure}{\measure_{0}}		
\safemath{\glength}{\measure_{\altfun}}	
\safemath{\lebmea}{\lambda}				
\safemath{\blebmea}{\lebmea_{0}}			
\safemath{\sfun}{s}						
\safemath{\absintspace}{\colL^{1}}		
\safemath{\sqintspace}{\colL^{2}}		
\safemath{\abssumspace}{l^{1}}		
\safemath{\sqsumspace}{l^{2}}		

\safemath{\sfield}{\setF}				
\safemath{\vectors}{\setV}				
\safemath{\vecspace}{(\vectors,\sfield)}	
\safemath{\linspace}{\setV}				
\safemath{\clinspace}{(\linspace,\sfield)} 
\safemath{\nspace}{\setU}				
\safemath{\metspace}{\setM}				
\safemath{\bspace}{\setB}				
\safemath{\ipspace}{\setG}				
\safemath{\hilspace}{\setH}				
\safemath{\blospace}{\setG}				
\safemath{\lop}{\opT}					
\safemath{\altlop}{\opS}					
\safemath{\nullsp}{\nullspace(\lop)}		
\safemath{\lfun}{l}						
\safemath{\altlfun}{g}					
\newcommand{\dual}[1]{#1^{'}}			
\safemath{\dsum}{\oplus}					
\safemath{\funspace}{\colL}				
\renewcommand{\adj}[1]{#1^{\times}}		
\safemath{\adjlop}{\adj{\lop}}			
\safemath{\hadjlop}{\hadj{\lop}}			
\safemath{\tow}{\xrightarrow{w}}			
\safemath{\tows}{\xrightarrow{w^{*}}}		
\safemath{\cparam}{\lambda}
\safemath{\lopl}{\lop_{\cparam}}		
\safemath{\iop}{\opI}					
\safemath{\resolop}{\opR}				
\safemath{\resolvent}{\resolop_{\cparam}(\lop)}	
\safemath{\reset}{\setQ}
\safemath{\spectrum}{\setS}
\safemath{\resolset}{\reset(\lop)}		
\safemath{\lopspec}{\spectrum(\lop)}		
\safemath{\pspec}{\spectrum_{p}(\lop)}	
\safemath{\cspec}{\spectrum_{c}(\lop)}	
\safemath{\rspec}{\spectrum_{r}(\lop)}	
\safemath{\ev}{\cparam}					
\newcommand{\specrad}[1]{r_{#1}}			
\safemath{\lopsrad}{\specrad{\lop}}		
\safemath{\pop}{\opP}					

\safemath{\specfam}{\colE}				
\safemath{\specop}{\opE_{\cparam}}		
\safemath{\altspecop}{\opE_{\mu}}		
\safemath{\vmulti}{\vecone}				
\safemath{\unitaryop}{\opU}				
\safemath{\sval}{\sigma}					
\safemath{\corrcoef}{\rho}				
\safemath{\sangle}{\theta}				

\let\time\undefined
\safemath{\iset}{\setI}				
\safemath{\shift}{\nu}
\safemath{\scale}{\alpha}
\safemath{\time}{t}
\safemath{\specfreq}{\theta}	
\newcommand{\transopgen}[1]{\opT_{#1}} 
\safemath{\transop}{\transopgen{\delay}}
\newcommand{\modopgen}[1]{\opM_{#1}}	
\safemath{\modop}{\modopgen{\shift}}
\newcommand{\dilopgen}[1]{\opD_{#1}}	
\safemath{\dilop}{\dilopgen{\scale}}
\safemath{\fram}{\setF}				
\safemath{\dfram}{\dual{\fram}}		
\safemath{\ufb}{B}					
\safemath{\lfb}{A}					
\safemath{\sop}{\hadj{\aop}}				
\safemath{\aop}{\opT}			
\safemath{\fop}{\opS}				
\safemath{\daop}{\tilde\opT}			
\safemath{\dsop}{\hadj{\tilde\opT}}				

\safemath{\ifop}{\inv{\fop}}			
\safemath{\rifop}{\fop^{-1/2}}			
\safemath{\transeq}{\setT}			
\safemath{\nfun}{\Phi}				
\safemath{\funvec}{\vecf}			
\safemath{\altfunvec}{\vecg}

\safemath{\samplespace}{\Omega}
\safemath{\probspace}{(\samplespace,\sfield,\Prob)}	
\safemath{\ccoef}{\rho}			

\safemath{\infstate}{\vecpi}				
\safemath{\typset}{\setA_{\epsilon}^{(N)}}	
\safemath{\expequal}{\doteq}				
\safemath{\code}{C}						
\safemath{\dstringset}{\setD^{\star}}		
\safemath{\cwlength}{l}					
\safemath{\elength}{L}					
\safemath{\extension}{C^{\star}}			
\safemath{\approaches}{\rightarrow}		
\safemath{\evnt}{\setA}					
\safemath{\altevnt}{\setB}					
\safemath{\rv}{\rndx}					
\safemath{\altrv}{\rndy}					
\safemath{\complexrv}{\rndu}					
\safemath{\altcrv}{\rndv}				
\safemath{\rvec}{\rvecx}					
\safemath{\altrvec}{\rvecy}				
\safemath{\crvec}{\rvecu}				
\safemath{\altcrvec}{\rvecv}				
\safemath{\variance}{\sigma^{2}}			
\safemath{\map}{T}						
\safemath{\jacobian}{\matJ}					
\safemath{\wvec}{\rvecw}					
\safemath{\wrv}{\rndw}					
\safemath{\orthmat}{\matQ}				
\safemath{\evmat}{\matLambda}			
\safemath{\identity}{\matidentity}		
\safemath{\innovec}{\vecv}				
\safemath{\convas}{\xrightarrow{\text{a.s.}}}	
\safemath{\convr}{\xrightarrow{\text{r}}}	
\safemath{\convp}{\xrightarrow{\text{P}}}	
\safemath{\convd}{\xrightarrow{\text{D}}}	
\safemath{\ltis}{\opL}				
\safemath{\ir}{h}					
\safemath{\tf}{\MakeUppercase{\ir}}	

\newcommand*{\fancyrefparlabelprefix}{par}		
\newcommand*{\fancyrefremlabelprefix}{rem}		
\newcommand*{\fancyrefchalabelprefix}{cha}		
\newcommand*{\fancyrefapplabelprefix}{app}		
\newcommand*{\fancyrefthmlabelprefix}{thm}		
\newcommand*{\fancyreflemlabelprefix}{lem}		
\newcommand*{\fancyrefcorlabelprefix}{cor}		
\newcommand*{\fancyrefdeflabelprefix}{def}		
\newcommand*{\fancyrefproplabelprefix}{prop}		

\frefformat{vario}{\fancyrefparlabelprefix}{Part~#1}
\frefformat{vario}{\fancyrefremlabelprefix}{Remark~#1}
\frefformat{vario}{\fancyrefchalabelprefix}{Chapter~#1}
\frefformat{vario}{\fancyrefseclabelprefix}{Section~#1}
\frefformat{vario}{\fancyrefthmlabelprefix}{Theorem~#1}
\frefformat{vario}{\fancyreflemlabelprefix}{Lemma~#1}
\frefformat{vario}{\fancyrefcorlabelprefix}{Corollary~#1}
\frefformat{vario}{\fancyrefdeflabelprefix}{Definition~#1}
\frefformat{vario}{\fancyreffiglabelprefix}{Fig.~#1}
\frefformat{vario}{\fancyrefapplabelprefix}{Appendix~#1}
\frefformat{vario}{\fancyrefeqlabelprefix}{(#1)}
\frefformat{vario}{\fancyrefproplabelprefix}{Property~#1}

\theoremstyle{plain}
\newtheorem{thm}{Theorem}
\newtheorem{lem}[thm]{Lemma}
\newtheorem{cor}[thm]{Corollary}
\theoremstyle{definition}

\theoremstyle{remark}
\newtheorem{rem}{Remark}

\newcommand{\iid}{i.i.d.\@\xspace}

\newcommand{\wrt}{w.r.t.\@\xspace}

\newcommand{\Pm}{\matP}						
\newcommand{\Qm}{\matQ}						
\newcommand{\pilotset}{\setP}   			
\newcommand{\nonpilotset}{\setJ}   			
\newcommand{\outremsize}{{\abs{\setI}}}   	
\newcommand{\poly}{p}						
\newcommand{\bijec}{\vecg}					
\safemath{\propspark}{\text{Property (A)}}
\safemath{\propsparkm}{\text{Property (A')}}

\newcommand{\RXant}{R}	 			    	
\newcommand{\shortant}{\tilde \RXant}		
\newcommand{\blocklength}{L}	 			
\newcommand{\rankcov}{Q}	 			    
\newcommand{\asconst}{\landauO(1)}	 		
\newcommand{\constalt}{c}	 			    
\newcommand{\prelog}{\chi}	 			    

\newcommand{\vinp}{\rvecx}	 			    
\newcommand{\nrvinp}{\vecx}	 			
\newcommand{\nrvinpinv}{\vecz}	 			
\newcommand{\vinpc}{\rvecxc}	 			
\newcommand{\nrvinpc}{\vecxc}	 			
\newcommand{\nrvinpinvc}{\veczc}	 		
\newcommand{\invinpc}{\rveczc}				
\newcommand{\minp}{\rmatX}	 			    
\newcommand{\nrminp}{\matX}	 			    

\newcommand{\vout}[1]{\rvecy_{#1}}	 		
\newcommand{\vectout}{\rvecy}	 			
\newcommand{\mout}{\rmatY}	 			    
\newcommand{\moutc}{\rndy}	 			    

\newcommand{\vnoise}[1]{\rvecw_{#1}}	 	
\newcommand{\mnoise}{\rmatW}	 			
\newcommand{\vectnoise}{\rvecw}	 			
\newcommand{\mnoisec}{\rndw}	 			

\newcommand{\vectoutnn}{\hat\rvecy}	 		
\newcommand{\vectoutnnc}{\hat\rvecyc}	 	

\newcommand{\sqrtcov}{\matD}	 			
\newcommand{\vsqrtcov}{\vecd}	 			
\newcommand{\rsqrtcov}{\bar\vecd}	 			
\newcommand{\sqrtcovc}{\matDc}	 			

\newcommand{\vchannel}[1]{\rvech_{#1}}	 	
\newcommand{\mchannel}{\rmatH}	 			
\newcommand{\mchannelc}{\rmatHc}	 		
\newcommand{\viid}[1]{\rvecs_{#1}}	 		
\newcommand{\nrviid}[1]{\vecs_{#1}}	 		
\newcommand{\miid}{\rmatS}	 			    
\newcommand{\rvmiid}{{\bar\rvecs}}	 		
\newcommand{\vectiid}{\rvecs}	 			
\newcommand{\nrvectiid}{\vecs}	 			
\newcommand{\vectiidc}{\rvecsc}	 			
\newcommand{\nrvectiidc}{\vecsc}	 		

\newcommand{\cube}{\setC}					
\newcommand{\manifold}{\setM}				
\newcommand{\mpoint}{\vecp}					
\newcommand{\vindx}{\vecm} 					
\newcommand{\indx}{m}						
\newcommand{\altvindx}{\vecn} 				
\newcommand{\blowup}{\psi}					
\newcommand{\coord}{\varphi}				

\newcommand{\degr}{K}						
\newcommand{\dimens}{K}						
\newcommand{\dimeni}{k}						

\safemath{\inpprobmeas}{\mathsf{Q}}
\safemath{\altinpprobmeas}{\widetilde{\inpprobmeas}}
\safemath{\outprobmeas}{\mathsf{R}}
\safemath{\chtran}{\mathsf{W}}

\begin{document}
	
	\acrodef{SISO}{single-input single-output}
	\acrodef{SIMO}{single-input multiple-output}
	\acrodef{MISO}{multiple-input single-output}
	\acrodef{MIMO}{multiple-input multiple-output}
	\acrodef{CSI}{channel state information}
	\acrodef{AWGN}{additive white Gaussian noise}
	\acrodef{SNR}{signal-to-noise ratio}
	\acrodef{JPG}{jointly proper Gaussian}
	\acrodef{PDF}{probability density function}
	\acrodef{DFT}{discrete Fourier transform}
	\acrodef{RHS}{right-hand side}
	\acrodef{LHS}{left-hand side}
	\acrodef{IO}{input-output}
	\acrodef{RV}{random variable}
	\acrodef{SVD}{singular-value decomposition}
	\acrodef{WSSUS}{wide-sense stationary uncorrelated scattering}
	
\IEEEoverridecommandlockouts

\title{Capacity Pre-Log of Noncoherent SIMO\\ Channels via Hironaka's Theorem}

\author{
Veniamin~I.~Morgenshtern, Erwin~Riegler,
\\   Wei~Yang, Giuseppe~Durisi,
\\ Shaowei~Lin, Bernd~Sturmfels, and Helmut~B\"olcskei 
\thanks{The results in this paper appeared in part at the 2010 and 2011 IEEE International Symposia on Information Theory~\cite{morgenshtern10-06},~\cite{riegler11-08}, and at the International Symposium on Wireless Communication Systems (ISWCS) 2011, Aachen, Germany~\cite{yang11-11}.}%
\thanks{V.~I.~Morgenshtern is with the Department of Statistics, Stanford University, CA, USA, Email: vmorgen@stanford.edu }
\thanks{E.~Riegler is with Vienna University of Technology,  Vienna, Austria, Email: erwin.riegler@nt.tuwien.ac.at}
\thanks{W.~Yang and G.~Durisi are with the Department of Signals and Systems, Chalmers University of Technology, Gothenburg, Sweden, Email: \{ywei, durisi\}@chalmers.se}
\thanks{S.~Lin is with the Institute for Infocomm Research, A*STAR, Singapore, Email: lins@i2r.a-star.edu.sg}
\thanks{B.~Sturmfels is with the Department of Mathematics, University of California Berkeley, CA, USA, Email:  bernd@math.berkeley.edu}
\thanks{H.~B\"olcskei is with the Dept. of IT \& EE, ETH Zurich, Switzerland, Email:  boelcskei@nari.ee.ethz.ch}}

\maketitle
\begin{abstract}
We find the capacity pre-log of a temporally correlated Rayleigh block-fading \ac{SIMO} channel in the noncoherent setting. 
It is well known that  for  block-length $\blocklength$ and rank of the channel covariance matrix equal to $\rankcov$,  the capacity pre-log in the \ac{SISO} case is given by $1-\rankcov/\blocklength$. Here, $\rankcov/\blocklength$ can be interpreted as the pre-log penalty incurred by channel uncertainty. Our main result reveals that, by adding only one receive antenna, this penalty can be reduced to $1/\blocklength$ and can, hence, be made to vanish for the block-length $\blocklength\to\infty$, even if $\rankcov/\blocklength$ remains constant as $\blocklength\to\infty$. 
Intuitively, even though the \ac{SISO} channels between the transmit antenna and the two receive
antennas are statistically independent, the transmit signal induces enough statistical
dependence between the corresponding receive signals for the second receive antenna to be able to resolve
the uncertainty associated with the first receive antenna's channel and thereby make the overall system appear coherent.
The proof of our main theorem is based on a deep result  from algebraic geometry known as Hironaka's Theorem on the Resolution of Singularities.

\end{abstract}
\IEEEpeerreviewmaketitle
\section{Introduction}
\label{sec:introduction}

It is well known that the capacity pre-log,  
 i.e., the asymptotic ratio between capacity and the logarithm of \ac{SNR}, as \ac{SNR} goes to infinity, of a \acf{SIMO} fading channel in the \emph{coherent setting} (i.e., when the receiver has perfect \ac{CSI}) is equal to $1$ and is, hence, the same as  that of a \acf{SISO} fading channel~\cite{telatar99-11}. This result holds under very general assumptions on the channel statistics.
Multiple antennas at the receiver only, hence, do not result in an increase of the capacity pre-log in the coherent setting~\cite{telatar99-11}.
In the \emph{noncoherent setting}, where neither transmitter nor receiver have \ac{CSI}, but both know the channel statistics, the effect of multiple antennas on the capacity\footnote{In the remainder of the paper, we consider the noncoherent setting only. Consequently, we will refer to capacity in the noncoherent setting simply as capacity.} pre-log is understood only for a specific simple channel model, namely, the Rayleigh  
\emph{constant block-fading} model. 
In this model the channel is assumed to remain constant over a block (of $\blocklength$ symbols) and to change in an independent fashion from block to block~\cite{marzetta99-01}. The corresponding \ac{SIMO} capacity pre-log is again equal to the \ac{SISO} capacity pre-log, but, differently from the coherent setting, is given by~$1-1/\blocklength$~\cite{hochwald00-05,zheng02-02}.

An alternative approach to 
 capturing channel variations in time is to assume that the fading process is \emph{stationary}. 
In this case, the capacity pre-log is known only in the \ac{SISO}~\cite{lapidoth05-02} and the \ac{MISO}~\cite[Thm. 4.15]{koch09} cases. 
The capacity bounds for the \ac{SIMO} stationary-fading channel available in the literature~\cite[Thm. 4.13]{koch09} do not allow  to determine whether the capacity pre-log in the \ac{SIMO} case equals that in the \ac{SISO} case.
Resolving this question for stationary fading seems elusive at this point.

A widely used channel model that can be seen as lying in between the stationary-fading model considered in~\cite{lapidoth05-02,koch09}, and the simpler constant block-fading model analyzed in~\cite{marzetta99-01,zheng02-02} is the \emph{correlated block-fading} model, which assumes that the fading process is temporally correlated within blocks of length~$\blocklength$ and independent across blocks. The $\blocklength\times\blocklength$ channel covariance matrix of rank $\rankcov\le\blocklength$ is taken to be the same for each block. 
This channel model is relevant as it captures channel variations in time in an accurate yet simple fashion: 
the rank~$\rankcov$ of the covariance matrix corresponds to the minimum number of channel coefficients per block  that need to be known at the receiver to perfectly reconstruct all channel coefficients within the same block. Therefore, larger $\rankcov/\blocklength$ corresponds to faster channel variations.

The \ac{SISO} capacity pre-log for correlated block-fading channels is given by~$1-\rankcov/\blocklength$~\cite{liang04-12}. 
In the \ac{SIMO} and the \ac{MIMO} cases the capacity pre-log is unknown. 
The main contribution of this paper is a full characterization of the capacity pre-log for \ac{SIMO}  correlated block-fading channels. Specifically, we prove that under a mild technical condition on the channel covariance matrix, the \ac{SIMO} capacity pre-log, $\prelog$, of a channel with $\RXant$ receive antennas and independent identically distributed (\iid) \ac{SISO} subchannels is given by 
\begin{align}
\label{eq:prelogans}
\prelog=\min[1-1/\blocklength,\RXant(1-\rankcov/\blocklength)].
\end{align}
This shows that even with $\RXant= 2$ receive antennas a capacity pre-log of~$1-1/\blocklength$ can be obtained in the \ac{SIMO} case (provided that $\blocklength\ge 2\rankcov-1$). This capacity pre-log  
is strictly larger than the capacity pre-log of the corresponding \ac{SISO} channel (i.e., the capacity pre-log of one of the component channels), given by $1-\rankcov/\blocklength$. Here $\rankcov/\blocklength$ can be interpreted as pre-log penalty due to channel uncertainty. 
Our result reveals that, by adding at least one receive antenna, this penalty can 
be made to vanish in the large block-length limit, $\blocklength\to\infty$, even if the amount of channel uncertainty scales linearly in the block-length.

A conjecture for the correlated block-fading channel model stated in~\cite{liang04-12} for the \ac{MIMO} case, when particularized to the \ac{SIMO} case, implies that the capacity pre-log in the \ac{SIMO} case would be the same as that in the \ac{SISO} case. As a consequence of~\fref{eq:prelogans} this conjecture is disproved.

In terms of the technical aspects of our main result, we 
sandwich capacity between an upper and a lower bound that turn out to be asymptotically (in \ac{SNR}) tight (in the sense of delivering the same capacity pre-log).
The upper bound is established by proving that 
the capacity pre-log of a correlated block-fading channel with $\RXant$ receive antennas can be upper-bounded by the capacity pre-log of a constant block-fading channel with $\RXant\rankcov$  receive antennas and the same \ac{SNR}.
The derivation of the capacity pre-log lower bound poses serious technical challenges. Specifically, after a change of variables argument applied to the integral expression for the differential entropy of the channel output signal,  the main technical difficulty lies in showing that 
the expected logarithm of the 
Jacobian  determinant corresponding to this change of variables is finite. As the Jacobian  determinant takes on a very involved form, a per pedes approach  appears infeasible. The problem is resolved by first distilling structural properties of the determinant through a suitable factorization and then introducing a powerful tool from  algebraic geometry, namely \cite[Th. 2.3]{watanabe09}, which is a consequence of Hironaka's Theorem on the Resolution of Singularities \cite{hironaka64-01,hironaka64-03}. 
Roughly speaking, this result allows to rewrite every real analytic function 
\cite[Def. 1.1.5, Def. 2.2.1]{krantz92} locally as a product of a monomial and a nonvanishing real analytic function. This factorization is then used to show 
that the integral of the logarithm of the absolute value of a real analytic function over a compact set is finite, provided that the real analytic 
function is not identically zero. 
This method is quite general and may be of independent interest when one tries to show that integrals of certain functions with singularities are finite, in particular, functions involving logarithms. In information theory such integrals often occur when analyzing differential entropy.

\subsubsection*{Notation}
Sets are denoted by calligraphic letters $\setA, \setB,\ldots$
Roman letters $\matA,\matB,\ldots$ and $\veca,\vecb,\ldots$ designate deterministic matrices and vectors, respectively. Boldface letters $\rmatA,\rmatB,\ldots$ and $\rveca,\rvecb,\ldots$ denote random matrices and random vectors, respectively. 
We let $\vecunit_{i}$ be the vector (of appropriate dimension) that has the $i$th entry equal to one and all other entries equal to zero, 
 and denote the $M\times M$ identity matrix as $\identity_{M}$.
The element in the $i$th row and $j$th column of a  deterministic matrix $\mat$ is $\matc_{ij}$ (italic letters), and the $i$th component of the deterministic vector~$\vectr$ is $\vectrc_i$ (italic letters); the element in the $i$th row and $j$th column of a random matrix $\rmatA$ is $\rmatAc_{ij}$ (sans serif letters), and the $i$th component of the random vector~$\rvectr$ is $\rvectrc_i$ (sans serif letters). 
For a vector $\vectr$, $\diag(\vectr)$ stands for the diagonal matrix that has the entries of $\vectr$ on its main diagonal. The linear subspace spanned by the vectors $\vectr_1,\ldots,\vectr_n$ is denoted by $\spn\{\vectr_1,\ldots,\vectr_n\}$.
 The superscripts~$\tp{}$ and~$\herm{}$ stand for transposition and Hermitian transposition, respectively.
For two matrices~$\mat$ and~$\altmat$, we designate their Kronecker product as~$\mat \kron \altmat$; 
to simplify notation, we use the convention that the ordinary matrix product precedes the Kronecker product, i.e., $\matA\matB\kron\matC\define (\matA\matB)\kron\matC$.
For a finite subset of the set of natural numbers,  $\setI\subset\naturals$, we write $\abs{\setI}$ 
for the cardinality of $\setI$.
For an $M\times N$ matrix $\mat$, and a set of indices $\setI\subset \natseg{1}{M}$, we use~$\matsegs{\mat}{\setI}$ to denote the  $\card{\setI}\times N$ submatrix of $\mat$ containing the rows of $\mat$ with indices in $\setI$.
For two matrices $\matA$ and $\matB$ of arbitrary size,  
$\diag(\matA,\matB)$ is the $2\times2$ block-diagonal matrix that has $\matA$ in the upper left corner and $\matB$ in the lower right corner. For $N$ matrices $\matA_1,\dots,\matA_N$, we let 
$\diag(\matA_1,\dots,\matA_N)\define \diag(\diag(\matA_1,\dots,\matA_{N-1}),\matA_N)$. 
The ordered eigenvalues of the $N\times N$ matrix $\mat$ are denoted by  $\eval_1(\mat)\ge\cdots\ge \eval_N(\mat)$.
For two functions~$\fun(\cdot)$ and~$\altfun(\cdot)$, the notation~$\fun(\cdot)=\landauO(\altfun(\cdot))$ means that~$\lim_{\genvar\to \infty} \abs{\fun(\genvar)/\altfun(\genvar)}$ is bounded.
 For a function $\fun(\cdot)$,
we say that $\fun(\cdot)$ is not identically zero and write $\fun(\cdot)\not\equiv 0$ if there exists at least one element $\vectr$ in the domain of 
$\fun(\cdot)$ such that $\fun(\vectr)\neq 0$. We say that a function $\fun(\cdot)$ is nonvanishing on a subset $\setS$ of its domain, if for all $\vectr\in\setS$, $\fun(\vectr)\neq 0$.
For two functions $\fun(\cdot)$ and $\altfun(\cdot)$, $(\fun\circ\altfun)(\cdot)$ denotes the composition $\fun(\altfun(\cdot))$.
For $x\in\reals$, 
$\lceil x\rceil\define \min\{m\in\integers\mid m\geq x\}$. 
 We use $\natseg{n}{m}$ to designate the set of natural numbers 
$\left\{n, n+1,\ldots,m\right\}$.
Let $\altfunvec:\complexset^{M}\to \complexset^{N},\ \vectr\mapsto \altfunvec(\vectr),$
 be a vector-valued function; then ${\partial \altfunvec}/{\partial \vectr}$ denotes the $N\times M$ Jacobian matrix~\cite[Def. 3.8]{fritzsche02} of the function $\altfunvec(\cdot)$, i.e., the matrix that contains the partial derivative ${\partial \altfun_i}/{\partial \vectrc_j}$ in its $i$th row and $j$th column.
The logarithm to the base 2 is written as $\log(\cdot)$.
For sets $\setA,\setB\subseteq\reals^M$, we define 
$\setA\pm \setB\define\{\veca\pm \vecb\mid \veca\in \setA, \vecb\in \setB\}$.  
If $\setA=\{\veca\}$, then $\veca\pm \setB\define \setA\pm \setB$. With $(-\epsilon,\epsilon)\define \{\genvar\in\reals\mid \abs{\genvar}<\epsilon\}$, 
we denote by $\cube(\vectr,\epsilon)\define \vectr+(-\epsilon,\epsilon)^M\subset\reals^M$ the open cube in $\reals^M$ with side length 
$2\epsilon$ centered at $\vectr\in \reals^M$. The set of natural numbers, including zero, is $\naturals_0$. For $\vectr\in\complexset^M$ and $\vindx\in\naturals_0^M$, we let 
$\vectr^{\vindx}\define \vectrc_1^{\indx_1}\dots \vectrc_M^{\indx_M}$. If $\setA$ is a subset of the image of a map $\fun(\cdot)$ then 
$\fun^{-1}(\setA)$ denotes the inverse image of $\setA$. 
The expectation operator is designated by~$\Ex{}{\cdot}$.  For random matrices $\rmatA$ and $\rmatB$, we write $\rmatA\stackrel{d}{\sim}\rmatB$ to indicate that $\rmatA$ and $\rmatB$ have the same distribution.
Finally, $\jpg(\vectr,\matC)$ stands for the distribution of a
\ac{JPG} random vector with mean~$\vectr$ and covariance
matrix~$\matC$.


\section{System Model} 
\label{sec:system_model}

We consider a \ac{SIMO} channel with $\RXant$ receive antennas. 
The fading in each  \ac{SISO} component  channel follows the correlated block-fading model described in the previous section. 
The \ac{IO} relation within any block of length $\blocklength$ for the $m$th \ac{SISO} component channel can be written as
\begin{equation}
	\label{eq:model1}
	\vout{m}=\sqrt{\SNR}\,\diag(\vchannel{m}) \vinp+\vnoise{m}, \quad m\in\natseg{1}{\RXant},
\end{equation}
where $\vinp=\tp{[\vinpc_1 \cdots\, \vinpc_\blocklength]}\in \complexset^\blocklength$ is the  signal vector transmitted in the given block, and the vectors $\vout{m},\vnoise{m}\in\complexset^\blocklength$ are  the corresponding received signal and additive noise, respectively, at the $m$th receive antenna. 
Finally, $\vchannel{m} \in \complexset^\blocklength$ contains the channel coefficients between the transmit antenna and the $m$th receive antenna. 
We assume that  
$\vchannel{m}\distas~\jpg(\veczero,\sqrtcov\herm\sqrtcov)$, for all $m\in\natseg{1}{\RXant}$, where
$\sqrtcov\in\complexset^{\blocklength\times\rankcov}$ (which is the same for all blocks and all component channels) has rank~$\rankcov\le\blocklength$. 
The entries of the vectors $\vchannel{m}$ are taken to be of unit variance, which implies that the main diagonal entries of $\sqrtcov\herm\sqrtcov$ are equal to 1 and the average received power is constant across time slots. 
It will turn out convenient to write the channel coefficient vector in whitened form as $\vchannel{m}=\sqrtcov \viid{m}$, where $\viid{m}\distas\jpg(\veczero,\matI_{\rankcov})$.
Further, we assume that~$\vnoise{m}\distas\jpg(\veczero,\matI_{\blocklength})$. 
As the noise vector has unit variance components, $\SNR$ in~\fref{eq:model1} can be interpreted as the \ac{SNR}. 
Finally, we assume that $\viid{m}$ and $\vnoise{m}$ are mutually independent, independent across $m$, and change in an independent fashion from block to block.
Note that for $\rankcov=1$ the correlated block-fading model reduces to the constant block-fading model as used in~\cite{hochwald00-05,zheng02-02}. 

With $\vectout\define \tp{[\tp{\vout{1}} \cdots\, \tp{\vout{\RXant}}]}$,
$\vectiid\define \tp{[\tp{\viid{1}} \cdots\, \tp{\viid{\RXant}}]}$, $\vectnoise\define \tp{[\tp{\vnoise{1}} \cdots \tp{\vnoise{\RXant}}]}$, and $\minp\define\diag(\vinp)$, 
we can write the \ac{IO} relation~\fref{eq:model1} in the following---more compact---form
\begin{equation}
	\label{eq:IOstacked}
	\vectout=\sqrt{\SNR}\left(\identity_{\RXant}\kron\minp\sqrtcov\right)\vectiid+\vectnoise.
\end{equation}
The capacity of the channel~\eqref{eq:IOstacked} is defined as
\begin{equation}
	\label{eq:capacitydef}
\capacity(\SNR)\define(1/\blocklength)\sup_{\pdf{\vinp}(\cdot)} \mi(\vinp;\vectout),
\end{equation}
where the supremum is taken over all input distributions $\pdf{\vinp}(\cdot)$ that satisfy the average-power constraint 
\begin{equation}
	\label{eq:apc}
	\Ex{}{\vecnorm{\vinp}^2}\le \blocklength.
\end{equation}
The capacity pre-log, the central quantity of interest in this paper, is defined as  
\begin{equation*}
	\prelog\define \lim_{\SNR\to\infty}\frac{\capacity(\SNR)}{\log(\SNR)}.
\end{equation*}

\section{Intuitive Analysis}
\label{sec:intuition}
We start with a simple ``back-of-the-envelope'' calculation that allows to develop some intuition on the main result in this paper, summarized in~\fref{eq:prelogans}. 
 The different steps in the intuitive analysis below will be seen to have rigorous counterparts in the formal proof of the capacity pre-log lower bound detailed in Section~\ref{sec:flwSIMO}.

The capacity pre-log characterizes the channel capacity behavior in the regime where additive noise can ``effectively'' be ignored. 
To guess the capacity pre-log, it therefore appears prudent to consider the problem of identifying the transmit symbols $\vinpc_{i},\ i\in\natseg{1}{\blocklength},$ from the \emph{noise-free} (and rescaled) observation
\begin{equation}
	\label{eq:IOnonoise}
	\vectoutnn\define\left(\identity_{\RXant}\kron\minp\sqrtcov\right)\vectiid.
\end{equation}
Specifically, we shall ask the question: ``How many symbols $\vinpc_{i}$ can be identified \emph{uniquely} from $\vectoutnn$ given that the vector of channel coefficients $\vectiid$ is unknown but the statistics of the channel, i.e., the matrix $\sqrtcov$, are known?'' 
The claim we make is that the capacity pre-log is given by the number of identifiable symbols divided by the block length $\blocklength$. 

We start by noting that the unknown variables in~\eqref{eq:IOnonoise} are $\vectiid$ and $\vinp$, which means that we have a quadratic system of equations. It turns out, however, that the simple change of variables 
\begin{equation}
\label{eq:changevar}
\invinpc_i\define 1/\vinpc_i,\ i\in\natseg{1}{\blocklength},	
\end{equation}
(we make the technical assumption $\abs{\vinpc_i}>0,\ i\in\natseg{1}{\blocklength}$, in the remainder of this section) transforms~\eqref{eq:IOnonoise} into a system of equations that is linear in $\vectiid$ and $\invinpc_i,\ i\in\natseg{1}{\blocklength}$. Since the transformation $\invinpc_i\define 1/\vinpc_i$ is invertible for $\abs{\vinpc_i}>0$, uniqueness of the solution of the linear system of equations in $\vectiid$ and $\invinpc_i,\ i\in\natseg{1}{\blocklength},$ is equivalent to uniqueness of the solution of the quadratic system of equations in $\vectiid$ and $\vinpc_i,\ i\in\natseg{1}{\blocklength}$.

 For concreteness and simplicity of exposition, we first consider the case $\blocklength=3$ and $\RXant=\rankcov=2$ and assume that $\sqrtcov$ satisfies the technical condition specified in Theorem~\ref{thm:mainLB}, stated in~\fref{sec:charprelog}. A direct computation reveals that upon change of variables according to~\fref{eq:changevar}, the quadratic system~\eqref{eq:IOnonoise} can be rewritten as the following linear system of equations:
\begin{equation}
	\label{eq:linearsystem1}
	\begin{bmatrix}
		\sqrtcovc_{11} & \sqrtcovc_{12} & 0 & 0 & \vectoutnnc_1 & 0 & 0 \\
		\sqrtcovc_{21} & \sqrtcovc_{22} & 0 & 0 & 0 & \vectoutnnc_2 & 0 \\
		\sqrtcovc_{31} & \sqrtcovc_{32} & 0 & 0 & 0 & 0 & \vectoutnnc_3 \\
		0 & 0 & \sqrtcovc_{11} & \sqrtcovc_{12} & \vectoutnnc_4 & 0 & 0 \\
		0 & 0 & \sqrtcovc_{21} & \sqrtcovc_{22} & 0 & \vectoutnnc_5 & 0 \\
		0 & 0 & \sqrtcovc_{31} & \sqrtcovc_{32} & 0 & 0 & \vectoutnnc_6 \\
	\end{bmatrix}
	\begin{bmatrix}
		\vectiidc_1\\
		\vectiidc_2\\
		\vectiidc_3\\
		\vectiidc_4\\
			-\invinpc_1\\
			-\invinpc_2\\
			-\invinpc_3\\
	\end{bmatrix}=\veczero.
\end{equation}
The solution of~\fref{eq:linearsystem1} can not be unique, as we have 6 equations in 7 unknowns.
The $\vinpc_i=1/\invinpc_i,\ i\in\natseg{1}{3},$ can, therefore, not be determined uniquely from $\vectoutnn$. 
We can, however, make the solution of~\fref{eq:linearsystem1} to be unique if we devote one of the data symbols $\vinpc_i$ to transmitting a pilot symbol (known to the receiver).
Take, for concreteness, $\vinpc_1=1$.  Then~\eqref{eq:linearsystem1} reduces to the following inhomogeneous system of 6 equations in 6 unknowns
\begin{equation}
	\underbrace{\begin{bmatrix}
		\sqrtcovc_{11} & \sqrtcovc_{12} & 0 & 0 & 0 & 0\\
		\sqrtcovc_{21} & \sqrtcovc_{22} & 0 & 0 & \vectoutnnc_2 & 0\\
		\sqrtcovc_{31} & \sqrtcovc_{32} & 0 & 0 & 0 & \vectoutnnc_3\\
		0 & 0& \sqrtcovc_{11} & \sqrtcovc_{12}& 0 & 0 \\
		0 & 0& \sqrtcovc_{21} & \sqrtcovc_{22} & \vectoutnnc_5 & 0  \\
		0 & 0& \sqrtcovc_{31} & \sqrtcovc_{32} & 0 & \vectoutnnc_6 \\
	\end{bmatrix}}_{\define\altmat}
	\begin{bmatrix}
		\vectiidc_1\\
		\vectiidc_2\\
		\vectiidc_3\\
		\vectiidc_4\\
		-\invinpc_2\\
		-\invinpc_3\\
	\end{bmatrix}=
	\begin{bmatrix}
		\vectoutnnc_1\\
		0\\
		0\\
		\vectoutnnc_4\\
		0\\
		0\\
	\end{bmatrix}.
	\label{eq:inhomeq}
\end{equation}
This system of equations has a unique solution if $\det\altmat\ne 0$. 
We prove in Appendix~\ref{app:bijection} 
that under the technical condition on $\sqrtcov$ specified in Theorem~\ref{thm:mainLB}, stated in \fref{sec:charprelog}, we, indeed, have that $\det\altmat\ne 0$ for almost all\footnote{Except for a set of measure zero.} $\vectoutnnc_2, \vectoutnnc_3, \vectoutnnc_5, \vectoutnnc_6$. 
It, therefore, follows that for almost all $\vectoutnn$, the linear system of equations~\eqref{eq:inhomeq} has a unique solution. 
As explained above, this implies uniqueness of the solution of the original quadratic system of equations~\fref{eq:IOnonoise}.
We can therefore recover $\invinpc_2$ and $\invinpc_3$, and, hence, $\vinpc_2=1/\invinpc_2$ and $\vinpc_3=1/\invinpc_3$ from $\vectoutnn$. 
Summarizing our findings, we expect that the capacity pre-log of the channel~\eqref{eq:IOstacked}, for the special case~$\blocklength=3$ and $\RXant=\rankcov=2$, is equal to $2/3$, which is larger than the capacity pre-log of the corresponding \ac{SISO} channel (i.e., one of the SISO component channels), given by $1-\rankcov/\blocklength=1/3$~\cite{liang04-12}.  This answer, obtained through the back-of-the-envelope calculation above,  coincides with the rigorous result in Theorem~\ref{thm:mainLB}.

We next generalize what we learned in the example above to $\blocklength, \RXant,$ and $\rankcov$ arbitrary,
and start by noting that if $(\minp,\vectiid)$ is a solution of $\vectoutnn=\left(\identity_{\RXant}\kron\minp\sqrtcov\right)\vectiid$ for fixed $\vectoutnn$, then $(a\minp,\vectiid/a)$ with $a\in \complexset$ is also a solution of this system of equations. It is therefore immediately clear that at least one pilot symbol is needed to make this system of equations uniquely solvable.

To guess the capacity pre-log for general parameters $\blocklength, \RXant,$ and $\rankcov,$ we first note that the homogeneous linear system of   equations corresponding to that in~\fref{eq:linearsystem1}, has $\RXant\blocklength$ equations for $\RXant\rankcov+\blocklength$ unknowns.
As the example above indicates, we need to seek conditions under which this homogeneous linear system of   equations can be converted into a linear system of  equations that has a unique solution. Provided that  $\sqrtcov$ satisfies the technical condition  specified in~\fref{thm:mainLB} below, this entails meeting the following two requirements:
\begin{inparaenum}[(i)]
	\item 
	at least one symbol is used as a pilot symbol to resolve the scaling ambiguity described in the previous paragraph;
	\item
	the number of unknowns in the system of equations corresponding to that in~\fref{eq:linearsystem1} must be smaller than or equal to the number of equations.
\end{inparaenum}
To maximize the capacity pre-log we want to use the minimum number of pilot symbols that guarantees (i) and (ii).
In order to identify this minimum, we have to distinguish two cases:
\begin{enumerate}
	\item
		When $\RXant\blocklength<\RXant\rankcov+\blocklength$ [in this case $\min[1-1/\blocklength,\RXant(1-\rankcov/\blocklength)]=\RXant(1-\rankcov/\blocklength)$] 		we will need at least $\RXant\rankcov+\blocklength-\RXant\blocklength$ pilot symbols to satisfy requirement (ii). Since $\RXant\rankcov+\blocklength-\RXant\blocklength\ge 1$, choosing exactly $\RXant\rankcov+\blocklength-\RXant\blocklength$ pilot symbols will satisfy both requirements.  The number of symbols left for communication will, therefore, be $\blocklength-(\RXant\rankcov+\blocklength-\RXant\blocklength)=\RXant(\blocklength-\rankcov)$. Hence, we expect the capacity pre-log to be given by $\RXant(1-\rankcov/\blocklength)$, which agrees with the result stated in~\fref{eq:prelogans}.
	\item 
	When $\RXant\blocklength\ge\RXant\rankcov+\blocklength$ [in this case $\min[1-1/\blocklength,\RXant(1-\rankcov/\blocklength)]=1-1/\blocklength$], we will need at least one pilot symbol to satisfy requirement (i). Since requirement (ii) is satisfied as a consequence of $\RXant\blocklength\ge\RXant\rankcov+\blocklength$, it suffices to choose exactly one pilot symbol. 
	The number of symbols left for communication will, therefore, be $\blocklength-1$ and  we  hence expect  the capacity pre-log  to equal  $1-1/\blocklength$, which again agrees with the result stated in~\fref{eq:prelogans}. 
	%
 	Note that the resulting inhomogeneous linear system of   equations has $\RXant\blocklength$ equations in $\RXant\rankcov+\blocklength-1$ unknowns. As there are  more equations than unknowns, $\RXant\blocklength-\RXant\rankcov-\blocklength+1$ equations are redundant and can be eliminated. 
\end{enumerate}

The proof of our main result, stated in the next section, will provide rigorous justification for the casual arguments put forward in this section.

\section{The Capacity Pre-Log}
\label{sec:charprelog}
The main result of this paper is the following theorem.

\begin{thm}
\label{thm:mainLB}
Suppose that $\sqrtcov$ satisfies the following 

\propspark: Every $\rankcov$ rows of $\sqrtcov$ are linearly independent.
	
	Then, the capacity pre-log of the \ac{SIMO} channel \eqref{eq:IOstacked} is given by
	\begin{equation}
		\label{eq:capprelog}
		\prelog= \min[1-1/\blocklength,\RXant(1-\rankcov/\blocklength)].
	\end{equation}
	
\end{thm}

\begin{rem}
\label{rem:}
	We will prove~\fref{thm:mainLB} by showing, in~\fref{sec:ub}, that the capacity pre-log of the \ac{SIMO} channel \eqref{eq:IOstacked} can be upper-bounded as
	\begin{equation}
		\label{eq:capUB}
		\prelog \leq \min[1-1/\blocklength,\RXant(1-\rankcov/\blocklength)]
	\end{equation}
	and by establishing, in~\fref{sec:flwSIMO}, the lower bound  
	\begin{equation}\label{eq:mainbound}
		\prelog\ge \min[1-1/\blocklength,\RXant(1-\rankcov/\blocklength)].
	\end{equation}
	While the upper bound~\fref{eq:capUB} can be shown to hold even if~$\sqrtcov$ does not satisfy \propspark, this property is crucial to establish the lower bound  
	\fref{eq:mainbound}.
\end{rem}

\begin{rem}
\label{rem:aprime}
	The lower bound \fref{eq:mainbound} continues to hold if \propspark is replaced by the following milder condition on~$\sqrtcov$. 
	
	\propsparkm: There exists a subset of indices $\setK\subseteq\natseg{1}{\blocklength}$ with cardinality 
	\begin{equation*}
	\abs{\setK}\define \min
	(\lceil(\RXant\rankcov-1)/(\RXant-1)\rceil,\blocklength)
	\end{equation*}
	such that every $\rankcov$ rows of $\sqrtcov_\setK$ are linearly independent.

We decided, however, to state our main result under the stronger \propspark as both \propspark and \propsparkm are very mild and the proof of the lower bound \fref{eq:mainbound}  under  \propsparkm is significantly more cumbersome and does not contain any new conceptual aspects. A sketch of the proof of the stronger result (i.e., under \propsparkm ) can be found in~\cite{riegler11-08}.
\end{rem}

We proceed to discussing the significance of~\fref{thm:mainLB}.

\subsection{Eliminating the prediction penalty} 

According to~\fref{eq:capprelog} the capacity pre-log of the \ac{SIMO} channel~\fref{eq:IOstacked} with $\RXant=2$ receive antennas is given by $\prelog= 1-1/\blocklength$, provided that \propspark holds, and  $\blocklength\ge 2\rankcov-1$. Comparing to the capacity pre-log $\prelog_{\mathrm{SISO}}= 1-\rankcov/\blocklength$ in the \ac{SISO} case\footnote{Note that the results in~\cite{liang04-12} are stated for general channel covariance matrix~$\sqrtcov$. }~\cite{liang04-12} (this result also follows from~\fref{eq:capprelog} with $\RXant=1$), we see that---under a mild condition on the channel covariance matrix $\sqrtcov$---adding only one receive antenna yields a reduction of the channel uncertainty-induced pre-log penalty from $\rankcov/\blocklength$ to $1/\blocklength$. How significant is this reduction?
Recall that $\rankcov$ is the number of uncertain channel parameters within each given block of length $\blocklength$.
Hence, the ratio between the rank of the covariance matrix and the block-length, $\rankcov/\blocklength$, is a measure that can be seen as quantifying the amount of channel uncertainty relative to the number of degrees of freedom for communication. 
It often makes sense to consider $\blocklength\to\infty$ with the amount of channel uncertainty $\rankcov/\blocklength$ held constant. 
 For concreteness, consider  $\blocklength,\rankcov\to\infty$ with $\blocklength=2\rankcov-1$ so that $\rankcov/\blocklength\to 1/2$. The capacity pre-log penalty due to channel uncertainty in the \ac{SISO} case is then given  by $1/2$.  \fref{thm:mainLB} reveals that, by adding a second receive antenna, this penalty can be reduced to $1/\blocklength$ and, hence, be made to vanish in the limit $\blocklength\to\infty$. 
Intuitively, even though the \ac{SISO} channels between the transmit antenna and the two receive
antennas are statistically independent, the transmit signal induces enough statistical
dependence between the corresponding receive signals for the second receive antenna to be able to resolve
the channel uncertainty associated with the first receive antenna's channel and thereby make the overall system appear coherent.

\subsection{Number of receive antennas} 
\label{sec:optnrec}

Note that for $\rankcov<\blocklength$, we can rewrite~\fref{eq:capprelog} as 
\begin{align}
	\prelog&=\min[1-1/\blocklength,\RXant(1-\rankcov/\blocklength)]\nonumber\\
	&=\begin{cases}
	1-1/\blocklength,\ &\text{if}\ \RXant\ge \lceil\frac{\blocklength-1}{\blocklength-\rankcov}\rceil\\
	\RXant(1-\rankcov/\blocklength),\ &\text{else.}\label{eq:car2}
	\end{cases}
\end{align}%
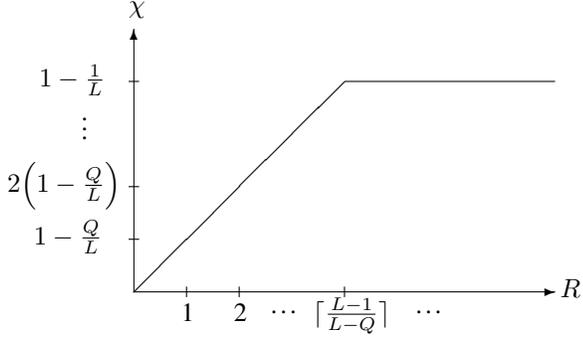
\begin{figure}
	\center{
	\setlength{\unitlength}{0.7mm}
	\begin{picture}(150,70)
	\put( 30,5){\vector(0,1){50}}
	\put( 30,5){\vector(1,0){80}}
	\multiput(40,4)(10,0){2}{\line(0,1){2}}
	\multiput(70,4)(10,0){1}{\line(0,1){2}}
	\put(39,-0.5){1}
	\put(49,-0.5){2}
	\put(56,1){\dots}
	\put(64,-0.5){$\lceil\frac{\blocklength-1}{\blocklength-\rankcov}\rceil$}
	\put(83.5,1){\dots}
	\put(111,4){$\RXant$} 
	\put(29,58){$\prelog$} 
	\multiput(29,15)(0,10){2}{\line(1,0){2}}
	\multiput(29,45)(0,10){1}{\line(1,0){2}}
	\put(11,14){$1-\frac{\rankcov}{\blocklength}$}
	\put(6,24){$2\Bigl(1-\frac{\rankcov}{\blocklength}\Bigr)$}
	\put(13,34){
	\put( 7,0){.}
	\put( 7,1.8){.}
	\put( 7,3.6){.}
	}
	\put(12,44){$1-\frac{1}{\blocklength}$}
	\put( 30,5){\line(1,1){40}}
	\put( 70,45){\line(1,0){40}}
	\end{picture}}
\caption{The capacity pre-log of the \ac{SIMO} channel~\fref{eq:IOstacked}.}
\label{fig:prelog}
\end{figure}%
As illustrated in~\fref{fig:prelog}, it follows from~\fref{eq:car2} that for fixed $\blocklength$ and $\rankcov$ with $\rankcov<\blocklength$ the capacity pre-log of the \ac{SIMO} channel~\fref{eq:IOstacked} grows linearly with $\RXant$ as long as $\RXant$ is smaller than the critical value $\lceil(\blocklength-1)/(\blocklength-\rankcov)\rceil$. Once $\RXant$ reaches this critical value, further increasing the number of receive antennas does not increase the capacity pre-log.

\subsection{\propspark is mild}
\propspark is not very restrictive and is satisfied by many practically relevant channel covariance matrices $\sqrtcov$. 
	For example, removing an arbitrary set of $\blocklength-\rankcov$ columns from an $\blocklength\times\blocklength$ \ac{DFT} matrix results in a matrix that satisfies \propspark when $\blocklength$ is prime~\cite{tao05}. 
(Weighted) \ac{DFT} covariance matrices arise naturally in so-called basis-expansion models for time-selective channels~\cite{liang04-12}.

\propspark can furthermore be shown to be satisfied by ``generic'' matrices $\sqrtcov$. Specifically, if the entries of $\sqrtcov$ are chosen randomly and independently from a continuous distribution~\cite[Sec. 2-3, Def. (2)]{grimmett01} (i.e., a distribution with a well-defined \ac{PDF}), then the resulting matrix $\sqrtcov$ will satisfy \propspark with probability one. The proof of this statement follows from a union bound argument together with the fact that 
$N$ independent $N$-dimensional vectors drawn independently from a continuous distribution are linearly independent with probability one.

\section{Proof of the Upper Bound~\fref{eq:capUB}}
\label{sec:ub}

The proof of~\fref{eq:capUB} consists of two parts. 
First, in~\fref{sec:firstpart}, we prove that $\prelog\le \RXant(1-\rankcov/\blocklength)$. This will be accomplished by generalizing---to the \ac{SIMO} case---the approach developed in~\cite[Prop.~4]{liang04-12} for establishing an upper bound on the \ac{SISO} capacity pre-log. 
Second, in~\fref{sec:secondpart}, we prove that $\prelog\le 1-1/\blocklength $ by showing that the capacity of a \ac{SIMO} channel with $\RXant$ receive antennas and  channel covariance matrix of rank $\rankcov$ can be upper-bounded by the capacity of a \ac{SIMO} channel with $\RXant\rankcov$ receive antennas, the same \ac{SNR}, and a rank-1 covariance matrix. The desired result, $\prelog\le 1-1/\blocklength$, then follows by application of~\cite[Eq. (27)]{zheng02-02}, \cite[Eq. (7)]{yang12-02} as detailed below.

\subsection{First part: $\prelog\le \RXant(1-\rankcov/\blocklength)$} 
To simplify notation, we first rewrite~\fref{eq:IOstacked} as
\begin{equation}
	\label{eq:model3}
	\mout=\sqrt{\SNR}\diag(\vinp)\sqrtcov \miid+\mnoise,
\end{equation}
where $\mout\define [\vout{1} \cdots\, \vout{\RXant}]$, $\mchannel\define [\vchannel{1} \cdots\, \vchannel{\RXant}]$, 
$\mnoise\define [\vnoise{1} \cdots\, \vnoise{\RXant}]$, and $\miid\define[\viid{1} \cdots\, \viid{\RXant}]$.

Recall that $\sqrtcov$ has rank $\rankcov$. Without loss of generality, we  assume, in what follows, that the first $\rankcov$ rows of
\label{sec:firstpart}
 $\sqrtcov$ are linearly independent. This can always be ensured by reordering the scalar \ac{IO} relations in~\fref{eq:model1}.  With $\setQ\define\natseg{1}{\rankcov}$ and $\setL\define\natseg{\rankcov+1}{\blocklength}$ we can write 
\begin{align}
	\mi(\mout;\vinp)
	&=\mi\lefto(\matsegs{\mout}{\setQ}, \matsegs{\mout}{\setL}; \vinp\right)\nonumber\\
	&\stackrel{(a)}{=}\mi\lefto(\matsegs{\mout}{\setQ}; \vinp\right)+\mi\lefto(\matsegs{\mout}{\setL}; \vinp\given \matsegs{\mout}{\setQ}\right)\nonumber\\
	&\stackrel{(b)}{=}\mi\lefto(\matsegs{\mout}{\setQ}; \vecseg{\vinp}{\setQ}\right)+\underbrace{\mi\lefto(\matsegs{\mout}{\setQ}; \vecseg{\vinp}{\setL}\given \vecseg{\vinp}{\setQ}\right)}_{0}+\mi\lefto(\matsegs{\mout}{\setL}; \vinp\given \matsegs{\mout}{\setQ}\right)\nonumber\\
	&\stackrel{(c)}{=}\mi\lefto(\matsegs{\mout}{\setQ}; \vecseg{\vinp}{\setQ}\right)+\mi\lefto(\matsegs{\mout}{\setL}; \vinp\given \matsegs{\mout}{\setQ}\right),
	\label{eq:UBRsimo1}
\end{align}
where (a) and (b) follow by the chain rule for mutual information and in (c) we used that $\matsegs{\mout}{\setQ}$  and $\vecseg{\vinp}{\setL}$ are independent  conditional on $\vecseg{\vinp}{\setQ}$.
Next, we upper-bound each term in~\fref{eq:UBRsimo1} separately. 

From~\cite[Thm. 4.2]{lapidoth03-10} we can conclude that the assumption of the first $\rankcov$ rows of $\sqrtcov$ being linearly independent implies that the first term on the RHS of \fref{eq:UBRsimo1} grows at most double-logarithmically with \ac{SNR} and hence does not contribute to the capacity pre-log. For the reader's convenience, we repeat the corresponding brief calculation from~\cite[Thm. 4.2]{lapidoth03-10} in \fref{app:repeatamos} and show that:
\begin{equation}
	\mi\lefto(\matsegs{\mout}{\setQ}; \vecseg{\vinp}{\setQ}\right)\le \rankcov \log\log(\SNR)+\asconst\label{eq:UB3}.
\end{equation} 
Here and in what follows, $\asconst$ refers to the limit $\SNR\to\infty$.  

For the second term in~\fref{eq:UBRsimo1} we can write
\begin{align}
	\mi\lefto(\matsegs{\mout}{\setL}; \vinp\given \matsegs{\mout}{\setQ}\right)
	&=\diffent\lefto(\matsegs{\mout}{\setL}\given \matsegs{\mout}{\setQ}\right)-\diffent\lefto(\matsegs{\mout}{\setL}\given \vinp, \matsegs{\mout}{\setQ}\right)\nonumber\\
	&\stackrel{(a)}{\le} \diffent\lefto(\matsegs{\mout}{\setL}\right)-\diffent\lefto(\matsegs{\mout}{\setL}\given \vinp, \matsegs{\mout}{\setQ}, \vectiid\right)\nonumber\\
	&=  \diffent\lefto(\matsegs{\mout}{\setL}\right)-\diffent\lefto(\matsegs{\mnoise}{\setL}\right)\nonumber\\
	&\stackrel{(b)}{\le} \sum_{l=\rankcov+1}^\blocklength\sum_{r=1}^\RXant \left(\diffent\lefto(\moutc_{lr}\right)-\diffent\lefto(\mnoisec_{lr}\right)\right)\nonumber\\
	&\stackrel{(c)}{\le} \sum_{l=\rankcov+1}^\blocklength\sum_{r=1}^\RXant \log\lefto(1+\SNR\Ex{}{\abs{\mchannelc_{lr}}^2} \Ex{}{\abs{\vinpc_{l}}^2}\right)\nonumber\\
	&\stackrel{(d)}{\le} \sum_{l=\rankcov+1}^\blocklength\sum_{r=1}^\RXant \log\lefto(1+\blocklength\SNR\Ex{}{\abs{\mchannelc_{lr}}^2} \right)\nonumber\\
	&\stackrel{(e)}{=} \RXant (\blocklength-\rankcov)\log(\SNR)+\asconst,\label{eq:UB4}
\end{align}
where in (a) we used the fact that conditioning reduces entropy; (b) follows from the chain rule for differential entropy and the fact that conditioning reduces entropy; (c) follows because Gaussian random variables are differential-entropy-maximizers for fixed variance and because $\mchannelc_{lr}$ and $\vinpc_{l}$ are independent; (d) is a consequence of the power constraint~\fref{eq:apc}; and (e) follows because $\Ex{}{\abs{\mchannelc_{lr}}^2}=1$. 

Combining~\fref{eq:UBRsimo1}, \fref{eq:UB3}, and~\fref{eq:UB4} yields 
\begin{equation}
	\label{eq:UB1}
\capacity(\SNR) \leq \RXant(1-\rankcov/\blocklength) \log(\SNR)+\left(\rankcov/\blocklength\right)\log\log(\SNR)+\asconst.	
\end{equation}
Since $\lim_{\SNR\to\infty}\log\log(\SNR)/\log(\SNR)=0,$ this completes the proof of the bound $\prelog\le \RXant(1-\rankcov/\blocklength)$.

It follows from \fref{eq:UB1} that for $\rankcov=\blocklength$, the capacity pre-log is zero and $\capacity(\SNR)$ can grow no faster than double-logarithmically in $\SNR$. 

Recall that $1-\rankcov/\blocklength$ is the capacity pre-log of the correlated block-fading \ac{SISO} channel~\cite{liang04-12}. As the proof of the upper bound  $\prelog\le\RXant(1-\rankcov/\blocklength)$ reveals, the capacity pre-log of the \ac{SIMO} channel \eqref{eq:IOstacked} can not be larger than $\RXant$ times the capacity pre-log of the corresponding \ac{SISO} channel (i.e., the capacity pre-log of one of the \ac{SISO} component channels). The upper bound $\RXant(1-\rankcov/\blocklength)$ may seem crude, but, surprisingly, it matches the lower bound for $\RXant< \lceil(\blocklength-1)/(\blocklength-\rankcov)\rceil$.

\subsection{Second part: $\prelog\leq 1-1/\blocklength$} 
\label{sec:secondpart}
The proof of $\prelog\leq 1-1/\blocklength$ will be accomplished in two steps. 
In the first step, we show that the capacity of a \ac{SIMO} channel with $\RXant$ receive antennas and rank-$\rankcov$ channel covariance matrix is upper-bounded by the capacity of a \ac{SIMO} channel  with $\RXant\rankcov$ receive antennas, the same \ac{SNR}, and rank-$1$ covariance matrix.
In the second step, we exploit the fact that the channel~\fref{eq:model3} with rank-$1$  covariance matrix (under the assumption that the rows of $\sqrtcov$ have unit norm) is a constant block-fading channel for which the capacity pre-log was shown in \cite{zheng02-02} to equal $1-1/\blocklength$. 
We now implement the proof program just outlined.

 Let $\vsqrtcov_1,\ldots,\vsqrtcov_\rankcov\in\complexset^{\blocklength}$ denote the columns of the $\blocklength\times\rankcov$ matrix $\sqrtcov$ so that $\sqrtcov=[\vsqrtcov_1\cdots\vsqrtcov_\rankcov]$. 
 Let $\rvmiid_1,\ldots,\rvmiid_\rankcov\in\complexset^{\RXant}$ denote the transposed rows of the $\rankcov\times\RXant$ matrix $\miid$ so that $\tp\miid=[\rvmiid_1\cdots\,\rvmiid_\rankcov]$. 
We can rewrite the IO relation~\eqref{eq:model3} in the following form that is more convenient for the ensuing analysis:
\begin{equation*}
\mout = \sqrt{\SNR}\sum\limits_{q=1}^{\rankcov}  \diag(\vsqrtcov_q) \vinp\tp{\rvmiid}_q  + \mnoise.
\end{equation*}    
Let $\mnoise_1, \ldots,\mnoise_\rankcov$ be independent random matrices of dimension $ \blocklength\times \RXant$, each with \iid $\jpg(0,1)$ entries.
As, by assumption, the rows of \sqrtcov have unit norm, we have that 
\begin{equation*}
\mnoise \stackrel{d}{\sim} \sum\limits_{q=1}^\rankcov  \diag(\vsqrtcov_q)\mnoise_q.
\end{equation*}
Hence, we can rewrite $\mout$ as   
\begin{equation}
	\label{eq:ytoyq}
\mout \stackrel{d}{\sim} \sum \limits_{q=1}^{\rankcov} \diag(\vsqrtcov_q)\mout_q,
\end{equation}
where
\begin{equation}
	\label{eq:yq}
\mout_q\define \sqrt{\SNR}\vinp\tp{\rvmiid}_q +\mnoise_q,\ q\in\natseg{1}{\rankcov}.
\end{equation}
Note now that each $\mout_q$ is the output of a \ac{SIMO} channel with $\RXant$ receive antennas, rank-$1$ channel covariance matrix, and \ac{SNR}~$\SNR$.  
Realizing that, by~\fref{eq:ytoyq} and~\fref{eq:yq}, $\vinp\to\{\mout_1,\ldots,\mout_{\rankcov}\}\to\mout$ forms a Markov chain, we conclude, by the data-processing inequality~\cite[Sec. 2.8]{cover06}, that                                                                                               
\begin{equation*}
\mi(\mout;\vinp)\leq\mi\lefto(\mout_1,\ldots,\mout_{\rankcov};\vinp\right)\!.
\end{equation*}
The claim now follows by noting that the $ \blocklength\times(\RXant\rankcov) $ matrix obtained by stacking the matrices $\mout_q$ next to each other can be interpreted as the output of a \ac{SIMO} channel with $\RXant\rankcov$ receive antennas, rank-$1$  covariance matrix, independent fading across receive antennas, and \ac{SNR} $\SNR$. The proof is completed by upper-bounding the capacity of this channel by means of the following lemma.

\begin{lem}\label{lem:SIMO_rank_one}
	The capacity of the \ac{SIMO} channel~\eqref{eq:model3} with $\RXant$ receive antennas, $\rankcov=1$, and $\blocklength\geq 2$  can be upper-bounded according to
	\begin{equation*}
		\capacity(\SNR)\le  \left(1-{1}/{\blocklength}\right)\log\SNR + \asconst, \quad\SNR\rightarrow \infty.
	\end{equation*}
\end{lem}                                                                                   
This result follows from~\cite[Eq.~(27)]{zheng02-02}. A simpler and more detailed proof can be found in~\cite[Eq.~(7)]{yang12-02}.

\section{Proof of the Lower Bound~\fref{eq:mainbound}}
\label{sec:flwSIMO}
To help the reader navigate through the proof of the lower bound~\fref{eq:mainbound}, we start by explaining the architecture of the proof.
\subsection{Architecture of the proof}
 The proof consists of the following steps, each of which corresponds to a subsection in this section:
\begin{enumerate} [Step 1:]
	\item Choose an input distribution; we will see that \iid $\jpg(0,1)$ input symbols allow us to establish the capacity pre-log lower bound~\fref{eq:mainbound}.  
	\item Decompose the mutual information between the input and the output of the channel according to $\mi(\vinp;\vectout)=\diffent(\vectout)-\diffent(\vectout\given \vinp)$.
	\label{step:condent}
	\item \label{step:3} Using standard information-theoretic bounds show that $\diffent(\vectout\given \vinp)$ is upper-bounded by
	 $\RXant\rankcov\log(\SNR)+\asconst$.
	\item Split $\diffent(\vectout)$ into three terms: a term that depends on \ac{SNR}, a differential entropy term that depends on the noiseless channel output $\vectoutnn$ only, and a differential entropy term that depends on the noise vector $\vectnoise$ only. Conclude that the last of these three terms is a finite constant\footnote{Here, and in what follows, whenever we say ``finite constant'', we mean  \ac{SNR}-independent and finite.}.
	\label{step:decomp}
	\item Conclude that the \ac{SNR}-dependent term obtained in Step~\ref{step:decomp} scales (in \ac{SNR}) as $\min[\RXant\rankcov +\blocklength-1,\RXant\blocklength]\log(\SNR)$.  Together with the decomposition from Step~\ref{step:condent} and the result from Step~\ref{step:3} this gives the desired lower bound~\fref{eq:mainbound} provided that the $\vectoutnn$-dependent differential entropy obtained in Step~\ref{step:decomp} can be lower-bounded by a finite constant.
	\item To show that the $\vectoutnn$-dependent differential entropy obtained in Step~\ref{step:decomp} can be lower-bounded by a finite constant, apply the  change of variables $\vectoutnn\to(\vinp,\vectiid)$ to rewrite the differential entropy as a sum of the differential entropy of $(\vinp, \vectiid)$ and the expected (w.r.t. $\vinp$ and $\vectiid$) logarithm of the Jacobian determinant corresponding to the transformation $\vectoutnn\to(\vinp,\vectiid)$. Conclude that the differential entropy of $(\vinp, \vectiid)$ is a finite constant.  It remains to show that the expected logarithm of  the Jacobian determinant is lower-bounded by a finite constant as well.
	\item Factor out the $\vinp$-dependent terms from the expected logarithm of the Jacobian determinant and conclude that these terms are  finite constants. It remains to show that the expected logarithm of the $\vectiid$-dependent factor in the Jacobian determinant is lower-bounded by a finite constant as well. This poses the greatest technical difficulties in the proof of the lower bound~\fref{eq:mainbound} and is addressed in the remaining steps. 
	\item Based on a deep result from algebraic geometry, known as Hironaka's Theorem on the Resolution of Singularities, conclude that the expected logarithm of the $\vectiid$-dependent factor in the Jacobian determinant is lower-bounded by a finite constant, provided that this factor is nonzero for at least one element in its domain.
	\item Prove by explicit construction that there exists at least one $\vectiid$, for which the $\vectiid$-dependent factor in the Jacobian determinant is nonzero.
\end{enumerate}

We next implement the proof program outlined above.

\subsection{Step 1: Choice of input distribution}
\label{sec:basicbounds}
First note that for $\rankcov=\blocklength$ the lower bound in \fref{eq:mainbound} is reduced to $\prelog\ge 0$ and is hence trivially satisfied. In the remainder of the paper we shall therefore assume that $\rankcov<\blocklength$.

We shall furthermore work under the assumption 
\begin{equation}
	\RXant\le \left\lceil\frac{\blocklength-1}{\blocklength-\rankcov}\right\rceil,
	\label{eq:RXass}
\end{equation}
which trivially leads to a capacity pre-log lower bound as  capacity is a nondecreasing function of $\RXant$ (one can always switch off receive antennas). 

A capacity lower bound is trivially obtained by evaluating the mutual information in \eqref{eq:capacitydef} for an appropriate 
input distribution. Specifically, we take \iid $\vinpc_i\sim \jpg(0,1)$, $i\in\natseg{1}{\blocklength}$.
This implies that $\diffent(\vinpc_i) > -\infty, i\in\natseg{1}{\blocklength},$ and, hence~\cite[Lem. 6.7]{lapidoth03-10}, 
\begin{align}\label{eq:propx}
\Ex{}{\log(\abs{\vinpc_i})}>-\infty,\quad i\in\natseg{1}{\blocklength}. 
\end{align}
We point out that every input vector with \iid, zero mean, unit variance entries $\vinpc_i$ that satisfy $\diffent(\vinpc_i) > -\infty, i\in\natseg{1}{\blocklength},$ would allow us to prove~\fref{eq:mainbound}. The choice $\vinpc_i\sim \jpg(0,1)$ is made for concreteness and convenience.

\subsection{Step 2: Mutual information decomposition}
Decompose
\begin{equation}
	\label{eq:midecomp}
	\mi(\vinp;\vectout)=\diffent(\vectout)-\diffent(\vectout\given \vinp)
\end{equation}
and separately bound the two differential entropy terms for the input distribution chosen in Step 1. 

\subsection{Step 3: Analysis of $\diffent(\vectout\given \vinp)$}
As~$\vectout$ conditioned on~$\vinp$ is  \ac{JPG}, the conditional differential entropy~$\diffent(\vectout\given \vinp)$ can be upper-bounded in a straightforward manner as follows:
\begin{align}
	\label{eq:UBcond}
		&\diffent(\vectout\given \vinp)=\RXant \blocklength \log(\pi e)\nonumber\\
		&+\Ex{\vinp}{\log\det \lefto(\identity_{\RXant \blocklength}+
		\SNR\left(\identity_{\RXant}\kron\minp\sqrtcov\right)\Ex{\vectiid}{\vectiid\herm\vectiid}\left(\identity_{\RXant}\kron\herm\sqrtcov\herm\minp\right)\right)}\nonumber	\\
							&=\RXant \blocklength \log(\pi e)+\RXant\Ex{\vinp}{\log\det \lefto(\identity_{ \blocklength}+
												\SNR\!\left(\minp\sqrtcov\herm\sqrtcov\herm\minp\right)\right)}\nonumber\\
										&=\RXant \blocklength \log(\pi e)+\RXant\Ex{\vinp}{\log\det \lefto(\identity_{ \rankcov}+
					\SNR\!\left(\herm\sqrtcov\herm\minp\minp\sqrtcov\right)\right)}\nonumber\\
					&\stackrel{(a)}{\le} \RXant \blocklength \log(\pi e)+\RXant\log\det \lefto(\identity_{ \rankcov}+
						\SNR\!\left(\herm\sqrtcov\Ex{\vinp}{\herm\minp\minp}\!\sqrtcov\right)\right)\nonumber\\
					&=\RXant \blocklength \log(\pi e)+\RXant\sum_{i=1}^{\rankcov}\log \lefto(1+
							\SNR\eval_i\lefto(\herm\sqrtcov\sqrtcov\right)\right)\nonumber\\
			&\stackrel{(b)}{\le}\RXant \rankcov\log \lefto(\SNR\right)+\asconst.
\end{align}
Here,~(a) follows from Jensen's inequality, and~(b)
holds because $\sqrtcov$ has rank~$\rankcov$  
and, therefore, $\eval_i\lefto(\herm\sqrtcov\sqrtcov\right)>0$ for all $i\in\natseg{1}{\rankcov}$.

\subsection{Step 4: Splitting $\diffent(\vectout)$ into three terms}
Finding an asymptotically (in \ac{SNR}) tight lower bound on $\diffent(\vectout)$ is the main technical challenge of the proof of~\fref{thm:mainLB}.
The back-of-the-envelope calculation presented in~\fref{sec:intuition} suggests that the problem can be approached by splitting $\diffent(\vectout)$ into a term that depends on the \emph{noiseless} channel output $\vectoutnn=\left(\identity_{\RXant}\kron\minp\sqrtcov\right)\vectiid$ only and a term that depends on  noise~$\vectnoise$ only.
 This can be realized as follows. 

Consider a set of indices $\setI\subseteq \natseg{1}{\blocklength\RXant}$ (we shall later discuss how to choose $\setI$) and 
define the following projection matrices
\begin{align*}
\Pm&\define {(\identity_{\blocklength\RXant})}_{\setI}\\
\Qm&\define {(\identity_{\blocklength\RXant})}_{\natseg{1}{\blocklength\RXant}\setminus\setI}. 
\end{align*}
We can lower-bound $\diffent(\vectout)$ according to
\begin{align}
\diffent\lefto(\vectout\right)\label{eq:lboundhy}
&=\diffent(\Pm\vectout,\Qm\vectout)\nonumber\\
&\stackrel{(a)}{=}\diffent\lefto(\Pm\vectout\right)+\diffent\lefto(\Qm\vectout\given\Pm\vectout\right)\nonumber\\
&\stackrel{(b)}{\geq} \diffent\lefto(\sqrt{\SNR}\Pm\vectoutnn+\Pm\vectnoise\given\Pm\vectnoise\right)+\diffent\lefto(\Qm\vectoutnn+\Qm\vectnoise\given\Qm\vectoutnn,\Pm\vectout\right)\nonumber\\
&\stackrel{(c)}{=} \diffent\lefto(\sqrt{\SNR}\Pm\vectoutnn\right)+\diffent\lefto(\Qm\vectnoise\given \Pm\vectout\right)\nonumber\\
&\stackrel{(d)}{=} \diffent\lefto(\sqrt{\SNR}\Pm\vectoutnn\right)+\diffent\lefto(\Qm\vectnoise\right)\nonumber\\
&\stackrel{(e)}{=} \outremsize\log(\SNR)+\diffent\lefto(\Pm\vectoutnn\right) +  \constalt.
\end{align}
Here, (a) follows by the chain rule for differential entropy; (b)~follows from~\fref{eq:IOstacked}, \fref{eq:IOnonoise}, and because conditioning reduces entropy; (c) follows because differential entropy is invariant under translations and because $\vectnoise$ and $\vectoutnn$ are independent; (d) follows because $\Qm\vectnoise$ and $\Pm\vectout$ are independent; and in (e) we used the fact that $\Pm\vectoutnn$ is a $\outremsize$-dimensional vector and  $\diffent\lefto(\Qm\vectnoise\right)=\constalt$, where $\constalt$ here and in what follows denotes a constant that is independent of $\SNR$ and can take a different value at each appearance.

Through this chain of inequalities, we disposed of noise~\vectnoise and isolated \ac{SNR}-dependence into a separate term. 
This corresponds to considering the noise-free \ac{IO} relation~\eqref{eq:IOnonoise} in the back-of-the-envelope calculation. 
Note further that we also rid ourselves of the components of $\vectoutnn$ indexed by 
$\natseg{1}{\blocklength\RXant}\setminus\setI$; this corresponds to eliminating unnecessary equations in the back-of-the-envelope calculation. The specific choice of the set $\setI$ is crucial and will be discussed next. 

\subsection{Step 5: Analysis of the \ac{SNR}-dependent term in \fref{eq:lboundhy}}
If $\diffent\lefto(\Pm\vectoutnn\right)>-\infty$, we can substitute~\fref{eq:lboundhy} and~\fref{eq:UBcond} into~\fref{eq:midecomp} which then yields 
a capacity lower bound of the form 
\begin{equation}
	\label{eq:cap1bound}
	\capacity(\SNR) \geq \frac{\outremsize-\RXant\rankcov}{\blocklength}\log(\SNR)+\asconst.
\end{equation}
This bound needs to be tightened by choosing the set $\setI$
 such that $\outremsize$ is as large as possible while guaranteeing $\diffent\lefto(\Pm\vectoutnn\right)>-\infty$. 
Comparing the lower bound~\fref{eq:cap1bound} to the upper bound~\fref{eq:capUB} we see that the bounds match if 
\begin{equation}
	\label{eq:Icond}
	\outremsize =\min[\RXant\rankcov +\blocklength-1,\RXant\blocklength].
 \end{equation}
Condition~\fref{eq:Icond} dictates that for $\RXant\blocklength\le \RXant\rankcov +\blocklength-1$
we must set $\setI=\natseg{1}{\RXant\blocklength}$, which yields $\Pm\vectoutnn=\vectoutnn$. When $\RXant\blocklength> \RXant\rankcov +\blocklength-1$
the set $\setI$ must be a proper subset of $\natseg{1}{\RXant\blocklength}$. Specifically, we shall choose $\setI$ as follows. Set 
\begin{equation}\label{eq:defshort}
	\shortant=\begin{cases}
	\RXant(\blocklength-\rankcov)-(\blocklength-1),\ &\text{if}\ \RXant\blocklength>\RXant\rankcov +\blocklength-1\\
	0,\  &\text{if}\ \RXant\blocklength\le \RXant\rankcov +\blocklength-1,
	\end{cases}
\end{equation}
let
\begin{equation*}
	\setI_r=\begin{cases}
	\natseg{(r-1)\blocklength+1}{r\blocklength-1},\ &1\le r\le \shortant\\
	\natseg{(r-1)\blocklength+1}{r\blocklength},\ &\shortant+1\le r \le \RXant,
	\end{cases}
\end{equation*}
and define $\setI\define\Union_{r=1}^{\RXant} \setI_r$. 

This choice can be verified to satisfy~\fref{eq:Icond}. 
Obviously, this is not the only choice for~$\setI$ that satisfies~\fref{eq:Icond}. The specific set $\setI$ chosen here will be seen to guarantee $\diffent\lefto(\Pm\vectoutnn\right)>-\infty$ and at the same time simplify the calculations in~\fref{sec:analdet}. 

Substituting~\fref{eq:Icond} into~\fref{eq:cap1bound}, we obtain the desired result~\fref{eq:mainbound}, provided that $\diffent\lefto(\Pm\vectoutnn\right)>-\infty$. Establishing that $\diffent\lefto(\Pm\vectoutnn\right)>-\infty$ is, as already mentioned, the major technical difficulty in the proof of \fref{thm:mainLB} and will be addressed next.

\subsection{Step 6: Analysis of $\diffent(\Pm\vectoutnn)$ through change of variables} 
\label{sec:diffentPv}

It is difficult to analyze $\diffent(\Pm\vectoutnn)$ directly since $\vectoutnn=\left(\identity_{\RXant}\kron\minp\sqrtcov\right)\vectiid$ depends on the pair of variables $(\vectiid,\vinp)$ in a nonlinear fashion. We have seen, in~\fref{sec:intuition}, that~\fref{eq:IOnonoise} has a unique solution in~$(\vectiid,\vinp)$, provided that the appropriate number of pilot symbols is used. This suggests that there must be a one-to-one correspondence between $\Pm\vectoutnn$ and the pair $(\vectiid,\vinp)$.
The existence of such a one-to-one correspondence allows us to locally linearize the equation  $\vectoutnn=\left(\identity_{\RXant}\kron\minp\sqrtcov\right)\vectiid$ and to relate $\diffent(\Pm\vectoutnn)$ to $\diffent(\vectiid,\vinp)= \diffent(\vectiid)+\diffent(\vinp)$. This idea is key to bringing $\diffent(\Pm\vectoutnn)$ into a form that eventually allows us to conclude that $\diffent(\Pm\vectoutnn)>-\infty$.

Formally, it is possible to relate the differential entropies of two random vectors of the same dimension that  
are related by a \emph{deterministic one-to-one} function (in the sense of~\cite[p.7]{rudin87}) according to the following lemma.
\begin{lem}[Transformation of differential entropy]
	\label{lem:Entrchange}
	Assume that $\altfunvec:\complexset^N\to\complexset^N$ is a continuous vector-valued function that is one-to-one and differentiable almost everywhere (a.e.) on $\complexset^N$. Let $\rvectr\in\complexset^N$ be a continuous~\cite[Sec. 2-3, Def. (2)]{grimmett01} random vector (i.e., it has a well-defined \ac{PDF}) and let $\raltvectr=\altfunvec(\rvectr)$.  
	Then
	\begin{equation*}
		\diffent(\raltvectr)=\diffent(\rvectr)+2 \Ex{\rvectr}{\log\abs{\det\lefto({\partial\altfunvec}/{\partial \rvectr}\right)}},
	\end{equation*}	
	where ${\partial\altfunvec}/{\partial \rvectr}$ is the Jacobian of the function $\altfunvec(\cdot)$.
\end{lem}
	The proof follows from the change-of-variables theorem for integrals~\cite[Thm. 7.26]{rudin87} and is given in \fref{app:Entrchange} for completeness since the version of the theorem for complex-valued functions does not seem to be well documented in the literature.

Note that $\Pm\vectoutnn\in\complexset^{\outremsize}$ with $\outremsize$ given in~\fref{eq:Icond}  and $\tp{[\tp\vectiid\ \tp\vinp]}\in\complexset^{\RXant\rankcov+\blocklength}$. Since $\outremsize<\RXant\rankcov+\blocklength$ (see \fref{eq:Icond}),  the vectors $\Pm\vectoutnn$ and $\tp{[\tp\vectiid\ \tp\vinp]}$ are of different dimensions and~\fref{lem:Entrchange} can therefore not be applied directly to relate $\diffent(\Pm\vectoutnn)$ to $\diffent(\vectiid,\vinp)$. 
This problem can be resolved by conditioning on a subset $\pilotset\subset\natseg{1}{\blocklength}$ (specified below) of components of $\vinp$ according to
\begin{equation}\label{eq:entPbound}
	\diffent\lefto(\Pm\vectoutnn\right)\ge \diffent\lefto(\Pm\vectoutnn\given  \vecseg{\vinp}{\pilotset}\right).
\end{equation}
The components $\vecseg{\vinp}{\pilotset}$ correspond to the pilot symbols in the back-of-the-envelope calculation. The set $\pilotset$ is chosen such that 
\begin{inparaenum}[(i)]
	\item 
	the set of remaining components in $\vinp$, $\nonpilotset=\natseg{1}{\blocklength}\setminus \pilotset$, is of appropriate size ensuring that $\Pm\vectoutnn$ and $\tp{[\tp\vectiid\ \vecseg{\tp\vinp}{\nonpilotset}]}$ are of the same dimension, and 
	\item
	$\Pm\vectoutnn$ and $\tp{[\tp\vectiid\ \vecseg{\tp\vinp}{\nonpilotset}]}$ are related by a deterministic bijection so that \fref{lem:Entrchange} can be applied to relate~$\diffent\lefto(\Pm\vectoutnn\given  \vecseg{\vinp}{\pilotset}\right)$ to $\diffent\lefto(\vectiid, \vinp_\nonpilotset \given  \vecseg{\vinp}{\pilotset}\right)$.
\end{inparaenum}
Specifically, set 
\begin{equation}
	\label{eq:alpha}
	\alpha=\max[1,\RXant\rankcov+\blocklength-\RXant\blocklength],
\end{equation}
let $\pilotset\define \natseg{1}{\alpha}$, which implies $\nonpilotset= \natseg{\alpha+1}{\blocklength}$.
Observe that~$\Pm\vectoutnn$ (conditioned on $\vecseg{\vinp}{\pilotset}$) depends  only on $\tp{[\tp\vectiid\ \vecseg{\tp\vinp}{\nonpilotset}]}$, and due to our choice of $\nonpilotset$ (it is actually the choice of $\abs{\nonpilotset}$ that is important here), 
the vectors $\Pm\vectoutnn$ and  $\tp{[\tp\vectiid\ \vecseg{\tp\vinp}{\nonpilotset}]}$ are of the same dimension.
Furthermore, these two vectors are related through a deterministic bijection:
Consider the vector-valued function $\bijec_{\nrvinp_{\pilotset}}: \complexset^{\outremsize}\to\complexset^{\outremsize}$
 \begin{equation}\label{eq:mapping1}
	\bijec_{\nrvinp_{\pilotset}}(\nrvectiid,\nrvinp_{\nonpilotset}) =\Pm(\identity_{\RXant}\otimes\nrminp\sqrtcov)\nrvectiid.
\end{equation}
Here, and whenever we refer to the function $\bijec_{\nrvinp_{\pilotset}}(\cdot)$ in the following, we use the convention that the parameter vector $\nrvinp_{\pilotset}\in\complexset^{\abs{\pilotset}}$ and the variable vector $\nrvinp_{\nonpilotset}\in\complexset^{\abs{\nonpilotset}}$ are stacked into the vector 
$\nrvinp\define \tp{[\tp\nrvinp_{\pilotset}\ \tp\nrvinp_{\nonpilotset}]}$ and we set $\nrminp\define\diag(\nrvinp)$.
\begin{lem}
	\label{lem:bijection}
If $\nrvinp_{\pilotset}$ has nonzero components only, i.e., $\nrvinpc_i\ne 0$ for all $i\in\pilotset$, then the function $\bijec_{\nrvinp_{\pilotset}}(\cdot)$ is one-to-one a.e. on $\complexset^{\outremsize}$.
\end{lem}
The proof of~\fref{lem:bijection} is given in~\fref{app:bijection} and is based on the results obtained later in this section.
We therefore invite the reader to first study the remainder of~\fref{sec:ub} and to return to~\fref{app:bijection} afterwards. 

Recall that $\Pm\vectoutnn=\Pm\left(\identity_{\RXant}\kron\minp\sqrtcov\right)\vectiid$ and hence $\Pm\vectoutnn=\bijec_{\vinp_{\pilotset}}(\vectiid,\vinp_{\nonpilotset})$. Therefore, it follows from \fref{lem:bijection} that as long as $\vinp_{\pilotset}=\nrvinp_{\pilotset}$ is fixed and satisfies $\nrvinpc_i\ne 0$, for all $i\in\pilotset$, $\Pm\vectoutnn$ and $\tp{[\tp\vectiid\ \vecseg{\tp\vinp}{\nonpilotset}]}$ are related through the bijection  $\bijec_{\nrvinp_{\pilotset}}(\cdot)$ as claimed.

\paragraph*{Comments}
A few comments on Lemma~\ref{lem:bijection} are in order.
For $\blocklength=3$ and $\RXant=\rankcov=2$ as in the simple example in Section~\ref{sec:intuition}, we see from~\fref{eq:Icond} that $\setI=\natseg{1}{\RXant\blocklength}$ so that $\Pm=\identity_{\RXant\blocklength}$ and $\Pm\vectoutnn=\vectoutnn$. Further, for this example, it follows from \fref{eq:alpha} that $\alpha=1$ and hence $\pilotset=\{1\}$ and $\nonpilotset=\{2,3\}$.
Therefore, \fref{lem:bijection} simply says that~\eqref{eq:IOnonoise} has a unique solution for fixed $\vinpc_1\ne 0$. 
As already mentioned, conditioning \wrt $\vinp_{\pilotset}=\vinpc_1$ in~\fref{eq:entPbound} in order to make the relation between $\Pm\vectoutnn$ and $\tp{[\tp\vectiid\ \vecseg{\tp\vinp}{\nonpilotset}]}$ be one-to-one corresponds to transmitting a pilot symbol, as was done in the back-of-the-envelope calculation by setting~$\vinpc_1=1$. 

We can now use \fref{lem:Entrchange} to relate $\diffent\lefto(\Pm\vectoutnn\given  \vecseg{\vinp}{\pilotset}\right)$ to 
$\diffent(\vectiid, \vinp_{\nonpilotset})$ as follows. Let $\pdf{\vinp_{\pilotset}}(\cdot)$ denote the \ac{PDF} of $\vinp_{\pilotset}$.
Then, we can write
\begin{equation}
	\label{eq:dec_change_var}
	\diffent(\Pm\vectoutnn\given \vinp_\pilotset)=\int\!\! \pdf{\vinp_{\pilotset}}(\nrvinp_{\pilotset}) \diffent\lefto(\Pm\vectoutnn\given \vinp_{\pilotset}=\nrvinp_{\pilotset}\right) d\nrvinp_{\pilotset}. 
\end{equation}
Let 
\begin{align}\label{eq:Jacobian2}
\jacobian(\nrvectiid,\nrvinp)&\define \frac{\partial \bijec_{\nrvinp_{\pilotset}}}{\partial (\nrvectiid,\nrvinp_{\nonpilotset})} 
\end{align}
be the Jacobian  of the mapping in \eqref{eq:mapping1} (where we again use the convention $\nrvinp= \tp{[\tp\nrvinp_{\pilotset}\ \tp\nrvinp_{\nonpilotset}]}$). Applying \fref{lem:Entrchange} to $\diffent\lefto(\Pm\vectoutnn\given \vinp_{\pilotset}=\nrvinp_{\pilotset}\right)$, we get for all $\nrvinp_{\pilotset}$ with $\nrvinpc_i\ne 0,$ $i\in\pilotset,$ that
\begin{align}
	\label{eq:condentchv}
	&\diffent\lefto(\Pm\vectoutnn\given \vinp_{\pilotset}=\nrvinp_{\pilotset}\right)
	=\diffent\lefto(\vectiid, \vinp_{\nonpilotset}\given \vinp_{\pilotset}=\nrvinp_{\pilotset}\right)\nonumber\\
	&\qquad\qquad\quad+2\Exop_{\vectiid,\vinp_{\nonpilotset}}\!\Bigl[\log\lefto(\abs{\det{\jacobian\lefto(\vectiid,\vinp\right)}}\right) \Big\vert\, \vinp_{\pilotset}=\nrvinp_{\pilotset}\Bigr].
\end{align}
Substituting~\fref{eq:condentchv} into~\fref{eq:dec_change_var}, we finally obtain
\begin{align}
	\label{eq:dec_change_var1}
&\diffent(\Pm\vectoutnn\given \vinp_\pilotset)\nonumber\\
&\stackrel{(a)}{=}\int\!\! \pdf{\vinp_{\pilotset}}(\nrvinp_{\pilotset})\diffent\lefto(\vectiid, \vinp_{\nonpilotset}\given \vinp_{\pilotset}=\nrvinp_{\pilotset}\right)d\nrvinp_{\pilotset}\nonumber\\
&\qquad +2\int\!\! \pdf{\vinp_{\pilotset}}(\nrvinp_{\pilotset})\Exop_{\vectiid,\vinp_{\nonpilotset}}\!\Bigl[\log\lefto(\abs{\det{\jacobian\lefto(\vectiid,\vinp\right)}}\right) \Big\vert\, \vinp_{\pilotset}=\nrvinp_{\pilotset}\Bigr]d\nrvinp_{\pilotset}\nonumber\\
&=	\diffent(\vectiid, \vinp_{\nonpilotset}\given \vinp_{\pilotset})+ 2\,\Ex{\vectiid,\vinp}
	{\log(\abs{\det(\jacobian(\vectiid,\vinp))})}.
\end{align}
Here, in (a), to be able to use~\fref{eq:condentchv}, we exclude the set $\{\nrvinp_{\pilotset}| \nrvinpc_i= 0 \text{ for at least one }  i\in\pilotset\}$ from the domain of integration. This is legitimate since that set has measure zero. 

The first term on the \ac{RHS} of~\eqref{eq:dec_change_var1} satisfies 
\begin{align}
	\diffent(\vectiid, \vecseg{\vinp}{\nonpilotset} \given \vecseg{\vinp}{\pilotset})
	&\stackrel{(a)}{=}\diffent(\vectiid \given \vecseg{\vinp}{\pilotset})+\diffent(\vecseg{\vinp}{\nonpilotset}\given \vectiid,\vecseg{\vinp}{\pilotset} )\nonumber\\
	&\stackrel{(b)}{=}\diffent(\vectiid)+\diffent(\vecseg{\vinp}{\nonpilotset})\stackrel{(c)}{=}\constalt,\label{eq:sxdecomp}
\end{align}
where (a) follows by the chain rule for differential entropy; in (b) we used that $\vinp$  is independent of $\vectiid$, and  $\vecseg{\vinp}{\pilotset}$ is independent of $\vecseg{\vinp}{\nonpilotset}$ because the $\vinpc_i,\, i\in\natseg{1}{\blocklength}$, are \iid and $\nonpilotset\intersect\pilotset=\emptyset$; and (c) follows because the $\vinpc_i,\, i\in\nonpilotset,$ and the $\vectiidc_i,\,i\in\natseg{1}{\RXant\rankcov},$ are \iid and have finite differential entropy, by assumption.  

Combining~\fref{eq:sxdecomp}, \fref{eq:dec_change_var1}, and~\fref{eq:entPbound}, we obtain
\begin{equation*}
	\diffent(\Pm\vectoutnn)\ge \constalt+2\Ex{\vectiid,\vinp}
	{\log(\abs{\det(\jacobian(\vectiid,\vinp))})}.
\end{equation*}
To show that $\diffent(\Pm\vectoutnn)>-\infty$, it therefore remains to 
prove that
\begin{equation}
	\label{eq:finitejac}
	\Ex{\vectiid,\vinp}
	{\log(\abs{\det(\jacobian(\vectiid,\vinp))})}>-\infty.
\end{equation}
This requires an in-depth analysis of the structure of $\abs{\det(\jacobian(\cdot))}$, which will be carried out in the next section. 

\subsection{Step 7: Factorization of $\det(\jacobian(\cdot))$ and analysis of $\vinp$-dependent terms}
\label{Jacobian}
The following lemma shows that the determinant of the Jacobian in \eqref{eq:Jacobian2} can be factorized into a product of simpler terms.

\begin{lem}\label{lem:lammaJac}
The determinant of the Jacobian in \eqref{eq:Jacobian2} factorizes as
\begin{equation*}
\det\lefto(\jacobian(\nrvectiid,\nrvinp)\right)
=\det\lefto(\jacobian_1(\nrvinp)\right)\det\lefto(\jacobian_2(\nrvectiid)\right)\det\lefto(\jacobian_3(\nrvinp_{\nonpilotset})\right),
\end{equation*}
where 
\begin{align}
\jacobian_1(\nrvinp)\nonumber
&\define \Pm(\identity_{\RXant}\otimes\nrminp)\tp{\Pm}\\
\jacobian_2(\nrvectiid)&\define\Pm[\identity_{\RXant}\otimes\sqrtcov\mid \veca_{\alpha+1}\mid \dots\mid \veca_{\blocklength}]\label{eq:jac2}\\
\jacobian_3(\nrvinp_{\nonpilotset})
&\define \diag(\identity_{\RXant\rankcov},\left(\diag(\nrvinp_{\nonpilotset})\right)^{-1})\nonumber
\end{align}
with 
\begin{align}\label{eq:ai}
\veca_i&\define (\identity_{\RXant}\otimes\diag(\vecunit_{i})\sqrtcov)\nrvectiid,\quad i\in\nonpilotset= [1\!:\!\blocklength].
\end{align}
\end{lem}

\begin{IEEEproof} 
First note that $\bijec_{\nrvinp_{\pilotset}}(\nrvectiid,\nrvinp_{\nonpilotset})$ in~\fref{eq:mapping1}
 can be written as
\begin{equation*}
	\bijec_{\nrvinp_{\pilotset}}(\nrvectiid,\nrvinp_{\nonpilotset})=\sum_{j\in [1:\blocklength]}x_j(\identity_{\RXant}\otimes\diag(\vecunit_{j})\sqrtcov)\nrvectiid
\end{equation*}
and, therefore,
\begin{align*}
\frac{\partial\bijec_{\nrvinp_{\pilotset}}}{\partial x_i}
&= \frac{\partial}{\partial x_i}\Big(\sum_{j\in \natseg{1}{\blocklength}}x_j(\identity_{\RXant}\otimes\diag(\vecunit_{j})\sqrtcov)\nrvectiid \Big)\\
&=\veca_i,\quad i\in \nonpilotset.
\end{align*}
With
\begin{equation*}
	\frac{\partial\bijec_{\nrvinp_{\pilotset}}}{\partial \nrvectiid}
	=\identity_{\RXant}\otimes\nrminp\sqrtcov	
\end{equation*}
we can now rewrite the Jacobian in~\fref{eq:Jacobian2} as
\begin{align}\label{eq:decoup}
\jacobian(\nrvectiid,\nrvinp)
&=\Pm[\identity_{\RXant}\otimes\nrminp\sqrtcov\mid \veca_{\alpha+1}\mid \dots\mid \veca_{\blocklength}]\nonumber\\
&=
(\Pm(\identity_{\RXant}\otimes\nrminp)\tp{\Pm})\, 
\jacobian_2(\nrvectiid)
\diag(\identity_{\RXant\rankcov}, \left(\diag(\nrvinp_{\nonpilotset})\right)^{-1}),
\end{align}  
which concludes the proof.
\end{IEEEproof}

Using \fref{lem:lammaJac}, we can rewrite the second term on the \ac{RHS} of~\fref{eq:dec_change_var1} according to
\begin{IEEEeqnarray}{rCl}
\!\!\!\! 	\Ex{\vectiid,\vinp}
 	{\log(\abs{\det(\jacobian(\vectiid,\vinp))})}
 	&=&\Ex{}{\log(\abs{\det\lefto(\jacobian_1(\vinp)\right)})}\nonumber\\
&&+\:\Ex{}{\log\lefto(\abs{\det\lefto(\jacobian_2(\vectiid)\right)}\right)}
 	\nonumber\\
  &&+\:\Ex{}{\log\lefto(\abs{\det\lefto(\jacobian_3(\vinp_{\nonpilotset})\right)}\right)}.
 \label{eq:sumexp}
 \end{IEEEeqnarray}
The first and the third term in \eqref{eq:sumexp} can be expanded as
\begin{align}\label{eq:expx}
\Ex{}{\log\lefto(\abs{\det\lefto(\jacobian_1(\vinp)\right)}\right)}
&=\shortant \sum_{j=1}^{\blocklength-1}\Ex{}{ \log(\abs{\vinpc_j})}\nonumber\\
&\qquad +(\RXant-\shortant) \sum_{j=1}^{\blocklength} \Ex{}{\log(\abs{\vinpc_j})}\\
\label{eq:expx2}
\Ex{}{\log\lefto(\abs{\det\lefto(\jacobian_3(\vinp_{\nonpilotset})\right)}\right)}
&=-\sum_{j\in\nonpilotset}
\Ex{}{\log(\abs{\vinpc_j})}.
\end{align}
Using \eqref{eq:propx}, \eqref{eq:apc}, and Jensen's inequality, we have
\begin{equation*}
	-\infty<\Ex{}{\log(\abs{\vinpc_j})}\le \log(\Ex{}{\abs{\vinpc_j}})<\infty,
\end{equation*}
which immediately implies that the terms on the \ac{LHS} of \fref{eq:expx} and \fref{eq:expx2} are finite.
It remains to show that $\Ex{}{\log\lefto(\abs{\det(\jacobian_2(\vectiid))}\right)}>-\infty$.  

\subsection{Step 8: Proving $\Ex{}{\log\lefto(\abs{\det(\jacobian_2(\vectiid))}\right)}>-\infty$ through resolution of singularities}
\label{sec:analdet}
This is the most technical part of the proof of Theorem \ref{thm:mainLB}. We need to show that 
\begin{align}
	\label{eq:j4lb}
	&\Ex{}{\log(\abs{\det(\jacobian_2(\vectiid))})}\nonumber\\
	&=\frac{1}{\pi^{\RXant\rankcov}} \int_{\complexset^{\RXant\rankcov}} \exp(-\vecnorm{\nrvectiid}^2)\log(\abs{\det(\jacobian_2(\nrvectiid))})d\nrvectiid> -\infty.
\end{align} 
Since $\jacobian_2(\cdot)$ is a large matrix with little structure to exploit, a direct evaluation of the integral in \fref{eq:j4lb} seems daunting. Note, however, that by \fref{eq:jac2}, \fref{eq:ai}, and \cite[4.2.1(2)]{lutkepohl96} it follows that $\det(\jacobian_2(\nrvectiid))$ is a homogeneous polynomial in $\nrvectiidc_1,\ldots,\nrvectiidc_{\RXant\rankcov}$; in other words $\det(\jacobian_2(\cdot))$ is a well-behaved function of its arguments. It turns out that this mild property is sufficient to prove the inequality in \fref{eq:j4lb}. 
The proof, however, requires powerful tools, which will be described next.

\begin{lem}
\label{lem:intoflogpoly}
Let $\poly(\vectr),\, \vectr\in\complexset^N,$ be a homogeneous polynomial in $\vectrc_1,\ldots,\vectrc_N.$
Then, $\poly(\cdot)\not\equiv 0$ implies that
\begin{equation*}
 \int_{\complexset^N}\exp(-\vecnorm{\vectr}^2)\log(\abs{\poly(\vectr)})d\vectr> -\infty.
\end{equation*}	
\end{lem}
\fref{lem:intoflogpoly} is proved in~\fref{app:intoflogpoly} using the following general result, which is a consequence of Hironaka's Theorem on the Resolution of Singularities~\cite[Theorem 2.3]{watanabe09}.
\begin{thm}\label{thm:TheoLin}
Let $f(\cdot)\not\equiv 0$ be a  real analytic function\footnote{Let $\Omega$ be an open subset of $\reals^\dimens$. A function $f(\cdot):\Omega\to\reals$ is real analytic  if for every $x_0\in \Omega$, $f(\cdot)$ can be represented by a convergent power series in some neighborhood of $x_0$.} \cite[Def. 2.2.1]{krantz92} on an open set $\Omega\subset\reals^\dimens$.
Then 
\begin{equation}\label{eq:Lineq}
\int_{\Delta} \abs{\log(|f(\vectr)|)} d\vectr < \infty
\end{equation}
for all compact sets $\Delta\subset \Omega$.
\end{thm}
For a formal proof of \fref{thm:TheoLin} see~\fref{app:resolution}. Here, we explain intuitively why this result holds. 
The only reason why the integral in~\fref{eq:Lineq} could diverge, is because $\abs{f(\cdot)}$ may take on the value zero and $\log(0)=-\infty$. Since $f(\cdot)$ is  a  real analytic function and since $f(\cdot)\not\equiv 0$, the zero set $f^{-1}(\{0\})$ 
has measure zero. To prove~\fref{eq:Lineq}, it remains to examine the detailed behavior of $f(\cdot)$ around the zero set $f^{-1}(\{0\})$.
The integral of $\abs{\log(|f(\cdot)|)}$ over a small enough neighborhood around each smooth (i.e. nonsingular) point in the zero set is bounded, but it is difficult to
determine what happens near the singularities. Hironaka's Theorem on the Resolution of Singularities ``untangles''
the singularities so that we can understand their structure. More
formally, Hironaka's Theorem states that in a small neighborhood around every point in
$f^{-1}(\{0\})$, the real analytic function $f(\cdot)$
behaves like a product of a monomial of finite degree and a \emph{nonvanishing} real analytic function.
The integral of the logarithm of the absolute value of this product over a small enough neighborhood around each point in $f^{-1}(\{0\})$ is then easily bounded and turns out to be finite. The union of the neighborhoods of the points in $f^{-1}(\{0\})$ forms an open cover for $f^{-1}(\{0\})$. Since $\Delta$ is a compact set, it is possible to find a finite subcover for $f^{-1}(\{0\})$. Summing up the integrals over the  elements of this subcover, each of which is finite as explained above, allows us to deduce that the integral in \fref{eq:Lineq} must be finite as well.

On account of  \fref{lem:intoflogpoly},  to show~\fref{eq:j4lb} it suffices to verify that  
$\det(\jacobian_2(\cdot))\not\equiv 0$. This is indeed the case as demonstrated next.

\subsection{Step 9: Identifying an $\nrvectiid$ for which $\det(\jacobian_2(\nrvectiid))\ne 0$}

\begin{lem}\label{lem:lemmazero}
	\begin{flushleft}
\propspark in Theorem \ref{thm:mainLB} implies that
$\det(\jacobian_2(\cdot))\not\equiv 0$.\end{flushleft}
\end{lem}
\begin{IEEEproof}
The proof is effected by showing that \propspark implies the existence of a vector $\nrvectiid\in\complexset^{\RXant\rankcov}$ such that $\det(\jacobian_2(\nrvectiid))\ne 0$.
To this end, we first note that $\jacobian_2(\nrvectiid)$ in \eqref{eq:jac2} can be written as 
	$\jacobian_2(\nrvectiid)=[\Pm\left(\identity_{\RXant}\otimes\sqrtcov\right)\ \ \matA]$ 
	with 
	\begin{equation}
		\matA\define \begin{bmatrix}\vspace{-1.2mm}
			\matA_1\\
			\vdots\\
			\matA_{\RXant}
		\end{bmatrix}
		\label{eq:matA}
	\end{equation}
	and 
	\begin{align*}
	\matA_i&\define 
	\begin{pmatrix}
	\veczero_{\alpha}&\cdots& \veczero_{\alpha}& \veczero_{\alpha}\\
	\tp{\rsqrtcov}_{\alpha+1}\nrviid{i}&\cdots&0&0\\
	\vdots&\ddots&0&0\\
	0&\cdots&\tp{\rsqrtcov}_{\blocklength-1}\nrviid{i}&0
	\end{pmatrix},\quad i\in \natseg{1}{\shortant},\\
	\matA_i&\define 
	\begin{pmatrix}
	\veczero_{\alpha}&\cdots&\veczero_{\alpha}&\veczero_{\alpha}&\\
	\tp{\rsqrtcov}_{\alpha+1}\nrviid{i}&\cdots&0&0\\
	\vdots&\ddots&0&0\\
	0&\cdots&\tp{\rsqrtcov}_{\blocklength-1}\nrviid{i}&0\\
	0&\cdots&0&\tp{\rsqrtcov}_{\blocklength}\nrviid{i}
	\end{pmatrix},\quad\ i\in \natseg{\shortant+1}{\RXant}.
	\end{align*} 
	Here, $\alpha$ was defined in~\fref{eq:alpha}; $\veczero_{\alpha}$ denotes an all-zero vector of dimension $\alpha$; $\rsqrtcov_1,\ldots,\rsqrtcov_\blocklength\in\complexset^{\rankcov}$ are the transposed rows of the $\blocklength\times\rankcov$ matrix $\sqrtcov$ so that $\tp\sqrtcov=[\rsqrtcov_1\cdots\,\rsqrtcov_\blocklength]$; and the $\nrviid{i}\in\complexset^{\rankcov},\, i\in\natseg{1}{\RXant},$ are defined  through $\nrvectiid\define \tp{[\tp{\nrviid{1}} \cdots\, \tp{\nrviid{\RXant}}]}$. 
	The calculations below are somewhat tedious but the idea is simple. Thanks to \propspark in~\fref{thm:mainLB}, it is possible to find vectors $\nrviid{i}\in\complexset^{\rankcov},\, i\in\natseg{1}{\RXant}$, such that each column of the matrix $\matA$ defined in~\fref{eq:matA} has exactly one nonzero element. For this choice of  $\nrviid{i}\in\complexset^{\rankcov},\, i\in\natseg{1}{\RXant},$ we can then conveniently factorize $\abs{\jacobian_2(\nrvectiid)}$ using the Laplace formula  \cite[p. 7]{horn85}; the resulting factors are easily seen to all be nonzero. We next detail the program just outlined.

Take an $i\in \natseg{1}{\RXant}$ and consider a set $\setK_i$ satisfying
\begin{equation}\label{eq:Kdef1}
\setK_i\subseteq \begin{cases} \natseg{\alpha+1}{\blocklength-1},\ &\text{ if } i\in \natseg{1}{\shortant},\\
\natseg{\alpha+1}{\blocklength},\ &\text{ if } i\in \natseg{\shortant+1}{\RXant},
\end{cases}
\end{equation}
with 
\begin{equation}\label{eq:Kdef}
	\abs{\setK_i}=\begin{cases}\rankcov-1,\ &\text{if}\ \RXant\blocklength>\RXant\rankcov +\blocklength-1\\
	(\RXant-1)(\blocklength-\rankcov),\  &\text{if}\ \RXant\blocklength\le\RXant\rankcov +\blocklength-1.
	\end{cases}
\end{equation}
The freedom in choice of the set $\setK_i$ will be used later to ensure 
that each column of the matrix $\matA$ has exactly one nonzero element.
We shall next show that the vector $\nrviid{i}\in\complexset^{\rankcov}$ can be chosen such that the entries of $\matA_i$ given by  $\tp{\rsqrtcov}_{j}\nrviid{i}$,   $j\in\setK_i$, equal zero and the entries $\tp{\rsqrtcov}_{j}\nrviid{i},\, j\notin\setK_i,$ are nonzero.
Since, by \fref{eq:Kdef}, $\abs{\setK_i}\le \rankcov-1$, \propspark in Theorem \ref{thm:mainLB} guarantees that the vectors $\{\rsqrtcov_{j}\}_{j\in \setK_i}$ are linearly independent. Furthermore, the vectors~$\rsqrtcov_{j},\, j\in \setK_i^c$, with
\begin{equation}\label{eq:Kdefcomp}
	\setK_i^c\define \begin{cases}\natseg{\alpha+1}{\blocklength-1}\setdiff \setK_i,\ &\text{ if } i\in \natseg{1}{\shortant},\\
	\natseg{\alpha+1}{\blocklength}\setdiff \setK_i,\ &\text{ if } i\in \natseg{\shortant+1}{\RXant},
	\end{cases}
\end{equation} 
do not belong to $\spn\{\rsqrtcov_{j}\}_{j\in \setK_i}$. 
Hence, we can find a vector $\nrviid{i}\in\complexset^{\rankcov}$ such that 
\begin{enumerate}[(a)]
\item $\tp{\rsqrtcov}_{j}\nrviid{i}=0$ for all $j\in\setK_i$;\vspace{1mm}
\item $\tp{\rsqrtcov}_{j}\nrviid{i}\neq 0$ for all $j\in\setK_i^c$. 
\end{enumerate}
Geometrically, this simply means that $\nrviid{i}$ must be chosen such that it is orthogonal to $\spn\{\rsqrtcov_{j}\}_{j\in \setK_i}$ (which is a subspace of $\complexset^{\rankcov}$ of dimension less than or equal to $\rankcov-1$) and, in addition, is not orthogonal to every vector in the set $\{\rsqrtcov_{j}\}_{j\in \setK_i^c}$ (see \fref{fig:geom}).
\begin{figure}
	\center{\includegraphics[width=60mm]{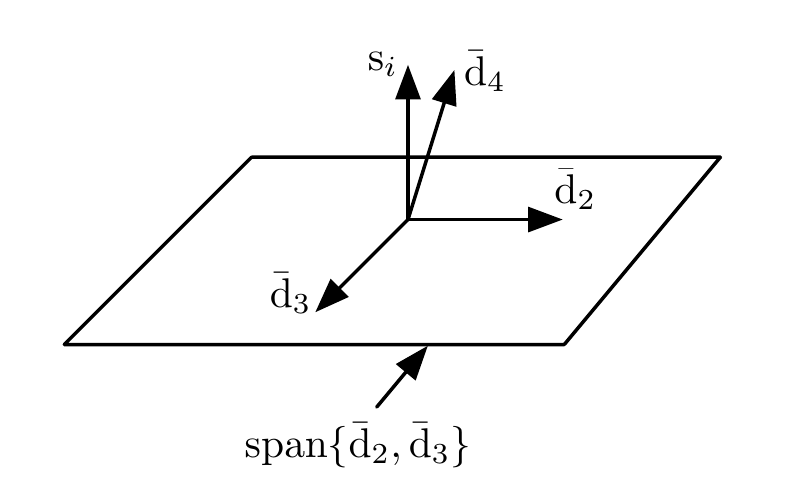}
	}
\caption{Choice of the vector $\nrviid{i}$ for $\blocklength=4,\ \rankcov=3,\ \alpha=1,\ \setK_i=\{2,3\},\ \setK_i^c=\{4\}.$}
\label{fig:geom}
\end{figure}
Note that if \propspark in Theorem \ref{thm:mainLB} were not satisfied, we could have a vector $\rsqrtcov_{j'},\, j'\in \setK_i^c,$ that belongs to the $\spn\{\rsqrtcov_{j}\}_{j\in \setK_i}$; in this case there would not exist a vector $\nrviid{i}$ that satisfies (a) and (b) simultaneously.
Based on \fref{eq:alpha}, \fref{eq:Kdef}, and~\fref{eq:Kdefcomp}, we can see that if the vector $\nrviid{i}$ is chosen such that conditions (a) and (b) above are satisfied, the number of  nonzero elements, $\abs{\setK_i^c}$, in the matrix $\matA_i$ is [see~\fref{eq:defshort}]
\begin{equation*}
\abs{\setK_i^c}=\begin{cases}
	\blocklength-\rankcov-1,\ &\text{ if }i\in \natseg{1}{\shortant},\\
	\blocklength-\rankcov,\ &\text{ if }i\in \natseg{\shortant+1}{\RXant}.
\end{cases}
\end{equation*}
Hence, applying the procedure described above to every $i\in \natseg{1}{\RXant}$ and choosing the corresponding vector $\nrviid{i}$ such that (a) and (b) are satisfied, we obtain  a matrix $\matA$ [see~\fref{eq:matA}] with total number of nonzero elements equal to the number of columns in $\matA$ and given by
\begin{equation*}
\sum_{i\in\natseg{1}{\RXant}}\abs{\setK_i^c}=\blocklength-\alpha.
\end{equation*} 
Now, recall that we have full freedom in our choice of $\setK_i,\ i\in \natseg{1}{\RXant},$ as long as \fref{eq:Kdef1} and \fref{eq:Kdef} are satisfied; this implies that we have control over the locations of the nonzero elements of $\matA$. Hence, by appropriate choice of the sets $\setK_i,\, i\in \natseg{1}{\RXant},$ we can ensure that each column of $\matA$ contains precisely one nonzero element.

Applying the Laplace formula \cite[p. 7]{horn85} iteratively, we then get 
\begin{equation}\label{eq:J2fact}
	\abs{\det(\jacobian_2(\nrvectiid))}
	=c
	\prod_{i\in [1:\RXant]}
	\abs{\det(\sqrtcov_{\setK_i\cup\natseg{1}{\alpha}})},
\end{equation}
where $c$ is a positive constant. Finally, since for every $i\in \natseg{1}{\RXant}$, $\sqrtcov_{\setK_i\cup\natseg{1}{\alpha}}$ is a $\rankcov\times\rankcov$ submatrix of $\sqrtcov$ [see \fref{eq:alpha} and \fref{eq:Kdef}], it follows from \propspark in \fref{thm:mainLB} that $\sqrtcov_{\setK_i\cup\natseg{1}{\alpha}}$ has linearly independent rows and hence $\abs{\det(\sqrtcov_{\setK_i\cup\natseg{1}{\alpha}})}>0$, for all $i\in \natseg{1}{\RXant}$, which by~\fref{eq:J2fact} concludes the proof.
\end{IEEEproof}

The proof of~\fref{thm:mainLB} is now completed as follows. Combining Lemmas \ref{lem:intoflogpoly} and \ref{lem:lemmazero}, we conclude that~\fref{eq:j4lb} holds. Substituting~\fref{eq:j4lb}  into~\fref{eq:sumexp} and using~\fref{eq:expx} and~\fref{eq:expx2}, we conclude that~\fref{eq:finitejac} holds. Therefore, by~\fref{eq:entPbound}, \fref{eq:dec_change_var1}, and~\fref{eq:sxdecomp}, it follows that $\diffent\lefto(\Pm\vectoutnn\right)>-\infty$.

\section{Conclusions and Future Work}

We characterized the  capacity pre-log of a temporally correlated block-fading \ac{SIMO} channel in the noncoherent setting under a mild assumption on the channel covariance matrix.
The most striking implication of this result is that the pre-log penalty in the \ac{SISO} case due to channel uncertainty 
can be made to vanish in the large block length regime by adding only one receive antenna.
 
It would be interesting to generalize the results in this paper to the~\ac{MIMO} case. Preliminary work in this direction was reported in~\cite{koliander12-10}, which establishes a lower bound on the  capacity pre-log of a temporally correlated block-fading \ac{MIMO} channel. This lower bound is not accompanied by a matching upper bound so that the problem of determining the capacity pre-log in the~\ac{MIMO} case remains open. It is also interesting to note that~\cite{koliander12-10} avoids the use of Hironaka's theorem through an alternative proof technique based on properties of subharmonic functions. 

Further interesting open questions include the generalization of the results in this paper to the stationary case and the development of coding schemes that achieve  the \ac{SIMO} capacity pre-log.

\appendices

\section{Proof of~\fref{eq:UB3}}
\label{app:repeatamos}
The following calculation repeats the steps in~\cite[Thm. 4.2]{lapidoth03-10} and is provided for the reader's convenience:
\begin{align*}
	&\mi\lefto(\matsegs{\mout}{\setQ}; \vecseg{\vinp}{\setQ}\right)
	=\sum_{q=1}^\rankcov \mi\lefto(\col{\mout}{q}; \vecseg{\vinp}{\setQ}\given \matsegs{\mout}{\natseg{1}{q-1}} \right)\nonumber\\
	&=\sum_{q=1}^\rankcov\left( \mi\lefto(\col{\mout}{q}; \matsegs{\mout}{\natseg{1}{q-1}}, \vecseg{\vinp}{\setQ} \right)- \mi\lefto(\col{\mout}{q}; \matsegs{\mout}{\natseg{1}{q-1}} \right)\right)\nonumber\\
	&\le\sum_{q=1}^\rankcov \mi\lefto(\col{\mout}{q}; \matsegs{\mout}{\natseg{1}{q-1}}, \vecseg{\vinp}{\setQ} \right)\nonumber\\
	&=\sum_{q=1}^\rankcov \left( \mi\lefto(\col{\mout}{q}; \matsegs{\mout}{\natseg{1}{q-1}}, \matsegs{\mchannel}{\natseg{1}{q-1}}, \vecseg{\vinp}{\setQ} \right)\right.\nonumber\\
	&\qquad\qquad\qquad-\left. \mi\lefto(\col{\mout}{q};  \matsegs{\mchannel}{\natseg{1}{q-1}}\given\matsegs{\mout}{\natseg{1}{q-1}}, \vecseg{\vinp}{\setQ} \right) \right)\nonumber\\
	&\le\sum_{q=1}^\rankcov \mi\lefto(\col{\mout}{q}; \matsegs{\mout}{\natseg{1}{q-1}}, \matsegs{\mchannel}{\natseg{1}{q-1}}, \vecseg{\vinp}{\setQ} \right)\nonumber\\
	&\stackrel{(a)}{=}\sum_{q=1}^\rankcov \mi\lefto(\col{\mout}{q}; \matsegs{\mchannel}{\natseg{1}{q-1}}, \vinpc_q \right)\nonumber\\
	&=\sum_{q=1}^\rankcov\left( \mi\lefto(\col{\mout}{q}; \matsegs{\mchannel}{\natseg{1}{q-1}}  \given \vinpc_{q}\right)+\mi\lefto(\col{\mout}{q}; \vinpc_{q} \right)\right)\nonumber\\
	&\stackrel{(b)}{\le}\sum_{q=1}^\rankcov\mi\lefto(\col{\mout}{q}; \matsegs{\mchannel}{\natseg{1}{q-1}}  \given \vinpc_{q}\right)+\rankcov \log\log(\SNR)+\asconst\nonumber\\
	&\stackrel{(c)}{=}\sum_{q=1}^\rankcov\mi\lefto(\col{\mout}{q},\vinpc_{q}; \matsegs{\mchannel}{\natseg{1}{q-1}} \right)+\rankcov \log\log(\SNR)+\asconst\nonumber\\
	&\le\sum_{q=1}^\rankcov\mi\lefto(\col{\mout}{q},\vinpc_{q},\col{\mchannel}{q}; \matsegs{\mchannel}{\natseg{1}{q-1}} \right)+\rankcov \log\log(\SNR)+\asconst\nonumber\\
	&\stackrel{(d)}{=}\sum_{q=1}^\rankcov\mi\lefto(\col{\mchannel}{q}; \matsegs{\mchannel}{\natseg{1}{q-1}} \right)+\rankcov \log\log(\SNR)+\asconst\nonumber\\
	&=\rankcov \diffent\lefto(\col{\mchannel}{1}\right)-\diffent\lefto(\matsegs{\mchannel}{\setQ}\right)+\rankcov \log\log(\SNR)+\asconst\nonumber\\
	&=\rankcov\sum_{i=1}^\RXant \diffent(\mchannelc_{i1})-\RXant\log\det\lefto(\matsegs{\sqrtcov}{\setQ}\herm{\matsegs{\sqrtcov}{\setQ}}\right)+\rankcov \log\log(\SNR)+\asconst\nonumber\\
	&\stackrel{(e)}{=}\rankcov \log\log(\SNR)+\asconst
\end{align*}
where (a) follows because $\col{\mout}{q}$ is conditionally independent of $\vecseg{\vinp}{\natseg{1}{q-1}}$ and of $\matsegs{\mout}{\natseg{1}{q-1}}$ given $\vinpc_{q}$ and $\matsegs{\mchannel}{\natseg{1}{q-1}}$;  (b) follows from \cite[Th. 4.2]{lapidoth03-10}; (c) follows because $\vinpc_{q}$ is independent of $\matsegs{\mchannel}{\natseg{1}{q-1}}$; (d) follows because $\matsegs{\mout}{q}$ and $\vinpc_{q}$ are conditionally independent of $\matsegs{\mchannel}{\natseg{1}{q-1}}$ given $\matsegs{\mchannel}{q}$; and (e) follows because the matrix $\matsegs{\sqrtcov}{\setQ}$ is full-rank and  $\diffent(\mchannelc_{i1})=\constalt$, $i\in\natseg{1}{\RXant}$.

\section{Proof of~\fref{lem:Entrchange}}
\label{app:Entrchange}
The lemma is based on the change of variables theorem for integrals, which we restate for the reader's convenience.
\begin{thm}{\cite[Thm. 7.26]{rudin87},\cite[p. 31, Thm. 7.2]{fritzsche02}}
\label{thm:changevarrud}
	Assume that $\altfunvec:\setU\subset\complexset^N\to\complexset^N$ is a continuous vector-valued function that is one-to-one and differentiable a.e. on $\setU$. Let $\setV=\altfunvec(\setU)$. Then,
	\begin{equation*}
	\int_{\setV} \fun(\altvectr) d\altvectr=\int_{\setU} \fun(\altfunvec(\vectr)) \abs{\det\lefto({\partial\altfunvec}/{\partial \vectr}\right)}^2 d\vectr	
	\end{equation*}
	for every measurable $\fun:\complexset^N\to[0,\infty]$.
\end{thm}	
To prove \fref{lem:Entrchange}, we let $\pdf{\rvecv}(\cdot)$ and $\pdf{\rvecu}(\cdot)$ denote the \acp{PDF} of random vectors $\rvecv$ and $\rvecu$, respectively. Then, according to \cite[(7-8)]{papoulis02} and \cite[p.31, Thm. 7.2]{fritzsche02}
\begin{equation}
	\label{eq:densitytrans}
	\pdf{\rvecv}(\altfunvec(\vecu))=\frac{\pdf{\rvecu}(\vecu)}{\abs{\det\lefto({\partial\altfunvec}/{\partial \vecu}\right)}^2}.
\end{equation}	

Next, let $\setU$ and $\setV$ denote the support of $\pdf{\rvecu}(\cdot)$ and $\pdf{\rvecv}(\cdot)$, respectively. Then, $\setV=\altfunvec(\setU)$ and, on account of~\fref{thm:changevarrud}, we have
\begin{align*}
		\diffent(\raltvectr)
		&=-\int_{\setV} 	\pdf{\rvecv}(\vecv)\log\lefto(	\pdf{\rvecv}(\vecv)\right) d\vecv\\
		&=-\int_{\setU} 	\pdf{\rvecv}(\altfunvec(\vecu))\log\lefto(	\pdf{\rvecv}(\altfunvec(\vecu))\right) \abs{\det\lefto({\partial\altfunvec}/{\partial \vecu}\right)}^2 d\vecu\\
	&\stackrel{(a)}{=}-\int_{\setU} \frac{\pdf{\rvecu}(\vecu)}{\abs{\det\lefto({\partial\altfunvec}/{\partial \vecu}\right)}^2} \times\nonumber\\	
	&\qquad\qquad\times\log\lefto(	\frac{\pdf{\rvecu}(\vecu)}{\abs{\det\lefto({\partial\altfunvec}/{\partial \vecu}\right)}^2}\right) \abs{\det\lefto({\partial\altfunvec}/{\partial \vecu}\right)}^2 d\vecu\\
		&=-\int_{\setU} 	\pdf{\rvecu}(\vecu)\log\lefto(	\pdf{\rvecu}(\vecu)\right) d\vecu\\
		&\qquad\qquad+2 \int_{\setU}\pdf{\rvecu}(\vecu) \log\lefto(\abs{\det\lefto({\partial\altfunvec}/{\partial \vecu}\right)} \right) d\vecu\\
		&=\diffent(\rvectr)+2 \Ex{\rvectr}{\log\abs{\det\lefto({\partial\altfunvec}/{\partial \rvectr}\right)}}
\end{align*}
where in (a) we used~\fref{eq:densitytrans}.
This concludes the proof.

\section{Proof of Lemma~\ref{lem:bijection}}
\label{app:bijection}

We  need to show that the function $\bijec_{\nrvinp_{\pilotset}}(\nrvectiid,\nrvinp_{\nonpilotset})$ is  one-to-one \emph{almost everywhere}. It is therefore legitimate to exclude sets of measure zero from its domain. 
In particular, we  consider the restriction of the function $\bijec_{\nrvinp_{\pilotset}}(\nrvectiid,\nrvinp_{\nonpilotset})$ to the set of pairs $(\nrvectiid,\nrvinp_{\nonpilotset})$ that satisfy 
\begin{enumerate}[(i)]
\item
\label{it:i1}
$\abs{\nrvinpc_i}>0$ for all $i\in\nonpilotset$;
\item
\label{it:i2}
$\det\jacobian_2(\nrvectiid)\ne 0$ with $\jacobian_2(\cdot)$ defined in~\fref{eq:jac2}.
\end{enumerate}
Condition (i) excludes those $\vecseg{\nrvinp}{\nonpilotset}$ from the domain of $\bijec_{\nrvinp_{\pilotset}}(\cdot)$ that have at least one component equal to zero; since the $\nrvinpc_i, i\in\nonpilotset,$ take on values in a continuum,  the excluded set has measure zero. Condition (ii) excludes those $\nrvectiid$ from the domain of $\bijec_{\nrvinp_{\pilotset}}(\cdot)$ that have $\det(\jacobian_2(\nrvectiid))=0$.
Remember that we proved in \fref{sec:analdet} (see~\fref{eq:j4lb}) that $\Ex{}{\log(\abs{\det(\jacobian_2(\vectiid))})}>-\infty$, which implies $\det(\jacobian_2(\cdot))\ne 0$ a.e. Therefore, the set excluded in (ii) must be a set of measure zero. 
We conclude that the set of pairs $(\nrvectiid,\vecseg{\nrvinp}{\nonpilotset})$ that violates at least one of the conditions (i) and (ii) 
is a set of measure zero. 

To show that the resulting restriction of the function $\bijec_{\nrvinp_{\pilotset}}(\cdot)$ [which, with slight abuse of notation we still call $\bijec_{\nrvinp_{\pilotset}}(\cdot)$] is one-to-one, we take two pairs 
 $(\tilde\nrvectiid,\vecseg{\tilde\nrvinp}{\nonpilotset})$ and $(\nrvectiid,\vecseg{\nrvinp}{\nonpilotset})$  from the domain of $\bijec_{\nrvinp_{\pilotset}}(\cdot)$ and show that if $\bijec_{\nrvinp_{\pilotset}}(\tilde\nrvectiid,\vecseg{\tilde\nrvinp}{\nonpilotset})=\bijec_{\nrvinp_{\pilotset}}(\nrvectiid,\vecseg{\nrvinp}{\nonpilotset})$, then necessarily $(\tilde\nrvectiid,\vecseg{\tilde\nrvinp}{\nonpilotset})=(\nrvectiid,\vecseg{\nrvinp}{\nonpilotset})$.

Indeed, assume that both $(\tilde\nrvectiid,\vecseg{\tilde\nrvinp}{\nonpilotset})$ and $(\nrvectiid,\vecseg{\nrvinp}{\nonpilotset})$ belong to the domain of $\bijec_{\nrvinp_{\pilotset}}(\cdot)$, i.e., both pairs satisfy conditions (i) and (ii) above. Suppose that $\bijec_{\nrvinp_{\pilotset}}(\tilde\nrvectiid,\vecseg{\tilde\nrvinp}{\nonpilotset})=\bijec_{\nrvinp_{\pilotset}}(\nrvectiid,\vecseg{\nrvinp}{\nonpilotset})$, or, equivalently,
\begin{equation}
 \Pm(\identity_{\RXant}\kron\tilde\nrminp\sqrtcov)\tilde\nrvectiid=\Pm(\identity_{\RXant}\kron\nrminp\sqrtcov)\nrvectiid 
	\label{eq:bijection1}
\end{equation}
where $\nrvinp= \tp{[\tp\nrvinp_{\pilotset}\ \tp\nrvinp_{\nonpilotset}]}$,  $\nrminp=\diag(\nrvinp)$, $\tilde\nrvinp= \tp{[\tp{\nrvinp_{\pilotset}}\ \tp{\tilde\nrvinp_{\nonpilotset}}]}$, and $\tilde\nrminp=\diag(\tilde\nrvinp)$.
We next consider \fref{eq:bijection1} as an equation parametrized by $(\nrvectiid,\vecseg{\nrvinp}{\nonpilotset})$ in the variables $(\tilde\nrvectiid,\vecseg{\tilde\nrvinp}{\nonpilotset})$  and show that this equation has a \emph{unique} solution. Since $(\tilde\nrvectiid,\vecseg{\tilde\nrvinp}{\nonpilotset})=(\nrvectiid,\vecseg{\nrvinp}{\nonpilotset})$ (trivially) satisfies \fref{eq:bijection1}, uniqueness then implies that $(\tilde\nrvectiid,\vecseg{\tilde\nrvinp}{\nonpilotset})=(\nrvectiid,\vecseg{\nrvinp}{\nonpilotset})$. 

To prove that~\eqref{eq:bijection1} has a unique solution, we follow the approach described in Section~\ref{sec:intuition} and convert~\eqref{eq:bijection1}  into a linear system of equations through a change of variables. 
In particular, thanks to constraint~\eqref{it:i1}, we can left-multiply both sides of~\eqref{eq:bijection1} by 
$\Pm[\identity_{\RXant}\kron\nrminp]^{-1}\tp\Pm\Pm[\identity_{\RXant}\kron\tilde\nrminp]^{-1}\tp\Pm$
to transform \fref{eq:bijection1}  into the equivalent equation
\begin{equation}
 \Pm\left(\identity_{\RXant}\kron\nrminp^{-1}\sqrtcov\right)\tilde\nrvectiid=\Pm\left(\identity_{\RXant}\kron\tilde\nrminp^{-1}\sqrtcov\right)\nrvectiid. 
	\label{eq:bijection12}
\end{equation}
Next, perform the substitutions  $\nrvinpinvc_i=1/\nrvinpc_i,\ \tilde\nrvinpinvc_i=1/\tilde\nrvinpc_i,\, i\in\natseg{1}{\blocklength},$ define $\nrvinpinv\define\tp{[\nrvinpinvc_1 \dots \nrvinpinvc_\blocklength]}$, and set $\matZ\define\diag(\nrvinpinv)$ so that \eqref{eq:bijection12} can be written as
\begin{equation}
 \Pm\left(\identity_{\RXant}\kron\matZ\sqrtcov\right)\tilde\nrvectiid=\sum_{i=1}^\blocklength \tilde\nrvinpinvc_i \Pm\veca_i,
	\label{eq:bijection13}
\end{equation} 
where $\veca_i= (\identity_{\RXant}\otimes\diag(\vecunit_{i})\sqrtcov)\nrvectiid,\, i\in [1:\blocklength],$ as defined in \fref{eq:ai}.
Finally, moving the terms containing the unknowns $\tilde\nrvinpinvc_i,\, i\in\nonpilotset,$ to the \ac{LHS} of~\fref{eq:bijection13} while keeping the terms containing the fixed parameters $\tilde\nrvinpinvc_i,\, i\in \pilotset,$ on the \ac{RHS}, we transform \fref{eq:bijection13} into the equivalent equation
\begin{equation}
 \Pm\left(\identity_{\RXant}\kron\matZ\sqrtcov\right)\tilde\nrvectiid-\sum_{i\in\nonpilotset} \tilde\nrvinpinvc_i \Pm\veca_i =\sum_{i\in \pilotset} \tilde\nrvinpinvc_i \Pm\veca_i.
	\label{eq:bijection14}
\end{equation}
Defining $\tilde\nrvinpinv_{\nonpilotset}\define\tp{[\tilde\nrvinpinvc_{\alpha+1} \dots \tilde\nrvinpinvc_\blocklength]}$ and using the expression for $\jacobian(\cdot)$ in \fref{eq:decoup}, we can write \fref{eq:bijection14} as
\begin{align}\label{eq:inhomogeneous}
\jacobian(\nrvectiid,\aaltvvrbl)
\begin{bmatrix}
\tilde\nrvectiid\\
-\tilde\aaltvvrbl_{\nonpilotset}
\end{bmatrix}
= \sum_{i\in\pilotset}\tilde \nrvinpinvc_i\Pm
\veca_i.
\end{align}

The solution of~\eqref{eq:inhomogeneous} is unique if and only if $\det\jacobian(\nrvectiid,\aaltvvrbl)\ne 0$. We use \fref{lem:lammaJac} to factorize $\det\jacobian(\nrvectiid,\aaltvvrbl)$ according to
\begin{align}
	\label{eq:factt}
\det\lefto(\jacobian(\nrvectiid,\aaltvvrbl)\right)
&=\det\lefto(\jacobian_1(\aaltvvrbl)\right)\det\lefto(\jacobian_2(\nrvectiid)\right)
\det\lefto(\jacobian_3(\aaltvvrbl_{\nonpilotset})\right).
\end{align}

The first and the third term on the \ac{RHS} of \fref{eq:factt} can be written as follows
\begin{align*}
\det\lefto(\jacobian_1(\nrvinpinv)\right)
&=\left(\prod_{j=1}^{\blocklength-1}\nrvinpinvc_j\right)^{\!\!\!\shortant }\left(\prod_{j=1}^{\blocklength} \nrvinpinvc_j\right)^{\!\!\!(\RXant-\shortant) }\\
&=\left(\prod_{j=1}^{\blocklength-1}\frac{1}{\nrvinpc_j}\right)^{\!\!\!\shortant }\left(\prod_{j=1}^{\blocklength} \frac{1}{\nrvinpc_j}\right)^{\!\!\!(\RXant-\shortant) }\\
\det\lefto(\jacobian_3(\nrvinpinv_{\nonpilotset})\right)
&=\prod_{j\in\nonpilotset}
\frac{1}{\nrvinpinvc_j}=\prod_{j\in\nonpilotset}
\nrvinpc_j
\end{align*}
and are nonzero due to constraint~\eqref{it:i1} stated at the beginning of this Appendix; $\det\lefto(\jacobian_2(\nrvectiid)\right)\ne 0$ due to constraint~\eqref{it:i2}. Hence $\det\jacobian(\nrvectiid,\aaltvvrbl)\ne 0$ and the solution of~\eqref{eq:inhomogeneous} in the variables $(\tilde\nrvectiid,\vecseg{\tilde\nrvinpinv}{\nonpilotset})$ is unique. Therefore, the solution of  \fref{eq:bijection1} [parametrized by $(\nrvectiid,\vecseg{\nrvinp}{\nonpilotset})$] in the variables $(\tilde\nrvectiid,\vecseg{\tilde\nrvinp}{\nonpilotset})$  is  unique.
This completes the proof.

We conclude this section by closing an issue that was left open in the back-of-the-envelope calculation in~\fref{sec:intuition}. Specifically, we will show that the matrix $\altmat$ in~\eqref{eq:inhomeq} is full-rank.
For $\blocklength=3$ and $\RXant=\rankcov=2$, the matrix $\altmat$ in~\eqref{eq:inhomeq} is related to $\jacobian(\cdot)$ in \fref{eq:decoup}  according to $\altmat=(\identity_2\kron\minp)\jacobian(\vectiid,\rvecz)$ with $\rvecz=\tp{[\invinpc_1\ldots\invinpc_\blocklength]}$. Hence $\det\lefto(\altmat\right)=\det\lefto(\identity_2\kron\minp\right)\det\lefto(\jacobian(\vectiid,\rvecz)\right)$. Since we assumed in~\fref{sec:intuition} that $\abs{\vinpc_i}>0,\ i\in\natseg{1}{\blocklength}$, we have $\det\lefto(\identity_2\kron\minp\right)\ne 0$. Together with $\det\lefto(\jacobian(\vectiid,\rvecz)\right)\ne 0$, a.e., as shown above, we can conclude that, indeed, $\det\lefto(\altmat\right)\ne 0$, a.e.,  as claimed in~\fref{sec:intuition}.

\section{Proof of~\fref{lem:intoflogpoly}}
\label{app:intoflogpoly}
Instead of working with 
\begin{equation}
	\label{eq:I1app}
	I\define\int_{\complexset^N}\exp(-\vecnorm{\vectr}^2)\log(\abs{\poly(\vectr)})d\vectr
\end{equation}
it will turn out convenient to consider $\abs{I}$ and to show that $\abs{I}<\infty$, which trivially implies $I>-\infty$.
As already mentioned, the proof of  $\abs{I}<\infty$ is based on~\fref{thm:TheoLin}. 
In order to be able to apply~\fref{thm:TheoLin} we will need to transform the  integration domain in~\fref{eq:I1app} into a compact set in $\reals^{2N}$, transform the complex-valued polynomial $\poly(\cdot)$ into a real-valued function, and 
 get rid of the term~$\exp(-\vecnorm{\vectr}^2)$. 
All this will be accomplished as follows. First, we bound $\abs{I}$  by a sum of two integrals over the set $\complexset^N$, then, we apply a change of variables to transform these two integrals into three new integrals. The first two of these three integrals are over the set $[0,\infty]$, which is still not compact,  but the resulting integrals are simple enough to be bounded directly. The third integral is over a compact set and can, thus, be bounded using \fref{thm:TheoLin}. We now implement the program just outlined. 

Let $\degr$ denote the degree of the homogeneous polynomial $\poly(\cdot)$. Then, by homogeneity of $\poly(\cdot)$,
\begin{equation*}
	\poly(\vectr)=\poly\lefto(\vecnorm{\vectr}\frac{\vectr}{\vecnorm{\vectr}}\right)=\vecnorm{\vectr}^\degr \poly\lefto(\frac{\vectr}{\vecnorm{\vectr}}\right)
\end{equation*}
and, therefore,
\begin{align*}
	I&= \int_{\complexset^N}\exp(-\vecnorm{\vectr}^2)\log(\abs{\poly(\vectr)})d\vectr\\
	&=\underbrace{\degr\int_{\complexset^N}\exp(-\vecnorm{\vectr}^2)\log(\vecnorm{\vectr})d\vectr}_{I_{1}}\\
	&\qquad\qquad+\underbrace{\int_{\complexset^N}\exp(-\vecnorm{\vectr}^2)\log(\abs{\poly(\vectr/\vecnorm{\vectr})})d\vectr}_{I_{2}}.
\end{align*}
We next change variables in $I_{1}$ and $I_{2}$ by first transforming the domain of integration from $\complexset^N$ to $\reals^{2N}$ and then using polar coordinates \cite[p. 55]{muirhead05}. Specifically, we introduce the function $\vectr:\reals^{2N}\to \complexset^N$ that acts according to 
\begin{equation}
	\vectr(\altvectr)\define \tp{[\altvectrc_1+\iu \altvectrc_2\ \cdots\ \altvectrc_{2N-1}+\iu \altvectrc_{2N}]},\label{eq:chvar1}
\end{equation}
and the function $\altvectr:\reals_+ \times \Delta\to \reals^{2N}$ with $\Delta\define [0,\pi]^{2N-2}\times[0,2\pi]$ defined through
\begin{equation}
	\altvectr(r,\vect)\define r \funvec(\vect)\label{eq:chvar2}
\end{equation}
with 
\begin{equation}\label{eq:polar}
	\funvec(\vect) \define\begin{bmatrix}
	\sin(\vectc_1)\sin(\vectc_2)\dots\sin(\vectc_{2N-2})\sin(\vectc_{2N-1})\\
	\sin(\vectc_1)\sin(\vectc_2)\dots\sin(\vectc_{2N-2})\cos(\vectc_{2N-1})\\
	\sin(\vectc_1)\sin(\vectc_2)\dots\cos(\vectc_{2N-2})\\
	\vdots&\\
	\sin(\vectc_1)\cos(\vectc_{2})\\
	\cos(\vectc_1)
	\end{bmatrix}.
\end{equation}
It follows from~\fref{eq:chvar1}--\fref{eq:polar} that 
\begin{equation*}
	\vecnorm{\vectr(\altvectr(r,\vect))}=\vecnorm{\altvectr(r,\vect)}=r
\end{equation*}
and therefore
\begin{equation*}
	\frac{\vectr(\altvectr(r,\vect))}{\vecnorm{\vectr(\altvectr(r,\vect))}}=\frac{\vectr(r\vecf(\vect)  )}{r}=\vectr(\vecf(\vect)  ).
\end{equation*}
The determinant of the Jacobian of the function $\altvectr(\cdot)$ 
is well-known and is given by \cite[p. 55]{muirhead05}
\begin{equation*}
	\det\frac{\partial \altvectr}{\partial (r,\vect)}=r^{2N-1}\underbrace{ \sin(\vectc_1)^{2N-2} \sin(\vectc_2)^{2N-3}\ldots\, \sin(\vectc_{2N-2})}_{\altfunvec(\vect)}.
\end{equation*}
Changing variables in $I_{1}$ and $I_{2}$ according to $\vectr\to\altvectr\to(r,\vect)$, we obtain
\begin{align*}
	I_{1}&=\degr\int_{r,\vect}\exp(-r^2)\log(r) r^{2N-1} \altfunvec(\vect)  d r d \vect\\
	I_{2}&=\int_{r,\vect}\exp(-r^2)\log(\abs{\poly(\vectr(\funvec(\vect)))})  r^{2N-1}\altfunvec(\vect) d r d \vect.
\end{align*}
By the triangle inequality we have
\begin{equation*}
	\abs{I}\le \abs{I_{1}}+\abs{I_{2}}.
\end{equation*} 
Using $\abs{\altfunvec(\vect)}<1$, we get
\begin{align}
	\abs{I_{1}}&\le \degr\, 2\pi^{2M-1}\int_{0}^\infty \exp(-r^2)\abs{\log(r)} r^{2N-1}  d r < \infty\nonumber\\
	\abs{I_{2}}&\le\int_{0}^\infty \exp(-r^2)  r^{2N-1} d r \times \int_{\Delta}\abs{\log(\abs{\poly(\vectr(\funvec(\vect)))})} d \vect\nonumber\\
	&\le \constalt\int_{\Delta}\abs{\log(\abs{\poly(\vectr(\funvec(\vect)))}^2)} d \vect.\label{eq:int12}
\end{align}
We hereby disposed of the integrals over unbounded domains and are left only with an integral over the compact set $\Delta$. Note also that by absorbing a factor $1/2$ into $\constalt$ we introduced a square in~\fref{eq:int12}, which will turn out useful later.
In order to prove that $\abs{I}<\infty$ it now remains to show that 
\begin{equation}
	I_3\define \int_{\Delta}\abs{\log(\abs{\poly(\vectr(\funvec(\vect)))}^2)} d \vect<\infty.
	\label{eq:I3}
\end{equation}

Note that $\abs{\poly(\vectr(\funvec(\cdot)))}^2:\Delta\to \reals_+$ is a real analytic function by \cite[Prop. 2.2.2]{krantz92}, because it is a composition of the polynomial $\abs{\poly(\vectr(\cdot))}^2:\reals^{2N}\to \reals_+$ and the function $\funvec(\cdot):\Delta\to\reals^{2N}$ that has real analytic components (trigonometric functions are real analytic on $\reals$).
Furthermore, by assumption, $\poly(\cdot)\not\equiv 0$ and hence $\abs{\poly(\vectr(\funvec(\cdot)))}^2\not\equiv 0$. Finally, $\Delta$ is a compact set. The inequality~\fref{eq:I3} now 
follows by application of \fref{thm:TheoLin}. This concludes the proof.

\section{Proof of~\fref{thm:TheoLin} via resolution of singularities}\label{app:resolution}
In order to prove~\fref{thm:TheoLin} note that $\int_{\Delta\subset\reals^M}\abs{\log\lefto(\abs{\fun(\vectr)}\right)}d\vectr$ would clearly be finite if the function $\fun(\cdot)$ were  bounded away from zero on the set $\Delta$. 
Unfortunately, this is not the case.
%
However, because $\fun(\cdot)$ is real analytic and $\fun(\cdot)\not\equiv 0$, it can take on the value zero only on a set of measure zero~\cite[Cor. 1.2.6]{krantz92}. 
Establishing whether the integral $\int_{\Delta\subset\reals^M}\abs{\log\lefto(\abs{\fun(\vectr)}\right)}d\vectr$ is finite, hence requires a fine analysis of the behavior of $\abs{\log\lefto(\abs{\fun(\cdot)}\right)}$ in the neighborhood of the zero-measure set $\fun^{-1}(\{0\})$. This can be accomplished using  Hironaka's Theorem on the Resolution of Singularities, which allows one to write $\fun(\cdot)$ as a product of a  monomial and a \emph{nonvanishing} real analytic function in the neighborhood of each point $\vectr$ where $\fun(\vectr)=0$.
The logarithm of this product can then easily be  bounded and shown to be finite. 
As the tools used in the following are non-standard, at least in the information theory literature, we review the main ingredients in some detail.
Formally, Hironaka's Theorem states the following:

\begin{thm}\cite[Theorem 2.3]{watanabe09}\label{thm:Theosing}
Let $\fun(\cdot)\not\equiv 0$ be a real analytic function \cite[Def. 1.1.5]{krantz92} from a neighborhood of the origin $\veczero$, denoted $\Omega\subseteq \reals^\dimens$,  to $\reals$, which satisfies $\fun(\veczero)=0$. Then, there exists a triple $(\setW, \manifold, \blowup(\cdot))$ such that
\begin{enumerate}[(a)]
\item $\setW\subset \Omega$ is an open set in $\reals^\dimens$ with $\veczero\in \setW$, \label{cond:res1}
\item $\manifold$ is a $\dimens$-dimensional real analytic manifold~\cite[Def. 2.10]{watanabe09} with coordinate charts $\{\manifold_{\mpoint},\coord_{\mpoint}:\cube(\veczero,\epsilon_{\mpoint})\to \manifold_{\mpoint}\}$ for each point $\mpoint\in \manifold$, where $\coord_{\mpoint}(\cdot)$ is an isomorphism\footnote{Let $\setU$ and $\setV$ be two real analytic manifolds. A real analytic map $\fun:\setU\to\setV$ is called an isomorphism between $\tilde\setU\subset\setU$ and $\tilde\setV\subset\setV$ if it is one-to-one and an onto map from $\tilde\setU$ to $\tilde\setV$ whose inverse on $\tilde\setV$ is also a real analytic map.} between $\cube(\veczero,\epsilon_{\mpoint})$ and $\manifold_{\mpoint}$ with $\coord_{\mpoint}(\veczero)=\mpoint$. \label{cond:res2}
\item $\blowup: \manifold\to \setW$ is a real analytic map, \label{cond:res3}
\end{enumerate}
that satisfies the following conditions:
\begin{enumerate}[(i)]
\item The map $\blowup(\cdot)$ is proper, i.e., the inverse image of every compact set under $\blowup(\cdot)$ is compact. \label{cond:res4}
\item\label{cond:res5} The map $\blowup(\cdot)$ is an isomorphism\cite[Def. 2.5]{watanabe09} between $\manifold\setminus (\fun\circ \blowup)^{-1}(\{0\})$ and $\setW\setminus \fun^{-1}(\{0\})$. 
\item \label{cond:res6} For every point $\mpoint\in \manifold\cap((\fun\circ \blowup)^{-1}(\{0\}))$, there exist $\vecm_{\mpoint},\altvindx_\mpoint\in\naturals_0^\dimens$ and a real analytic function $\altfun_{\mpoint}(\cdot)$ that is bounded and nonvanishing on $\cube(\veczero,\epsilon_{\mpoint})$ such that
\begin{equation*}
	\abs{(\fun\circ \blowup\circ\coord_\mpoint)(\altvectr)} 
	= \altvectr^{\vecm_{\mpoint}},
	\ \text{ for all }\ \altvectr\in \cube(\veczero,\epsilon_{\mpoint})
\end{equation*}
and the determinant of the Jacobian of the mapping 
$(\blowup\circ\coord_\mpoint)(\cdot)$ satisfies
\begin{equation*}
	\det\lefto(\frac{\partial (\blowup\circ\coord_\mpoint)}{\partial \altvectr}\right)
	= \altfun_{\mpoint}(\altvectr)\altvectr^{\altvindx_\mpoint},\ \text{ for all }\ \altvectr\in \cube(\veczero,\epsilon_{\mpoint}).
\end{equation*}
\end{enumerate}
\end{thm}

Thanks to \fref{thm:Theosing}, in the \emph{neighborhood of zero}, 
every real analytic function that satisfies $\fun(\cdot)\not\equiv 0$ and $\fun(\veczero)=0$ can be written as  a product of a monomial and a \emph{nonvanishing} real analytic function. In order to bound the integral in~\fref{eq:Lineq}, we will need to represent $\fun(\cdot)$ in this form  in the neighborhood of \emph{every} point in the domain of integration. 
 This representation can be obtained by analyzing two cases separately.
For points $\vecx$ such that $\fun(\vecx)\ne 0$, by real-analyticity and, hence, continuity, it follows that $\fun(\cdot)$ is already nonvanishing in the neighborhood of $\vecx$ and is hence  trivially representable as  a product of a monomial and a \emph{nonvanishing} real analytic function. For points $\vecx$ such that $\fun(\vecx)=0$, the desired representation can be obtained by appropriately shifting the origin in \fref{thm:Theosing}. 
The following straightforward corollary to~\fref{thm:Theosing} conveniently formalizes these statements in a unified fashion.

\begin{cor}\label{cor:Lemmasing}
Let $\fun(\cdot)\not\equiv 0$ be a  real analytic function from a neighborhood of $\vectr\in\reals^\dimens$, denoted $\Omega\subseteq \reals^\dimens$,   to $\reals$. 
Then, there exists a triple $(\setW, \manifold, \blowup(\cdot))$, such that
\begin{enumerate}[(a)]
\item $\setW\subset \Omega$ is an open set in $\reals^\dimens$ with $\vectr\in \setW$, \label{cond:rescor1}
\item $\manifold$ is a $\dimens$-dimensional real analytic manifold \cite[Def. 2.10]{watanabe09} with coordinate charts $\{\manifold_{\mpoint},\coord_{\mpoint}:\cube(\veczero,\epsilon_{\mpoint})\to \manifold_{\mpoint}\}$ for each point $\mpoint\in \manifold$, where $\manifold_{\mpoint}$ is an open set with $\mpoint\in\manifold_{\mpoint}$ and $\coord_{\mpoint}(\cdot)$ is an isomorphism between $\cube(\veczero,\epsilon_{\mpoint})$ and $\manifold_{\mpoint}$ with $\coord_{\mpoint}(\veczero)=\mpoint$. \label{cond:rescor2}
\item $\blowup: \manifold\to \setW$ is a real analytic map, \label{cond:rescor3}
that satisfies the following conditions:
\end{enumerate}
\begin{enumerate}[(i)]
\item \label{cond:rescor4} The map $\blowup(\cdot)$ is proper, i.e., the inverse image of any compact set under $\blowup(\cdot)$ is compact.
\item \label{cond:rescor7}
The map $(\blowup\circ\coord_{\mpoint})(\cdot)$ is an isomorphism between $\cube(\veczero,\epsilon_\mpoint)\setminus (\fun\circ \blowup\circ\coord_{\mpoint})^{-1}(\{0\})$ and $\blowup(\manifold_\mpoint)\setminus\fun^{-1}(\{0\})$.
\item \label{cond:rescor6} For every point $\mpoint\in \manifold$, there exist $\vecm_{\mpoint},\altvindx_\mpoint\in\naturals_0^\dimens$ and real analytic functions $\aaltfun_\mpoint(\cdot)$ and $\altfun_\mpoint(\cdot)$ that are bounded  and nonvanishing on $\cube(\veczero,\epsilon_{\mpoint})$ such that  
\begin{equation}
	\abs{(\fun\circ\blowup\circ\coord_\mpoint)(\altvectr)} 
	= \aaltfun_\mpoint(\altvectr)\altvectr^{\vecm_\mpoint},
	\ \text{ for all } \ \altvectr\in \cube(\veczero,\epsilon_\mpoint)
	\label{eq:resolcor1}
\end{equation}
and the determinant of the Jacobian of the  mapping\\
$(\blowup\circ\coord_\mpoint)(\cdot)$ satisfies 
\begin{equation*}
	\det\lefto(\frac{\partial (\blowup\circ\coord_\mpoint)}{\partial \altvectr}\right)
	= \altfun_\mpoint(\altvectr)\altvectr^{\altvindx_\mpoint},\ \text{ for all }\ \altvectr\in \cube(\veczero,\epsilon_\mpoint).
\end{equation*}
\end{enumerate}
\end{cor}
\begin{IEEEproof} 
First consider $\vectr$ such that $\fun(\vectr)\ne 0$. As already mentioned, in this case the statement of the corollary is a pure formality since $\fun(\cdot)$ itself is a nonvanishing real analytic function in the neighborhood of $\vectr$. Formally, since $\fun(\cdot)$ is real analytic and, hence, continuous, there exists an open cube $\cube(\vectr,\epsilon)$ on which $\fun(\cdot)$ is uniformly bounded and satisfies $\fun(\altvectr)\ne 0$ for all $\altvectr\in\cube(\vectr,\epsilon)$. In this case, the corollary, therefore, follows immediately by choosing $\manifold\define\cube(\vectr,\epsilon)$, $\setW\define\cube(\vectr,\epsilon)$, setting $\blowup(\cdot)$ to be the identity map, defining $\manifold_\mpoint\define\manifold$ for all $\mpoint\in\manifold$, and setting $\coord_\mpoint(\altvectr)\define \altvectr+\mpoint$ for all $\altvectr\in\cube(\veczero,\epsilon)$.

Next, consider the more complicated case $\fun(\vectr)= 0$. The main idea is to apply \fref{thm:Theosing} to the function 
$\tilde \fun(\aaltvectr)\define \fun(\aaltvectr+\vectr),\ \aaltvectr\in\Omega-\vectr$. 
\fref{thm:Theosing} implies that there exists a triple $(\tilde\setW,\tilde \manifold, \tilde \blowup)$ that satisfies  \eqref{cond:res1}--\eqref{cond:res3} and \eqref{cond:res4}--\eqref{cond:res6} in \fref{thm:Theosing} for $\tilde \fun(\cdot)$. Now let 
\begin{align*}
\setW&\define \tilde \setW+\vectr\\
\manifold&\define \tilde \manifold\\
\blowup(\cdot)&\define\tilde\blowup(\cdot)+\vectr.
\end{align*}
Then \eqref{cond:rescor1}--\eqref{cond:rescor3} and \eqref{cond:rescor4} in the statement of~\fref{cor:Lemmasing} follow immediately from \eqref{cond:res1}--\eqref{cond:res3} and \eqref{cond:res4} 
in \fref{thm:Theosing}. 

Condition \eqref{cond:rescor7} in the statement of~\fref{cor:Lemmasing} follows from \eqref{cond:res5} in \fref{thm:Theosing} and the fact that $\coord_\mpoint(\cdot)$ is an isomorphism between $\cube(\veczero,\epsilon_{\mpoint})$ and $\manifold_\mpoint$.

To verify \eqref{cond:rescor6} in the statement of~\fref{cor:Lemmasing}, consider the following two cases separately.
 First, let $\mpoint\in \manifold$ such that $(\fun\circ\blowup)(\mpoint)=0$. Then \eqref{cond:rescor6} in the statement of~\fref{cor:Lemmasing} follows from \eqref{cond:res6} in \fref{thm:Theosing}  and the fact that 
\begin{align*} 
(\fun\circ\blowup\circ\coord_\mpoint)(\altvectr)
&=(\tilde f\circ\tilde\blowup\circ\coord_\mpoint)(\altvectr),\ \text{ for all } \ \altvectr\in \cube(\veczero,\epsilon_{\mpoint})\\
\det\lefto(\frac{\partial (\blowup\circ\coord_\mpoint)}{\partial \altvectr}\right) 
&=\det\lefto(\frac{\partial (\tilde\blowup\circ\coord_\mpoint)}{\partial \altvectr}\right) 
,\ \text{ for all } \ \altvectr\in \cube(\veczero,\epsilon_{\mpoint}).
\end{align*}
Second, let $\mpoint\in \manifold$ with $(\fun\circ\blowup)(\mpoint)\ne0$. As $(\tilde \fun\circ \tilde \blowup)(\mpoint)=(\fun\circ\blowup)(\mpoint)$,
 this implies that $(\tilde \fun\circ \tilde \blowup)(\mpoint)\neq 0$. Since $\tilde \fun(\cdot)$ is a continuous function (as  a translation of $\fun(\cdot)$ that is real analytic and hence continuous), there exists an $\epsilon_{\mpoint}>0$ such that $\tilde \fun(\cdot)$ is bounded and nonvanishing  on the open cube $\cube(\tilde \blowup(\mpoint),\epsilon_{\mpoint})$. 
Now \eqref{cond:res5} in \fref{thm:Theosing} implies that $\tilde\blowup(\cdot)$ is an isomorphism, i.e.,
\begin{align*}
\tilde\blowup: \tilde\blowup^{-1}(\cube(\tilde \blowup(\mpoint),\epsilon_{\mpoint}))\to \cube(\tilde \blowup(\mpoint),\epsilon_{\mpoint}).
\end{align*}
Define 
$\coord_\mpoint(\altvectr)\define \tilde\blowup^{-1}(\altvectr+\tilde\blowup(\mpoint)) \text{ for } \altvectr\in \cube(\veczero,\epsilon_{\mpoint})$. 
Then $\coord_\mpoint(\veczero)=\mpoint$ and
\begin{align*}
f(\blowup\circ\coord_\mpoint)(\altvectr)
&=(\tilde \fun\circ \tilde\blowup\circ\coord_\mpoint)(\altvectr)\\
&=\tilde \fun( \altvectr+\tilde\blowup(\mpoint)), \ \text{ for all } \ \altvectr\in \cube(\veczero,\epsilon_{\mpoint}).
\end{align*}
Therefore, we can simply set $\aaltfun_\mpoint(\altvectr)\define\tilde\fun(\altvectr+\tilde\blowup(\mpoint))$ and the representation~\fref{eq:resolcor1} is obtained. 
Furthermore, since $\blowup(\coord_\mpoint(\altvectr))=\tilde\blowup(\coord_\mpoint(\altvectr))+\vectr=\tilde\blowup(\tilde\blowup^{-1}(\altvectr+\tilde\blowup(\mpoint)))+\vectr=\altvectr+\tilde\blowup(\mpoint)+\vectr$, we have
\begin{equation*}
	\det\lefto(\frac{\partial (\blowup\circ\coord_\mpoint)}{\partial \altvectr}\right)=1
	,\ \text{ for all } \ \altvectr\in \cube(\veczero,\epsilon_{\mpoint}).	
\end{equation*}
\end{IEEEproof}

We now have all the ingredients required to prove \fref{thm:TheoLin}.

\begin{IEEEproof}
For each $\vectr\in\Delta$,   
\fref{cor:Lemmasing} 
implies that there exists a triple
 $(\setW_{\vectr},\manifold_{\vectr},\blowup_{\vectr})$ such that 
$\setW_{\vectr}\subseteq\Omega$ is an open set containing $\vectr$, $\manifold_{\vectr}$ is a real analytic manifold, and 
$\blowup_{\vectr}:\manifold_{\vectr}\to \setW_{\vectr}$ is a proper map. Furthermore, 
for each $\mpoint\in \manifold_{\vectr}$ there exists a  coordinate chart $\{\manifold_{\vectr,\mpoint},\coord_{\vectr,\mpoint}:\cube(\veczero,\epsilon_{\vectr,\mpoint})\to \manifold_{\vectr,\mpoint}\}$, where $\manifold_{\vectr,\mpoint}$ is an open set with $\mpoint\in\manifold_{\vectr,\mpoint}$ and $\coord_{\vectr,\mpoint}(\cdot)$ is an isomorphism between $\cube(\veczero,\epsilon_{\vectr,\mpoint})$ and $\manifold_{\vectr,\mpoint}$ with $\coord_{\vectr,\mpoint}(\veczero)=\mpoint$, such that $(\blowup_\vectr\circ\coord_{\vectr,\mpoint})(\cdot)$ is a real analytic map~\cite[p.49]{watanabe09} on  $\cube(\veczero,\epsilon_{\vectr,\mpoint})$ and
\begin{equation*}
\begin{split}
\abs{(\fun\circ\blowup_{\vectr}\circ\coord_{\vectr,\mpoint})(\altvectr)}=\aaltfun_{\vectr,\mpoint}(\altvectr)\altvectr^{\vindx_{\vectr,\mpoint}}\\
\det\lefto(\frac{\partial (\blowup_{\vectr}\circ\coord_{\vectr,\mpoint})}{\partial \altvectr}\right) = \altfun_{\vectr,\mpoint}(\altvectr)\altvectr^{\altvindx_{\vectr,\mpoint}}
\end{split}
\end{equation*}
for all $\altvectr\in  \cube(\veczero,\epsilon_{\vectr,\mpoint})$, where $\altfun_{\vectr,\mpoint}(\cdot)$ and $\aaltfun_{\vectr,\mpoint}(\cdot)$ are real analytic functions that are nonvanishing
on $\cube(\veczero,\epsilon_{\vectr,\mpoint})$. 
Now, for each $\vectr\in\Delta$ we choose an open neighborhood of $\vectr$, denoted as $\setW'_{\vectr}$, and a compact neighborhood of $\vectr$, denoted $\Delta_{\vectr}$, such that $\vectr\in \setW'_{\vectr}\subset \Delta_{\vectr} \subset \setW_{\vectr}$.
Since $\Delta$ is a compact set~\cite[2.31]{rudin76} there exists a finite set of vectors $\{\vectr_{1},\dots,\vectr_{N}\}$ with 
$\vectr_{i}\in\Delta$ such that 
\begin{align*}
\Delta\ \ \subset\! \bigcup_{i\in \natseg{1}{N}}\! \setW'_{i}\ \ \subset\! \bigcup_{i\in \natseg{1}{N}}\!\Delta_{i},
\end{align*}
where we set $\setW'_{i}\define \setW'_{\vectr_{i}}$ and $\Delta_{i}\define \Delta_{\vectr_{i}}$ for  $i\in \natseg{1}{N}$.
Take an  $i\in\natseg{1}{N}$ and set $\manifold_{i}\define \manifold_{\vectr_{i}}$, $\setW_{i}\define \setW_{\vectr_{i}}$, and 
$\blowup_{i}\define\blowup_{\vectr_{i}}$. Since 
the mapping $\blowup_{i}:\manifold_{i}\to \setW_{i}$ is proper,  
the set 
${\blowup_{i}}^{-1}(\Delta_{i})\subset \manifold_{i}$ is a compact set. Therefore, there exists a finite number $M_i$ of points 
$\mpoint_{1},\dots,\mpoint_{M_{i}}\in \manifold_{i}$ such that 
\begin{align}\label{eq:compactinU}
{\blowup_{i}}^{-1}(\Delta_{i})\ \ \subset\!\bigcup_{j\in\natseg{1}{M_{i}}}\!\manifold_{i,j} 
\end{align}
with $\manifold_{i,j}\define \manifold_{\vectr_{i},\mpoint_{j}}$. Since \eqref{eq:compactinU} holds for all $i\in\natseg{1}{N}$, 
we can upper-bound the integral in \eqref{eq:Lineq} as follows:
\begin{align}
&\int_{\Delta} \abs{\log (\abs{f(\vectr)})}d\vectr\nonumber\\ 
&\qquad\leq \sum_{i\in \natseg{1}{N}}\int_{\Delta_{i}} \abs{\log (\abs{f(\vectr)})}d\vectr\nonumber\\ 
&\qquad\leq
\sum_{i\in \natseg{1}{N}}
\sum_{j\in \natseg{1}{M_{i}}}
\phantom{\leq}
\int_{\Delta_{i}\cap \blowup_i(\manifold_{i,j})} \abs{\log (\abs{\fun(\vectr)})}d\vectr\nonumber\\ 
&\qquad\leq
\sum_{i\in \natseg{1}{N}}
\sum_{j\in \natseg{1}{M_{i}}}
\int_{\blowup_i(\manifold_{i,j})} \abs{\log (\abs{\fun(\vectr)})}d\vectr.\label{eq:firstpartintbound}
\end{align}
Since $\fun(\cdot)$ is a real analytic function and, hence, $\fun^{-1}(\{0\})$ is a set of measure zero, we have
\begin{equation}
	\int_{\blowup_i(\manifold_{i,j})}\!\!\! \abs{\log (\abs{\fun(\vectr)})}d\vectr=\int_{\blowup_i(\manifold_{i,j})\setminus\fun^{-1}(\{0\})}\!\!\! \abs{\log (\abs{\fun(\vectr)})}d\vectr. \label{eq:boundintint}
\end{equation}
Next, recall that according to \eqref{cond:rescor7} in \fref{cor:Lemmasing} $(\blowup_i\circ\coord_{\mpoint_j})(\cdot)$ is an isomorphism between $\cube_{i,j}\define\cube(\veczero,\epsilon_{\vectr_i,\mpoint_j})\setminus (\fun\circ \blowup_i\circ\coord_{\mpoint_j})^{-1}(\{0\})$ and $\blowup_i(\manifold_{i,j})\setminus\fun^{-1}(\{0\})$.
Therefore, we can apply the change of variables theorem~\cite[Theorem 7.26]{rudin87} to get
\begin{align}
&\int_{\blowup_i(\manifold_{i,j})\setminus\fun^{-1}(\{0\})} \abs{\log (\abs{\fun(\vectr)})}d\vectr\nonumber\\
& =
\int_{\cube_{i,j}}
\abs{\altfun_{i,j}(\altvectr)\altvectr^{\altvindx_{i,j}}
\log\big(\abs{\aaltfun_{i,j}(\altvectr)\altvectr^{\vindx_{i,j}}}\big)}d\altvectr\nonumber\\
&\le
\underbrace{
\sup_{\altvectr\in\cube_{i,j}}(\abs{\altfun_{i,j}(\altvectr)\altvectr^{\altvindx_{i,j}}})}_{\constalt_{i,j}} 
\int_{\cube_{i,j}}
\abs{\log\big(\abs{\aaltfun_{i,j}(\altvectr)\altvectr^{\vindx_{i,j}}}\big)}d\altvectr\nonumber\\
&\stackrel{(a)}{=}
\constalt_{i,j}
\int_{\cube_{i,j}}
\abs{\log\big(\abs{\altvectr^{\vindx_{i,j}}}\big)}d\altvectr\nonumber\nonumber\\
&\quad+\constalt_{i,j}
\int_{\cube_{i,j}}
\abs{
\log\big(\abs{\aaltfun_{i,j}(\altvectr)}\big)}d\altvectr\nonumber\\
&\stackrel{(b)}{\le}
\constalt_{i,j}
\int_{-\epsilon_{i,j}}^{\epsilon_{i,j}}\!\!\!\!\dots\int_{-\epsilon_{i,j}}^{\epsilon_{i,j}} \Big|\sum_{\dimeni=1}^\dimens[\vindx_{i,j}]_\dimeni
 \log\big(\abs{\altvectrc_\dimeni}\big)\Big|d \altvectrc_1\dots d\altvectrc_\dimens\nonumber\nonumber\\
&\quad+\underbrace{\constalt_{i,j}
\sup_{\altvectr\in\cube_{i,j}}\lefto(\abs{\log\big(\abs{\aaltfun_{i,j}(\altvectr)}\big)}\right)
(2\epsilon_{i,j})^\dimens}_{\hat\constalt_{i,j}}\nonumber\\
&\stackrel{(c)}{\le}
\constalt_{i,j}
\sum_{\dimeni=1}^\dimens[\vindx_{i,j}]_\dimeni \int_{-\epsilon_{i,j}}^{\epsilon_{i,j}}\!\!\!\!\dots\int_{-\epsilon_{i,j}}^{\epsilon_{i,j}} 
 \abs{\log\big(\abs{\altvectrc_\dimeni}\big)}d\altvectrc_1\dots d\altvectrc_\dimens+\hat\constalt_{i,j}\nonumber\\
&=
\underbrace{\constalt_{i,j}
\sum_{\dimeni=1}^\dimens[\vindx_{i,j}]_\dimeni (2\epsilon_{i,j})^{(\dimens-1)}\int_{-\epsilon_{i,j}}^{\epsilon_{i,j}}
 \abs{\log\big(\abs{\altvectrc}\big)}d\altvectrc}_{\tilde\constalt_{i,j}}+\hat\constalt_{i,j}\nonumber\\
&\stackrel{(d)}{<}\infty.\label{eq:boundintfinal}
\end{align}
Here, $c_{i,j},\tilde c_{i,j}, \hat c_{i,j}>0,\,  i\in \natseg{1}{N},\, j\in \natseg{1}{M_{i}},$ are finite constants;
in (a) we used the fact that $\altfun_{i,j}(\cdot)$ is bounded and nonvanishing on $\cube_{i,j}$; in (b) $[\vindx_{i,j}]_\dimeni$ denotes the $\dimeni$th component of the vector $\vindx_{i,j}$; in (c) we used the triangle inequality to bound the first term,  the second term is finite because $\aaltfun_{i,j}(\cdot)$ is bounded and nonvanishing on $\cube_{i,j}$; and in (d) we used $\int_{-\epsilon_{i,j}}^{\epsilon_{i,j}}\log\big(\abs{\altvectrc}\big)d\altvectrc<\infty$.
Combining~\fref{eq:firstpartintbound}, \fref{eq:boundintint}, and~\fref{eq:boundintfinal}, we complete the proof.
\end{IEEEproof}

\bibliographystyle{IEEEtran}
\bibliography{IEEEabrv,publishers,confs-jrnls,vebib}

\end{document}